\documentclass{article}
\usepackage[english]{babel}
\usepackage[utf8]{inputenc}
\usepackage{color}

\usepackage{arxiv}

\usepackage{graphicx} 
\usepackage{amsmath} 
\usepackage{amssymb} 
\usepackage{amsthm} 
\usepackage{tabularx} 
\usepackage{siunitx} 
\usepackage{algorithm}
\usepackage{float} 
\usepackage{caption} 
\usepackage{subcaption} 
\usepackage{filecontents} 
\usepackage{epstopdf}
\usepackage[makeroom]{cancel} 
\usepackage[noend]{algpseudocode}
\usepackage{gensymb} 
\usepackage{comment}

\usepackage[T1]{fontenc}    
\usepackage{hyperref}       
\usepackage{url}            
\usepackage{booktabs}       
\usepackage{amsfonts}       
\usepackage{nicefrac}       
\usepackage{microtype}      
\usepackage{lipsum}		
\usepackage[square,numbers]{natbib}
\usepackage{doi}

\usepackage[sc,osf]{mathpazo}   
\everymath{\displaystyle}
\graphicspath{{img/}{Figures/}}

\theoremstyle{remark}

\title{Comparative study and limits of different level-set formulations for the modeling of anisotropic grain growth}

\author{ \href{https://orcid.org/0000-0002-6513-7505}{\includegraphics[scale=0.06]{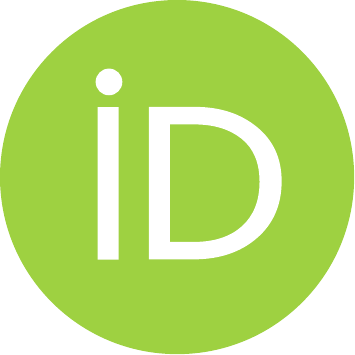}\hspace{1mm}Brayan ~Murgas} 
   \thanks{Corresponding author: brayan.murgas@mines-paristech.fr} $^1$,
	\href{https://orcid.org/0000-0002-5962-7700}{\includegraphics[scale=0.06]{orcid.pdf}\hspace{1mm}Sebastian ~Florez}$^1$,
	\href{https://orcid.org/0000-0002-8963-977X}{\includegraphics[scale=0.06]{orcid.pdf}\hspace{1mm}Nathalie ~Bozzolo}$^1$,
	\href{https://orcid.org/0000-0001-9911-1859}{\includegraphics[scale=0.06]{orcid.pdf}\hspace{1mm}Julien ~Fausty}$^1$,
	\href{https://orcid.org/0000-0002-6677-2850}{\includegraphics[scale=0.06]{orcid.pdf}\hspace{1mm}Marc ~Bernacki}$^1$ \\
	\\
	$^1$ Mines-ParisTech, PSL-Research University, CEMEF – Centre de mise en forme des mat\'{e}riaux, CNRS UMR 7635\\
	CS 10207 rue Claude Daunesse, 06904 Sophia Antipolis Cedex, France \\
}

\renewcommand{\headeright}{}
\renewcommand{\undertitle}{}
\renewcommand{\shorttitle}{Comparative study of FE-LS formulations for the modeling of anisotropic GG}

\begin{document}
\maketitle

\begin{abstract}

  Four different finite element level-set (FE-LS) formulations are compared for the modeling of grain growth in the context of polycrystalline structures and, moreover, two of them are presented for the first time using anisotropic grain boundary (GB) energy and mobility. Mean values and distributions are compared using the four formulations. First, we present the strong and weak formulations for the different models and the crystallographic parameters used at the mesoscopic scale. Second, some Grim Reaper analytical cases are presented and compared with the simulation results, here the evolutions of individual multiple junctions are followed. Additionally, large scale simulations are presented. Anisotropic GB energy and mobility are respectively defined as functions of the misorientation/inclination and disorientation. The evolution of the disorientation distribution function (DDF) is computed and its evolution is in accordance with prior works. We found that the formulation called "Anisotropic" is the more physical one but it could be replaced at the mesoscopic scale by an Isotropic formulation for simple microstructures presenting an initial Mackenzie-type DDF. 

\end{abstract}

\keywords{Heterogeneous Grain Growth \and Grain Boundary Energy \and Grain Boundary Mobility \and Finite Element Analysis \and Level-Set method }

\section{Introduction}

The study of GB \textit{Thermodynamics} and \textit{Kinetics} are two fundamental topics in materials science. The study of thermodynamics provides information about a system at equilibrium; its extrapolation, under the assumption of local equilibrium, provides the basis for kinetic theories. Additionally, kinetics approaches study the evolution of systems out of equilibrium involving changes in the microstructure. Determining the kinetics of recovery, grain growth (GG), recrystallization, solidification and other metallurgical mechanisms is necessary to predict and optimize material properties \cite{humphreys2012recrystallization}. The need for high-performance materials demands a better knowledge and control of the behavior of GBs under thermomechanical loads. This topic became a strong issue of materials science and gave rise to a branch called GB engineering \cite{watanabe2011grain}. 

In the context of GG, the evolution of GB is driven by the reduction of interfacial energy and its velocity is classically described, at the mesoscopic scale, by the well known equation $v=\mu P$ where $\mu$ is the GB mobility and $P=-\gamma\kappa$ is the curvature flow driving pressure with $\gamma$ the GB energy and $\kappa$ the mean curvature (i.e. the trace of the curvature tensor in 3D). If this kinetic equation is a simplification of lower scale phenomena in constant discussions \citep{PhysRevLett.119.246101,zhu2019situ}, it constitutes at the polycrystalline scale and in metal forming state-of-the-art a kinematically accepted physical framework. In the discussion that this kinetic equation is a reasonable approximation \cite{wang2019} and is the reduced mobility ($\mu\gamma$ product) could be only considered as defined by the temperature and macroscopic properties of the interface as misorientation and inclination? A clear and univocal answer seems complicated today.
First of all, the answer at the few interfaces scale and at the homogenized polycrystal scale can be contradictory by the statistical effects. 
Moreover, a bias in the reduced mobility field discussion lies today in the real capacity of full-field methods to take into account a reduced mobility defined properly in the 5D space defined by the misorientation and inclination in representative 2D or 3D simulations. As detailed below, such a capacity is typically unclear in the current state of the art.
Thus the discussion between experimental data and anisotropic full-field simulations is to be treated with extreme caution. \medbreak

If numerical modeling by considering heterogeneous values of GB mobility and GB energy remains a complex discussion, it has been in fact widely studied at the polycrystalline scale with a large variety of numerical approaches: multi phase-field \citep{garcke1999multiphase, miyoshi2017multi, moelans2009comparative}, Monte Carlo \citep{gao1996real, upmanyu2002boundary}, molecular dynamics \citep{hoffrogge2017grain}, vertex \citep{BarralesMora2010}, front-tracking Lagrangian or Eulerian formulations in a FE context \citep{wakai2000,Florez2020,Florez2020b} and FE-LS \citep{bernacki2011level, miessen2015advanced, Fausty2020}, to cite some examples. During annealing two properties have been widely studied: the GB energy and mobility. The first models proposed in the literature define the GB mobility and energy as constants, carrying the name of Isotropic, \cite{anderson1984computer, gao1996real, lazar2011more, bernacki2011level, garcke1999multiphase}, this category shows good agreement in terms of mean quantities and distributions, nevertheless, they are restrictive in terms of the grains morphology and texture predictions. GB energy and mobility have been early reported as anisotropic by Smith \citep{Smith1948introduction} and Kohara \citep{kohara1958anisotropy}. Hence, the models have evolved in order to reproduce more complex microstructures or local heterogeneities, such as twin boundaries. Heterogeneous models were proposed, in which each boundary has its own energy and mobility \citep{rollett1989simulation, hwang1998simulation, upmanyu2002boundary, Fausty2018, zollner2019texture, miyoshi2016validation, chang2019effect, miyoshi2019accuracy, miessen2015advanced, Fausty2020, holm2001misorientation}. For instance, every grain could be related with an orientation, thus the mobility and energy can be computed in terms of the disorientation \citep{miyoshi2017multi, Fausty2020} but the misorientation axis and inclination dependence are frequently not taken into account. Finally, general frameworks in which the five parameters, misorientation and inclination, are discussed have been proposed, these models could be categorized as fully anisotropic \citep{kazaryan2002grain, fausty2020geo, hallberg2019modeling}. 

However, it must be highlighted that the distinction between 3-parameters and 5-parameters full-field frameworks is not straightforward. In the literature heterogeneous values of GB properties have been often categorized as anisotropic. For instance, in \citep{holm2001misorientation, elsey2013simulations, chang2019effect, miyoshi2017multi} heterogeneous GB energy and a constant GB mobility to model polycrystal evolution during GG are considered, and the models are categorized as anisotropic even if it is assumed that the GB energy does not depend on the GB normal direction and the GB mobility is not heterogeneous. In \citep{hallberg2019modeling}, the proposed level-set formulation in context of regular grids includes the effect of anisotropic GB energy into the driving force term ($P$) using both the effect of the misorientation and the inclination in a GB energy gradient. However, the GB energy dependence on the normal direction is defined without inquiring if additional torque terms in solved equations are needed. 

Due to the wide variety of formulations, this paper aims to compare four different formulations within a FE-LS approach, the former is an isotropic formulation frequently used in different contexts such as GG, recrystallization, GG with second phase particles \citep{Bernacki2008, Bernacki2009, Bernacki2011, scholtes2015new, maire2018fullfield}, which will be referred as the Isotropic model in the following. The second one is a simple extension of the isotropic formulation by considering non-homogeneous values of the reduced mobility.  The third formulation was firstly proposed in \citep{Fausty2018} and extended to polycrystals using different models of GB energy in \citep{Fausty2020}. The last formulation is based on a more robust thermodynamics and differential geometry framework but was only applied, at yet, to a bicrystal-like geometry \citep{fausty2020geo}. Another particularity of the discussed approaches is to be usable on unstructured finite element mesh and in the context of large deformations and displacements. The goals of this work are to criticize these existing formulations but also to consider the enrichment of GB mobility in FE-LS framework. First, some crystallographic definitions, level-set treatments and the formulations are introduced in section~\ref{sec:FrameworkDevelopment}. In section~\ref{sec:GrimReaper}, simulation results are compared with analytical solutions in the context of simple triple junction geometries. In section~\ref{sec:PX}, polycrystalline simulations are studied. Mean values and statistical quantities are compared with two different initial textures and using heterogeneous GB energy and mobility. Finally the last section is dedicated to the inclination dependence discussions.

\section{The numerical framework} \label{sec:FrameworkDevelopment}

Before formulating the equations related to GG, the constituents of polycrystalline materials and specially GB structures must be defined.

\subsection{Crystallographic definitions}

Let us consider a domain $\Omega$ of dimension $d$ filled by $n$ grains $G_i \in \Omega$, being open spaces of $\Omega$ and defining the set of grains $\mathbf{ \mathcal{G} } = \{ G_i,  i=1,\dotsc,N_G \} $. The interface between two neighboring grains $G_i$ and $G_j$ constitutes a GB $B_{ij}$, the whole set of boundaries form the GB network $\Gamma$. A boundary $B_{ij}$ is characterized by its morphology and its crystallographic properties, which are described using five variables: $2$ shape properties, describing the interfaces by the unitary-outward normal direction $n_{ij}$, and $3$ crystallographic properties describing the orientation relationship between the two adjacent grains, $O_i$ and $O_j$, known as the misorientation $M_{ij}$. As such, at the mesoscopic scale each boundary may be characterized by a tuple:            

\begin{align*}
    B_{ij} = (M_{ij},n_{ij}).
\end{align*}

\begin{figure}[h]
  \centering
  \includegraphics[scale=0.15]{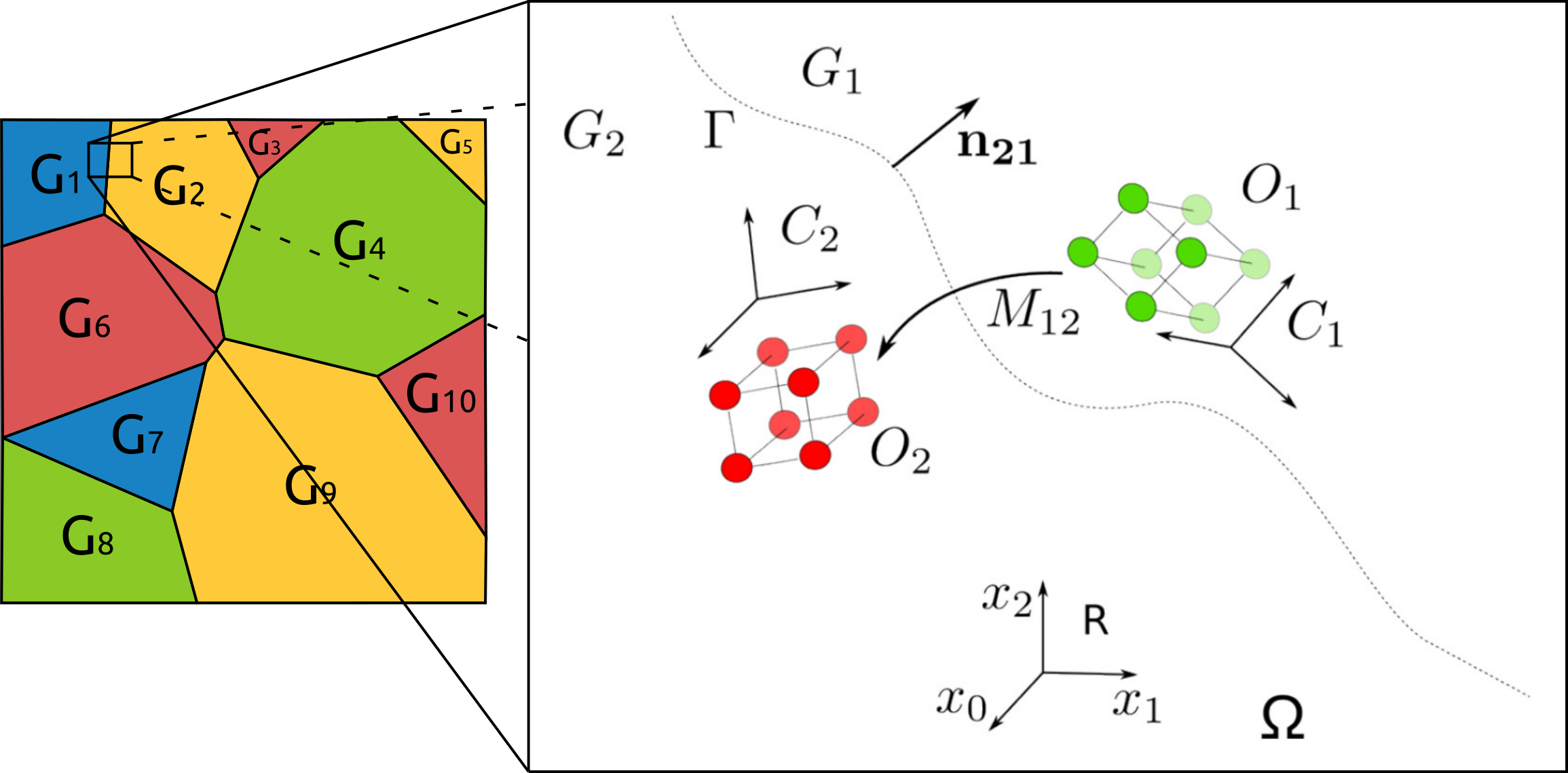}
  \caption{Scheme depicting one GB and its parameters.}\label{fig:GBparameters}
\end{figure}

The GB space $\mathcal{B}$ parameterized by the misorientation and the normal direction is illustrated in figure~\ref{fig:GBparameters}. The two quantities of interest, the GB energy $\gamma$ and the mobility $\mu$, are maps from $\mathcal{B}$ to $\mathbb{R}^{+}$. 

\subsection{Level-set finite element formulation}

The LS method is a powerful tool firstly proposed by Osher and Sethian \citep{Osher1988} to describe curvature flow of interfaces, enhanced later for evolving multiple junctions \citep{Merriman1994, Zhao1996}, and considered in recrystallization and grain growth problems in \citep{Bernacki2008, Bernacki2011}. The principle for modeling polycrystals is the following: the grain interfaces are defined through scalar fields called level-set functions $\phi$ in the space $\Omega$ and more precisely by the zero-isovalue of the $\phi$ functions. Level-set functions to the interfaces are classically initialized as the sign euclidean distance functions to these interfaces:  

\begin{align}
  \left\{
  \begin{array}{l}
    \phi(X) = \pm d(X,\Gamma), \quad X \in \Omega,\quad \Gamma=\partial G  \\
    \phi(X \in \Omega) = 0 \quad \rightleftharpoons \quad X \in \Gamma.
\end{array}
\label{eqn:LS}
\right .
\end{align}
with $d$ the Euclidean distance and $\phi$ generally defined as positive inside the grain and negative outside. The dynamics of the interface is studied by following the evolution of the level-set field. The interface may be subjected to an arbitrary velocity field $\vec{v}$ and its movement is described by solving the transport equation:     

\begin{align}
    \dfrac{\partial \phi}{\partial t} + \vec{v} \cdot \vec{\nabla} \phi = 0.
    \label{eqn:transport}
\end{align}

The flexibility of this method lies in the ability to define different physical phenomena encapsulated in the velocity field. This equation is solved to describe the movement of every grain. When the number of grains increases one may use a graph coloring/recoloring strategy \citep{scholtes2015new} in order to limit drastically the number of involved LS functions $\Phi = \{ \phi_i, i=1, \dotsc, N \}$ with $N \ll N_G$ being $N_G$ the number of grains. Additionally, two more treatments are necessary. Firstly, the LS functions are reinitialized at each time step to keep the metric property of a distance function: 

\begin{align}
    \| \nabla \phi \| = 1.
    \label{eqn:normLS}
\end{align}
Secondly, the evolution may not preserve the impenetrability constraints of the LS functions leading to overlaps and voids between grain interfaces. These events are corrected after solving the transport equation by resolving Eq.~\ref{eqn:voidsLS} as proposed in \citep{Merriman1994} and classically used in LS framework \cite{Bernacki2011,Scholtes2016}.

\begin{align}
    \phi_i(X) = \dfrac{1}{2} \left[  \phi_i(X) - \max_{j \ne i} \phi_j(X)  \right], \quad \forall i=\{ 1, \dotsc, N \}. 
    \label{eqn:voidsLS}
\end{align}

Several formulations using the LS framework exist in the literature. The initial GG formulation uses a homogeneous grain boundary energy and mobility, i.e. $\gamma ( \cancel{M}, \cancel{n} )$ and $\mu ( \cancel{M}, \cancel{n} ) $ \citep{bernacki2011level}, the velocity field is thus defined as :

\begin{align}
    \vec{v} = \mu P \vec{n}=- \mu \gamma \kappa \vec{n},
    \label{eqn:VelFieldHomo}
\end{align}
with $P=-\gamma \kappa$ the capillarity pressure and $n$ the outward unitary normal to the interface. When dealing with recrystallization, supplemental terms could be added to the velocity as proposed in \cite{bernacki2011level}. If $\phi$ is defined as positive inside the grain and remains a distance function, the mean curvature and the normal may be defined as: 
\begin{align}
    \kappa = - \Delta \phi,\quad \vec{n} = - \vec{\nabla} \phi,
    \label{eqn:curvLS}
\end{align}
then the velocity in Eq.\ref{eqn:VelFieldHomo} may also be defined as:
\begin{align}
    \vec{v} = - \mu \gamma \Delta \phi \vec{\nabla} \phi.
    \label{eqn:VelFieldHomoLS}
\end{align}

Four different formulations will be studied. In the first one, an Isotropic formulation is considered by introducing Eq.~\ref{eqn:VelFieldHomoLS} into Eq.~\ref{eqn:transport}, thus the Isotropic transport equation may be defined as a pure diffusive problem: 

\begin{align}
    \dfrac{\partial \phi}{\partial t} - \mu \gamma \Delta \phi = 0.
    \label{eqn:transportClassic}
\end{align}
This formulation has shown good agreement with experimental data regarding GG predictions concerning the mean grain size and even the grain size distribution (GSD). However, this approach is limited to reproduce complex grain morphology (non-equiaxed ones), described special grain boundaries  and to respect textures. This formulation could be slightly modified in a second one with the introduction of heterogeneous GB properties leading to a Heterogeneous formulation:
\begin{align}
    \dfrac{\partial \phi}{\partial t} - \mu(M) \gamma(M) \Delta \phi = 0.
    \label{eqn:transportClassicHet}
\end{align}
With this formulation, it is expected to obtain more physical grain shapes. Indeed, some GBs can evolve faster thanks to higher grain boundary mobility values, and triple junctions may have different dihedral angles thanks to different GB energy values. This strategy classically used in full-field formulations (not only in LS ones) can lead to confusion when it is named as ``heterogeneous''. Indeed, \textit{stricto sensu}, the heterogeneity shape of $\mu$ and $\gamma$ can lead to additional terms in the driving pressure of the kinetic equation (Eq.~\ref{eqn:VelFieldHomo}) but also in the weak formulation derived to solve the GB motion. However, the term ``heterogeneous'' will be used in the following to distinguish this formulation from the purely isotropic model. 

Such discussion is described in \citep{Fausty2018} where an additional term capturing the local heterogeneity of the multiple junctions is added to the velocity equation such that: 
\begin{align}
    \vec{v} = \mu ( \vec{\nabla} \gamma \cdot \vec{\nabla} \phi - \gamma \Delta \phi ) \vec{\nabla} \phi.
    \label{eqn:VelFieldFull}
\end{align}
Inserting this term into the transport equation (Eq. \ref{eqn:transport}) leads to the, hereafter called, ``Heterogeneous with Gradient'' formulation \citep{Fausty2018}: 
\begin{align}
    \dfrac{\partial \phi}{\partial t} + \mu \vec{\nabla} \gamma \cdot \vec{\nabla} \phi - \mu \gamma \Delta \phi = 0.
    \label{eqn:transportFull}
\end{align}
The introduction of the term $\vec{\nabla} \gamma \cdot \vec{n}$ only acts at multiple junctions because these are the only places where this term does not vanish. This formulation is equivalent to the Isotropic one if no heterogeneity is added.

Finally, in \citep{fausty2020geo} using differential geometry and thermodynamics, a new relation for the velocity was proposed taking into account the five crystalline parameters, with an intrinsic torque term which leads to the following transport equation:
\begin{align}
    \dfrac{\partial \phi}{\partial t} + \mu \mathbb{P} \vec{\nabla} \gamma \cdot \vec{\nabla} \phi - \mu \dfrac{\partial^2 \gamma}{\partial (\vec{\nabla} \phi)^2 } \Delta \phi - \mu \gamma \Delta \phi = 0,
    \label{eqn:transportAniso}
\end{align}
where $\mathbb{P}=Id-\vec{n}\otimes\vec{n}$ is the tangential projection tensor and $\mu \dfrac{\partial^2 \gamma}{\partial (\vec{\nabla} \phi)^2 } \Delta \phi$ an additional torque term. With this formulation the 5D-GB space $\mathcal{B}$ is fully described and is referred as ``Anisotropic-5''. If the torque term is neglected, the formulation used could be simplified as:
\begin{align}
    \dfrac{\partial \phi}{\partial t} + \mu \mathbb{P} \vec{\nabla} \gamma \cdot \vec{\nabla} \phi - \mu \gamma \Delta \phi = 0.
    \label{eqn:transportHetero}
\end{align}
This equation is hereafter called ``Anisotropic'' and is not equivalent to the "Heterogeneous with Gradient" formulation (Eq.\ref{eqn:transportFull}). The strong formulations used in this work are finally the ones defined by the equations \ref{eqn:transportClassic}, \ref{eqn:transportClassicHet}, \ref{eqn:transportFull}, and \ref{eqn:transportHetero}. Moreover, the effect of heterogeneous GB mobility is take into account in the weak formulations in the form of a GB mobility gradient in the Heterogeneous with Gradient and Anisotropic formulations. The weak formulations of equations \ref{eqn:transportClassic}, \ref{eqn:transportClassicHet}, \ref{eqn:transportFull} and \ref{eqn:transportHetero}, with $\varphi \in H_0^1(\Omega)$, can be summarized as :
\begin{align}
    \int_{\Omega} \dfrac{\partial \phi}{\partial t} \varphi d \Omega + \int_{\Omega} \mu \gamma \vec{\nabla}  \varphi \cdot \vec{\nabla} \phi d \Omega - \int_{\partial \Omega} \mu \gamma \varphi \vec{\nabla} \phi \cdot \vec{n} d (\partial \Omega) = 0,
\label{eqn:WeakFormClassic}
\end{align}

\begin{align}
    \int_{\Omega} \dfrac{\partial \phi}{\partial t} \varphi d \Omega + \int_{\Omega} \mu (M) \gamma (M) \vec{\nabla}  \varphi \cdot \vec{\nabla} \phi d \Omega - \int_{\partial \Omega} \mu (M) \gamma (M) \varphi \vec{\nabla} \phi \cdot \vec{n} d (\partial \Omega) = 0,
\label{eqn:WeakFormClassicHet}
\end{align}

\begin{align}
\begin{split}
    \int_{\Omega} \dfrac{\partial \phi}{\partial t} \varphi d \Omega + \int_{\Omega} \mu \gamma \vec{\nabla}  \varphi \cdot \vec{\nabla} \phi d \Omega - \int_{\partial \Omega} \mu \gamma \varphi \vec{\nabla} \phi \cdot \vec{n} d (\partial \Omega) + \\
    +2 \int_{\Omega} \mu \vec{\nabla} \gamma \cdot \vec{\nabla} \phi \varphi d \Omega + \int_{\Omega} \gamma \vec{\nabla} \mu \cdot \vec{\nabla} \phi \varphi d \Omega = 0, 
\end{split}
\label{eqn:WeakFormFull}
\end{align}

and 
\begin{align}
\begin{split}
    \int_{\Omega} \dfrac{\partial \phi}{\partial t} \varphi d \Omega + \int_{\Omega} \mu \gamma \vec{\nabla}  \varphi \cdot \vec{\nabla} \phi d \Omega + \\
    \int_{\Omega} \mu ( \mathbb{P} \cdot \vec{\nabla} \gamma + \vec{\nabla} \gamma ) \varphi \vec{\nabla} \phi d \Omega + \int_{\Omega} \vec{\nabla} \mu \gamma \cdot \vec{\nabla} \phi \varphi d \Omega = 0, 
\end{split}
\label{eqn:WeakFormAnisoSimp}
\end{align}
 \medbreak

All the formulations presented are equivalent if the properties are homogeneous, but the main question remains to test their capacity otherwise. In the next sections, a comparative study is presented. In the following, the ``Isotropic'', ``Heterogeneous'', ``Heterogeneous with Gradient'' and ``Anisotropic'' formulations will be referred as Iso, Het, HetGrad and Aniso. It must be highlighted that the formulations proposed in Eq.\ref{eqn:WeakFormFull} and Eq.\ref{eqn:WeakFormAnisoSimp} are slightly more general than those proposed in \citep{Fausty2018} and \citep{fausty2020geo}, respectively, as $\mu$ is here considered also as heterogeneous.

\section{The Grim Reaper case \citep{Garcke1999,Elsey2013}}
\label{sec:GrimReaper}
\subsection{Description of the test case}
\begin{figure}[H]
  \centering
  \includegraphics[scale=0.25]{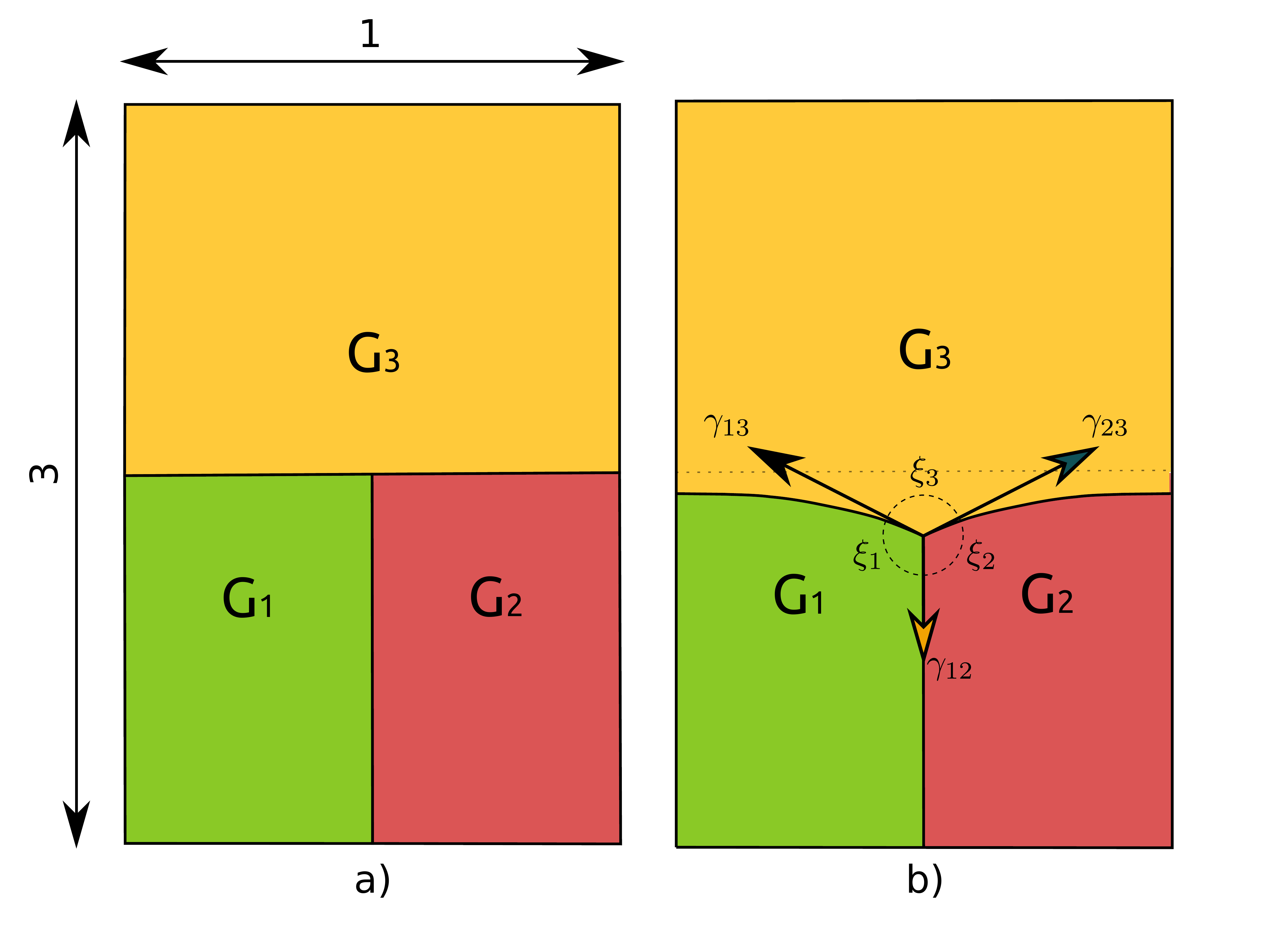}
  \caption{T-shape triple junction (a) and an illustration of the triple junction migration showing the dihedral angles and GB energies (b).}
  \label{fig:TJ}
\end{figure}

In this section, simulation results obtained with the Het, HetGrad and Aniso formulations are compared for a 2D-triple junction configuration proposed in \citep{Garcke1999} and described by Fig.\ref{fig:TJ}. The initial microstructure is a dimensionless T-shape triple junction with $L_x=1$ and $L_y=3$. This geometry is chosen because after a transient-state, a quasi-steady-state is reached where analytical relations, depending on the reduced mobility, are available for the triple junction velocity and equilibrium angles. 

When the quasi-steady-state is reached, the triple junction move with a constant velocity towards the bottom of the domain with a stable triple junction profile which respects the conditions imposed by the Herring's equation \citep{herring1999surface} :

\begin{equation}
  \sum_{j>i} \gamma_{ij} \vec{\tau}_{ij} + \dfrac{ \partial \gamma_{ij} }{ \partial \vec{\tau}_{ij} } = 0, \label{eqn:Herring}
\end{equation}
being $\gamma_{ij}$ the GB energy and $\vec{\tau}_{ij}$ the inward pointing tangent vectors of the three boundaries at the triple junction. In the present example the grain boundary energy are constant per interface ($\gamma(M)$) and the above equation may as well be expressed by the Young's law (no torque terms):

\begin{equation}
  \sum_{j>i} \gamma_{ij} \vec{\tau}_{ij} = 0, \label{eqn:Young}
\end{equation}
which may be expressed in terms of the angles $\xi_i$ of the grain $i$, through the Young's equilibrium (see figure \ref{fig:TJ}):

\begin{equation}
  \dfrac{ sin \xi_1 }{ \gamma_{23} } = \dfrac{ sin \xi_2 }{ \gamma_{13} } = \dfrac{ sin \xi_3 }{ \gamma_{12} }.    \label{eqn:YoungA}
\end{equation}

By considering an axially symmetric configuration where $\gamma_{13} = \gamma_{23} = \gamma_{top}$ and $\gamma_{12} = \gamma_{bot}$ and by defining the ratio of grain boundary energies as $r = \frac{\gamma_{top}}{\gamma_{bot}}$, an analytical value for the angle $\xi_3$ can be obtained:

\begin{equation}
  \xi_3^{ana} = 2 arccos\left( \dfrac{1}{2r} \right). \label{eqn:xi3}
\end{equation}

Moreover, the stationary transported profile takes the form of the ''Grim Reaper'' profile defined as
\begin{equation}
  \left\{
  \begin{array}{l}
    y(x,t) = g(x) + v_{TJ}^{ana} t \\
    g(x) = -\dfrac{ \mu \gamma_{top} }{ v_{TJ}^{ana} } ln \left( cos \left( \dfrac{ v_{TJ}^{ana} }{ \mu \gamma_{top} } x \right) \right) + y_0
\end{array}
\right .
\end{equation}
where $v_{TJ}^{ana}$ is the magnitude of the stationary velocity, $y_0$ the initial y-value and $(x,y)$ are the Cartesian coordinates. By using Neumann boundary conditions, the stationary velocity could be related to the x-size of the domain:
\begin{equation}
  v_{TJ}^{ana} = - \dfrac{ 2 \mu \gamma_{top} }{ L_x } \left( \dfrac{ \pi }{ 2 } - \dfrac{ \xi_3^{ana} }{ 2 } \right). \label{eqn:c}
\end{equation}

In order to focus on a considerable level of heterogeneity in the system, $r$ is initially fixed as equal to 10 ($\gamma_{top} = 1$ and $\gamma_{bot} = 0.1$). $\mu$ is defined as unitary. Several simulations were carried out and compared with the analytical values of $\xi_3^{ana}=174.27^\circ$ and $v_{TJ}^{ana} = -0.100042$. These variables are computed as follows: 

\begin{itemize}
\item The velocity of the triple junction is computed using the relation $v_{TJ} = \left(y_{TJ}^{t+\Delta t} - y_{TJ}^{t}\right)/\Delta t $ with $y_{TJ}^t$ the y-position of the triple point  at time $t$ and $\Delta t$ the time step. 

\item The dihedral angles are computed using the methodology presented in \cite{Fausty2018}: one may define, at each time, a circle of radius $\varepsilon$ with circumference $\mathcal{C}_{\varepsilon}$ and divide it into arcs which pass through grain $G_i$ with length $\mathcal{L}_{\varepsilon}^i$. The angle of the arc, $\xi_i$, could be approximated thanks to the relation $\xi_i = 2 \pi \mathcal{L}_{\varepsilon}^i / \mathcal{C}_{\varepsilon}$.  
\end{itemize}
Hence, these variables are affected by the spatial discretization of the domain and the choice of $\varepsilon$ which must be close enough to the multiple junction while containing a sufficient number of finite elements as illustrated in Fig.\ref{fig:TJdihedral} where different values of $\epsilon$ are tested. Here the value $\varepsilon=0.05$ is adopted. $v_{TJ}$ and $\xi_i$ are compared using relative errors which are defined as: 

\begin{equation*}
  e_X= \left\lvert \dfrac{X^{ana} - X}{X^{ana}} \right\rvert , \label{eqn:RelError}
\end{equation*}
where $X^{ana}$ is the analytical value of the variable to be compared. Another discussed quantity is the interfacial energy, calculated using:

\begin{equation}
  E_{\Gamma}= \sum_i \sum_{e \in \mathcal{T}} \dfrac{1}{2} \gamma l_e (\phi_i), \label{eqn:IntEnergy}
\end{equation}
where $\mathcal{T}$ is the set of all elements in the FE mesh, $l_e$ the length of the zero iso-value existing in the element $e$ and $i$ refers to the number of level-set functions, the $\dfrac{1}{2}$ is necessary due to the duplicity of the level-set functions in the interfaces defining a grain boundary. This variable is frequently studied and it may be seen as a energetic measure of how fast the system reaches equilibrium.

\subsection{Numerical strategy}
\begin{figure}[H]
  \centering
  \includegraphics[scale=0.6]{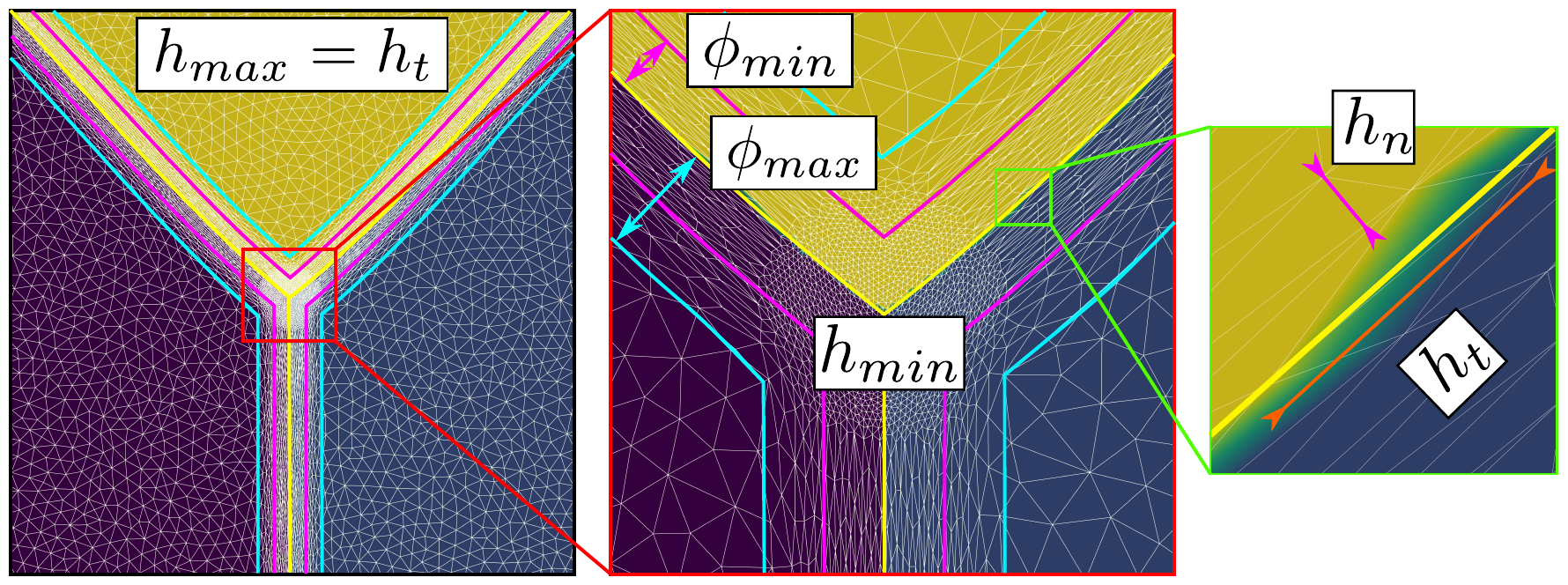}
  \caption{Illustration of the anisotropic mesh refinement \citep{Bernacki2009}.}
  \label{fig:TJMesh}
\end{figure}

The simulations presented here were carried out with unstructured triangular meshes, a P1 interpolation and using an implicit backward Euler time scheme for the time discretization. The system is assembled using typical P1 FE elements with a Streamline Upwind Petrov-Galerkin (SUPG) stabilization for the convective term \cite{Brooks1982}. The boundary conditions (BCs) are classical null-von Neumann BCs applied to the all LS functions. This choice imposes the orthogonality between the LS functions and the boundary domain (each plane of the boundary domain can be seen as a symmetric plane). By considering a minimal and maximal mesh size (resp. $h_{min}$ and $h_{max}$), an optimized anisotropic remeshing strategy developed by Bernacki et al. \cite{Bernacki2009,ROUX201332}, used in the DIGIMU$^{\mbox{\scriptsize{\textregistered}}}$ software \cite{Micheli2019} and illustrated in Fig.\ref{fig:TJMesh}, is adopted here. The mesh is finely and anisotropically refined close to the interfaces ($\phi < \phi_{min}$) and becomes isotropic when $\phi > \phi_{max}$ with a linear evolution of the normal mesh size between $\phi_{min}$ and $\phi_{max}$. A homogeneous tangential mesh size ($h_t = h_{max}$) is considered everywhere and the normal mesh size is then defined as:

\begin{equation}
  \left\{
  \begin{array}{l}
    h_n = h_{min}, \quad \phi < \phi_{min},  \\
    h_n = m (\phi - \phi_{min}) + h_{min}, \quad m=\dfrac{h_{max} - h_{min}}{\phi_{max} - \phi_{min}} \quad \phi_{min} \leq \phi \leq \phi_{max}, \\
    h_n = h_t = h_{max}, \quad \phi > \phi_{max}.
\end{array}
\label{eqn:Mesh}
\right .
\end{equation}
By generalizing this approach at the multiple junctions, a fine isotropic ($h_n = h_t = h_{min}$) remeshing is automatically performed (see \cite{Bernacki2009} for more details). During grain boundary migration, thanks to a topological mesher/remesher, anisotropic remeshing operations are performed periodically to follow the grain interfaces. Typically, a remeshing operation is considered each time a level-set is about to leave the fine mesh area set by $\phi_{min}$.

\subsection{Results and analysis}

\begin{figure}[h]
  \centering
  \includegraphics[scale=1.0]{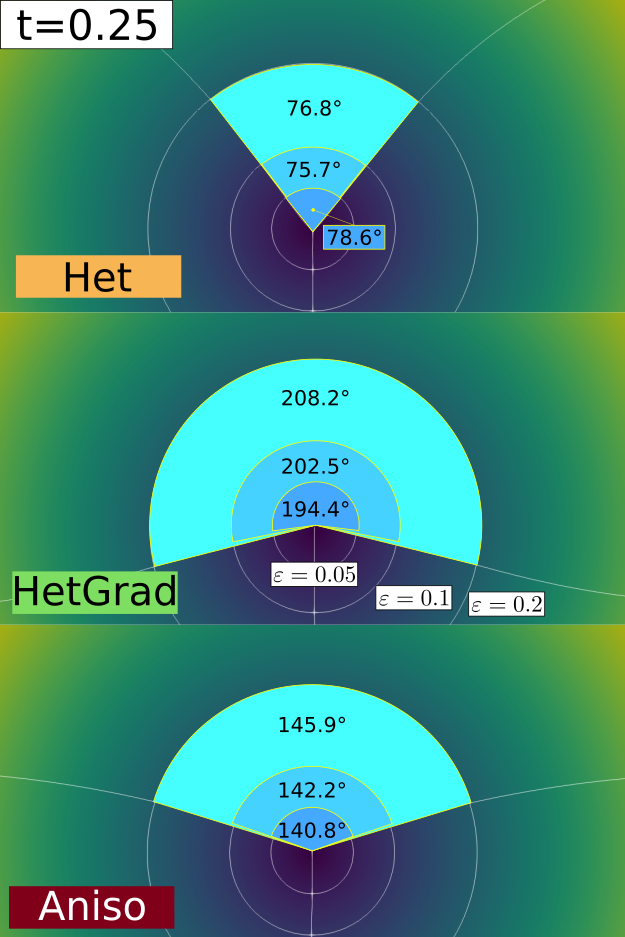}
  \caption{The triple junction at $t=0.25$ using $h_{min}=5e-4, \quad \Delta t=1e-5$. The three white circles represent the radius $\varepsilon=\{0.05, 0.1, 0.3\}$ used to compute the top dihedral angle. One can see the impact of the curved interfaces around the triple junctions in the $\xi_3$ estimation. In the following, the value $\varepsilon=0.05$ is adopted.}
  \label{fig:TJdihedral}
\end{figure}

\begin{figure}[h]
  \centering
  \begin{subfigure}[c]{0.495\textwidth}
    \centering
    \includegraphics[scale=0.25]{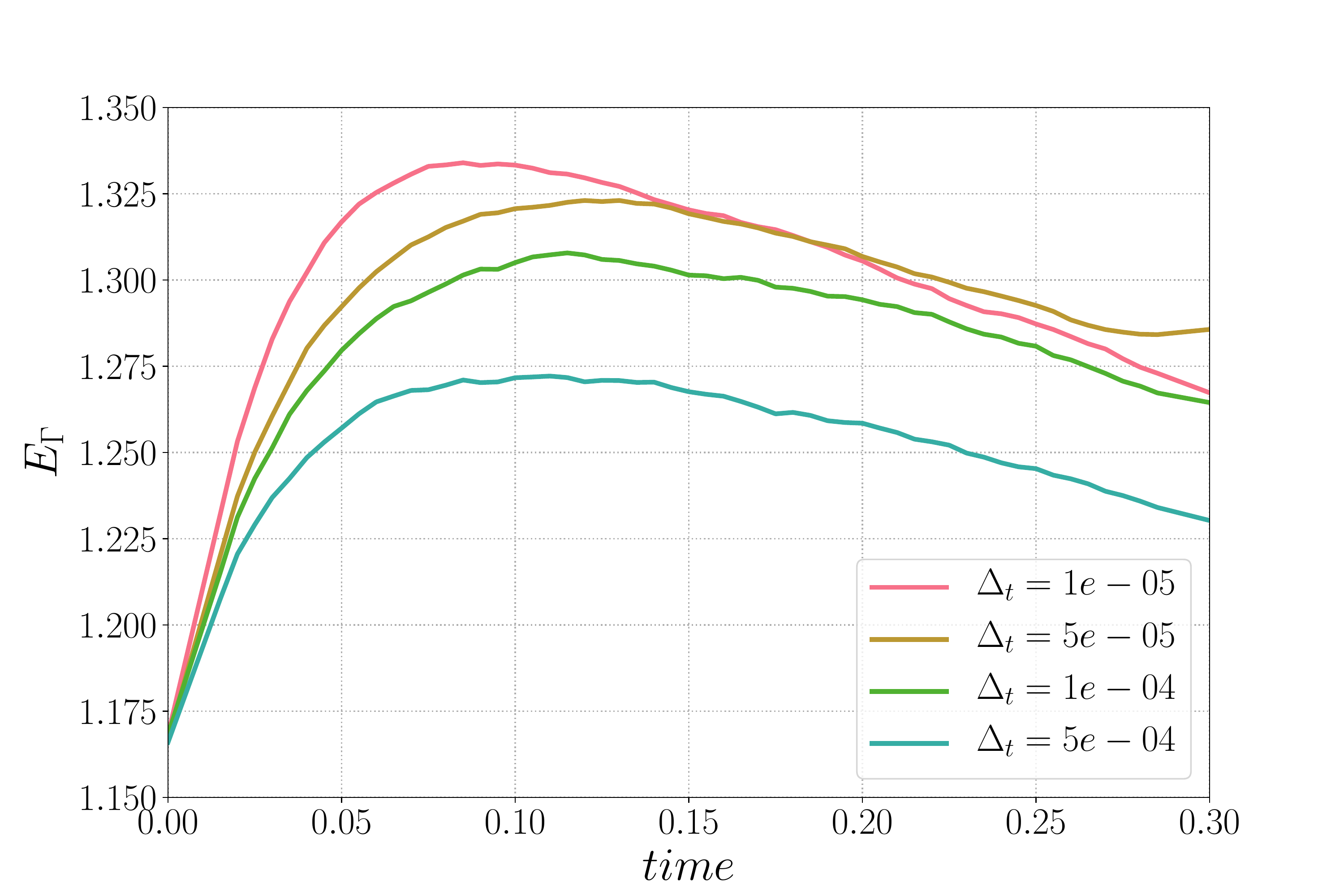}
  \end{subfigure}
  \begin{subfigure}[c]{0.495\textwidth}
    \centering
    \includegraphics[scale=0.25]{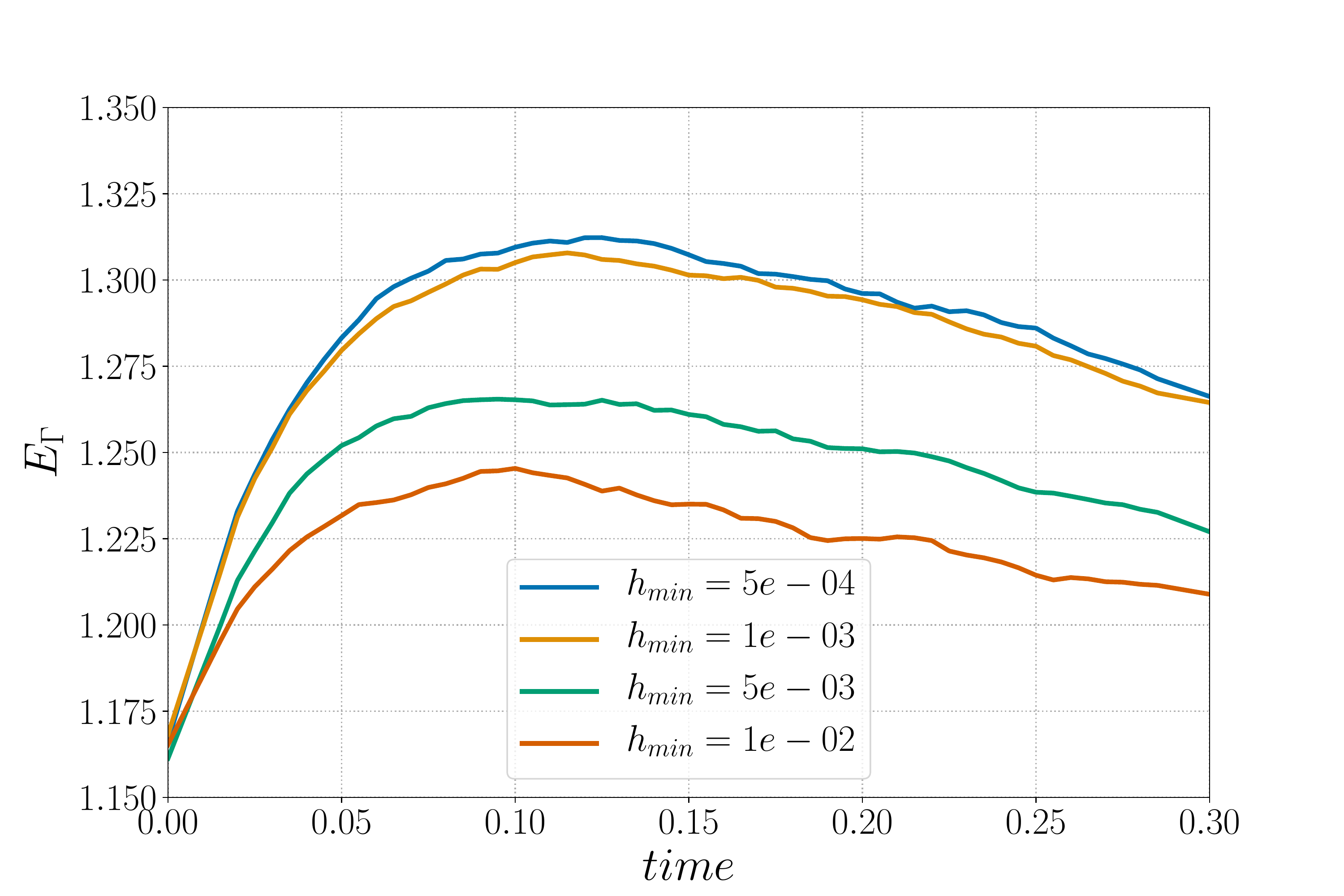} \\
  \end{subfigure}
  \begin{subfigure}[c]{0.495\textwidth}
    \centering
    \includegraphics[scale=0.25]{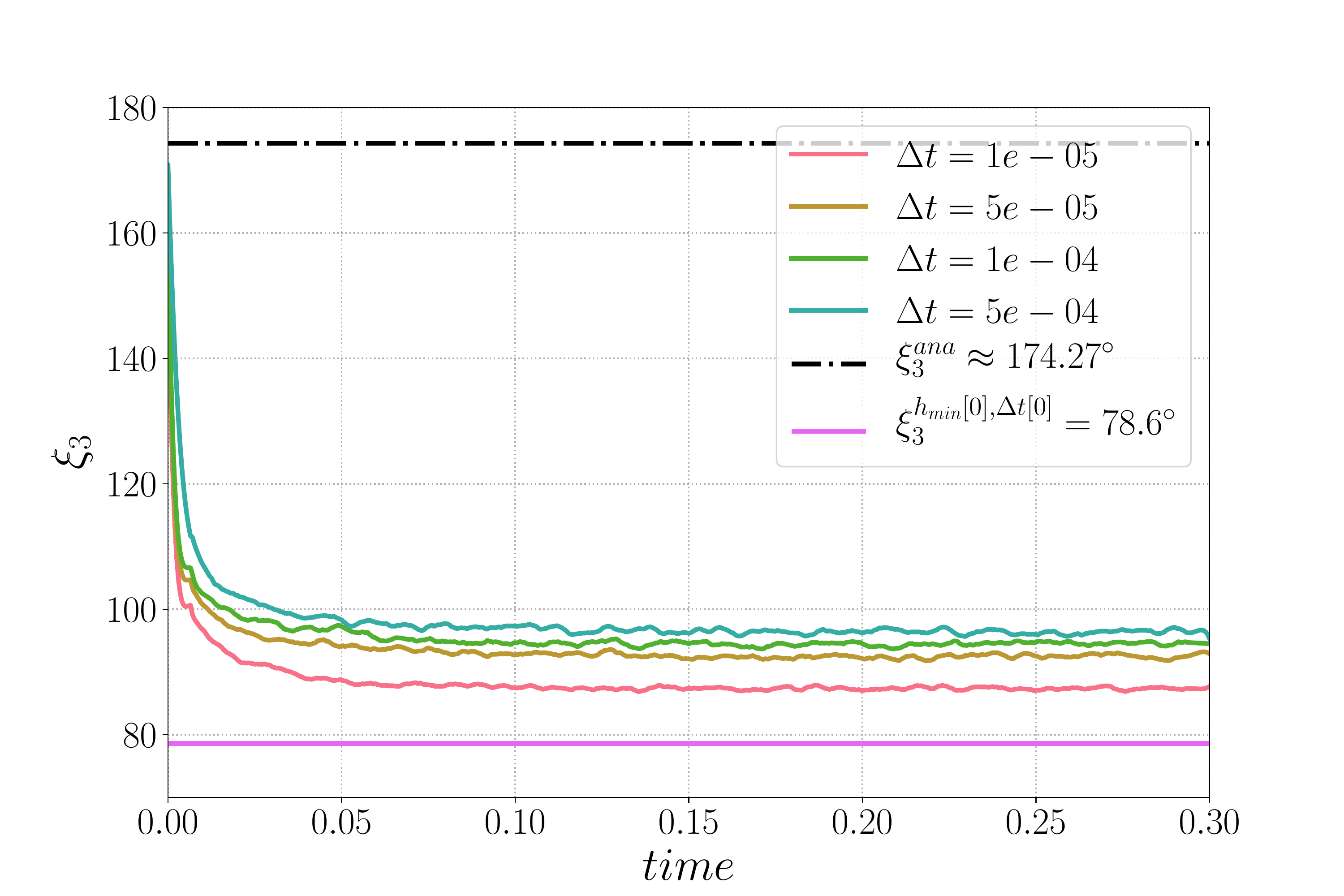}
  \end{subfigure}
  \begin{subfigure}[c]{0.495\textwidth}
    \centering
    \includegraphics[scale=0.25]{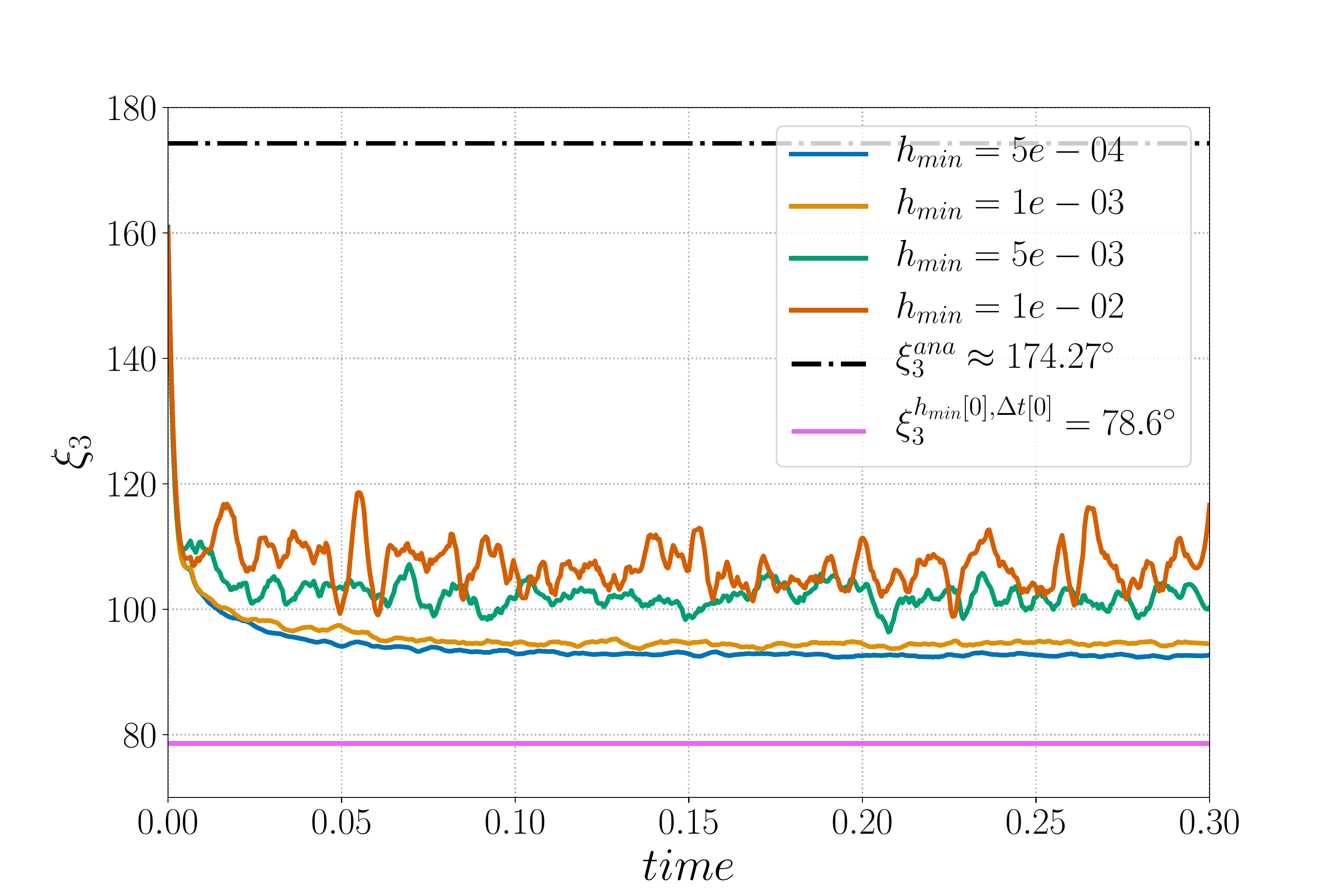} \\
  \end{subfigure}
  \begin{subfigure}[c]{0.495\textwidth}
    \centering
    \includegraphics[scale=0.25]{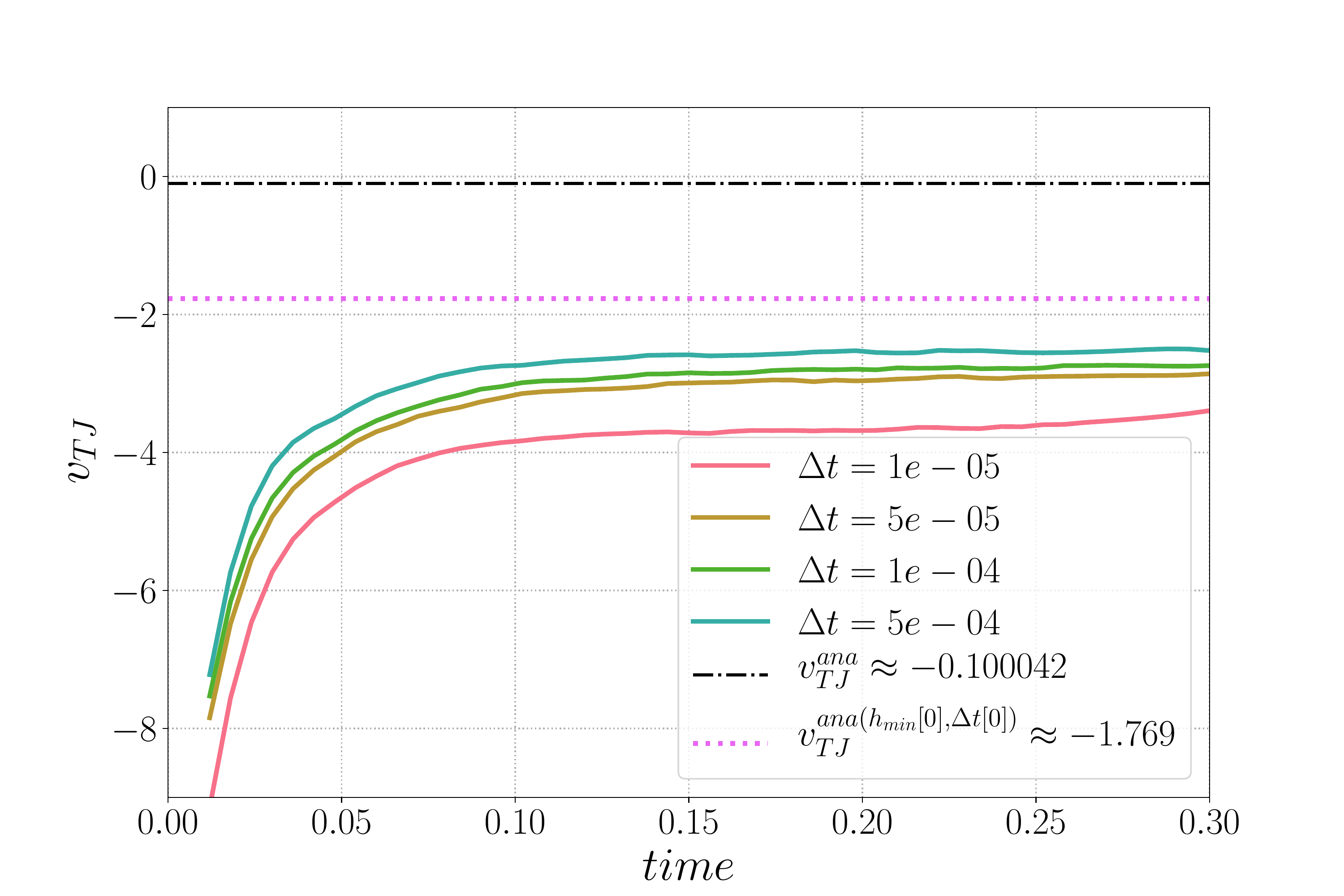}
  \end{subfigure}
  \begin{subfigure}[c]{0.495\textwidth}
    \centering
    \includegraphics[scale=0.25]{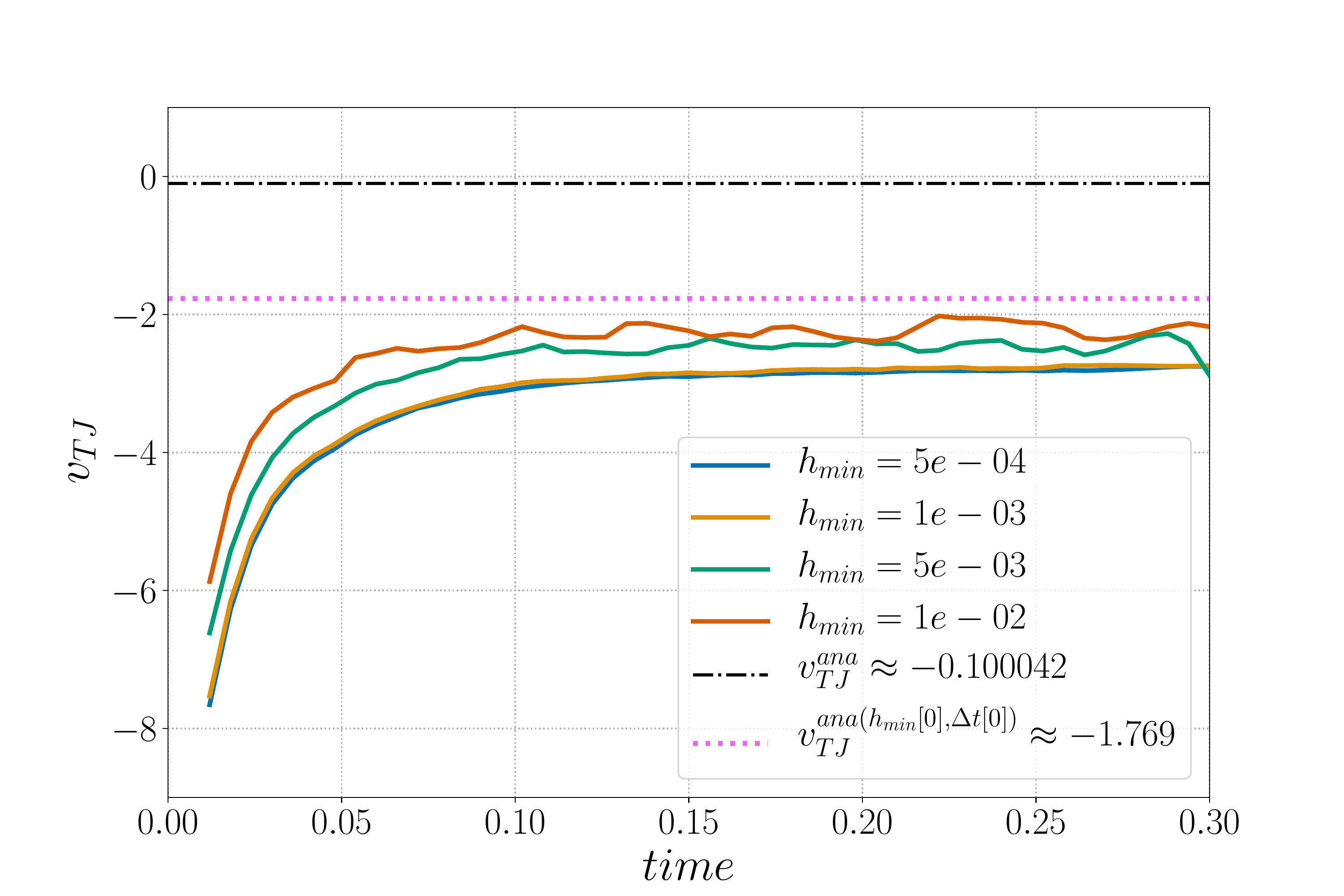}
  \end{subfigure}
  \caption{Sensibility analysis for the Het formulation in (left) time for $h_{min}=1e-3$ and in (right) $h_{min}$ size for $\Delta t=1e-4$: (Top) Interfacial energy sensibility, (middle) Triple junction angle $\xi_3$ sensibility analysis, and (bottom) Triple junction velocity $v_{TJ}$ sensibility analysis.}
  \label{fig:TJHetAnalysis}
\end{figure}

\begin{figure}[h]
  \centering
  \begin{subfigure}[c]{0.495\textwidth}
    \centering
    \includegraphics[scale=0.25]{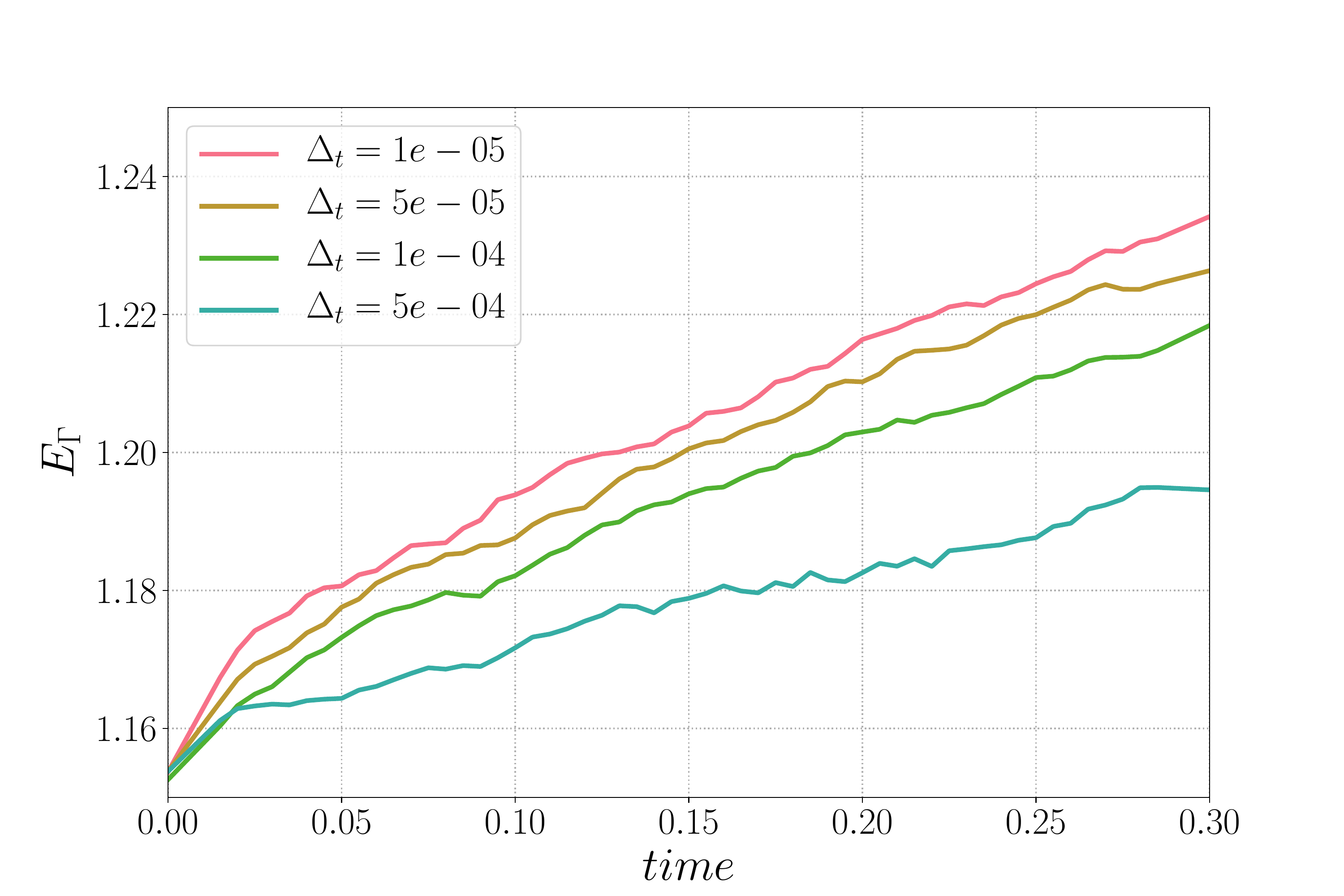}
  \end{subfigure}
  \begin{subfigure}[c]{0.495\textwidth}
    \centering
    \includegraphics[scale=0.25]{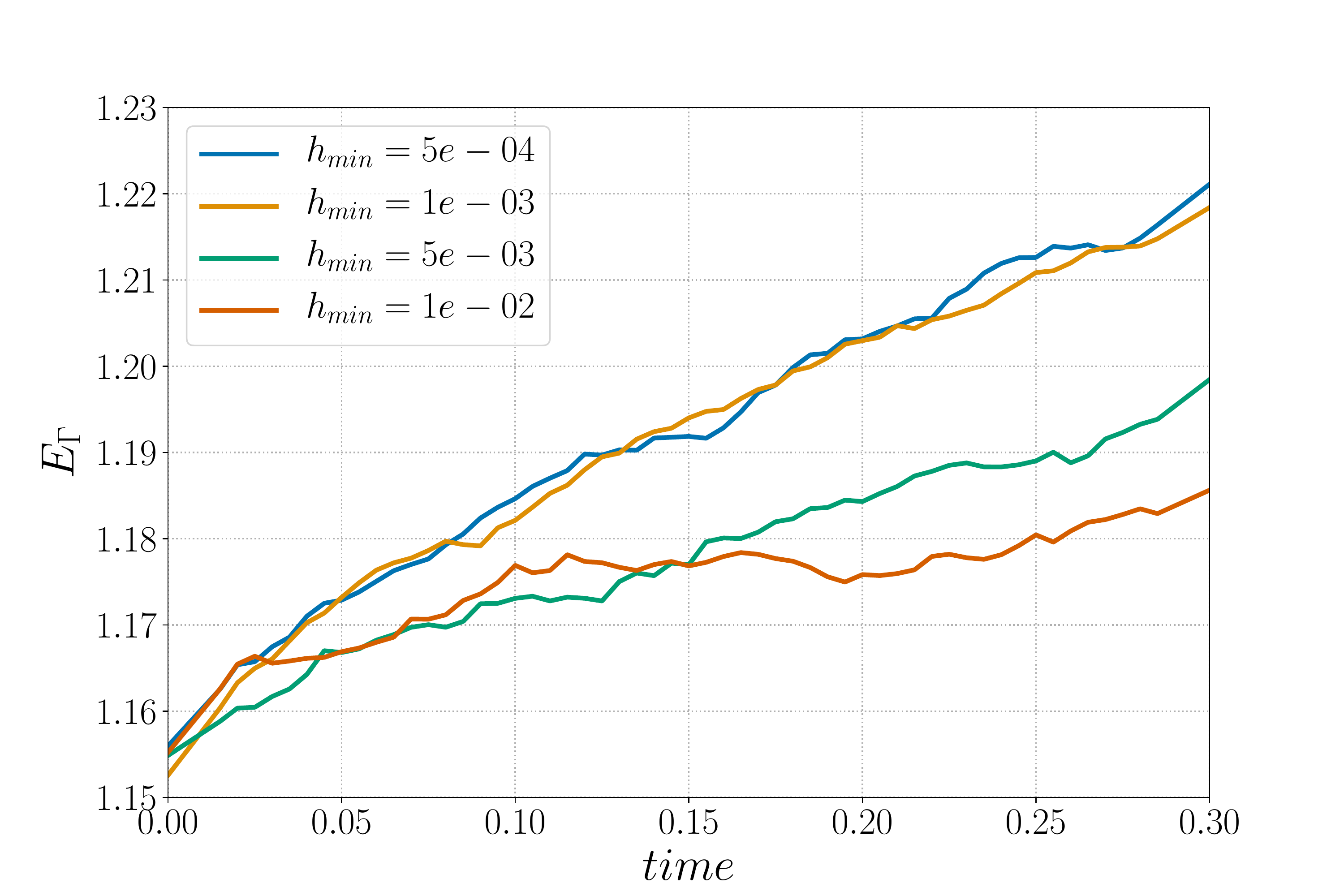} \\
  \end{subfigure}
  \begin{subfigure}[c]{0.495\textwidth}
    \centering
    \includegraphics[scale=0.25]{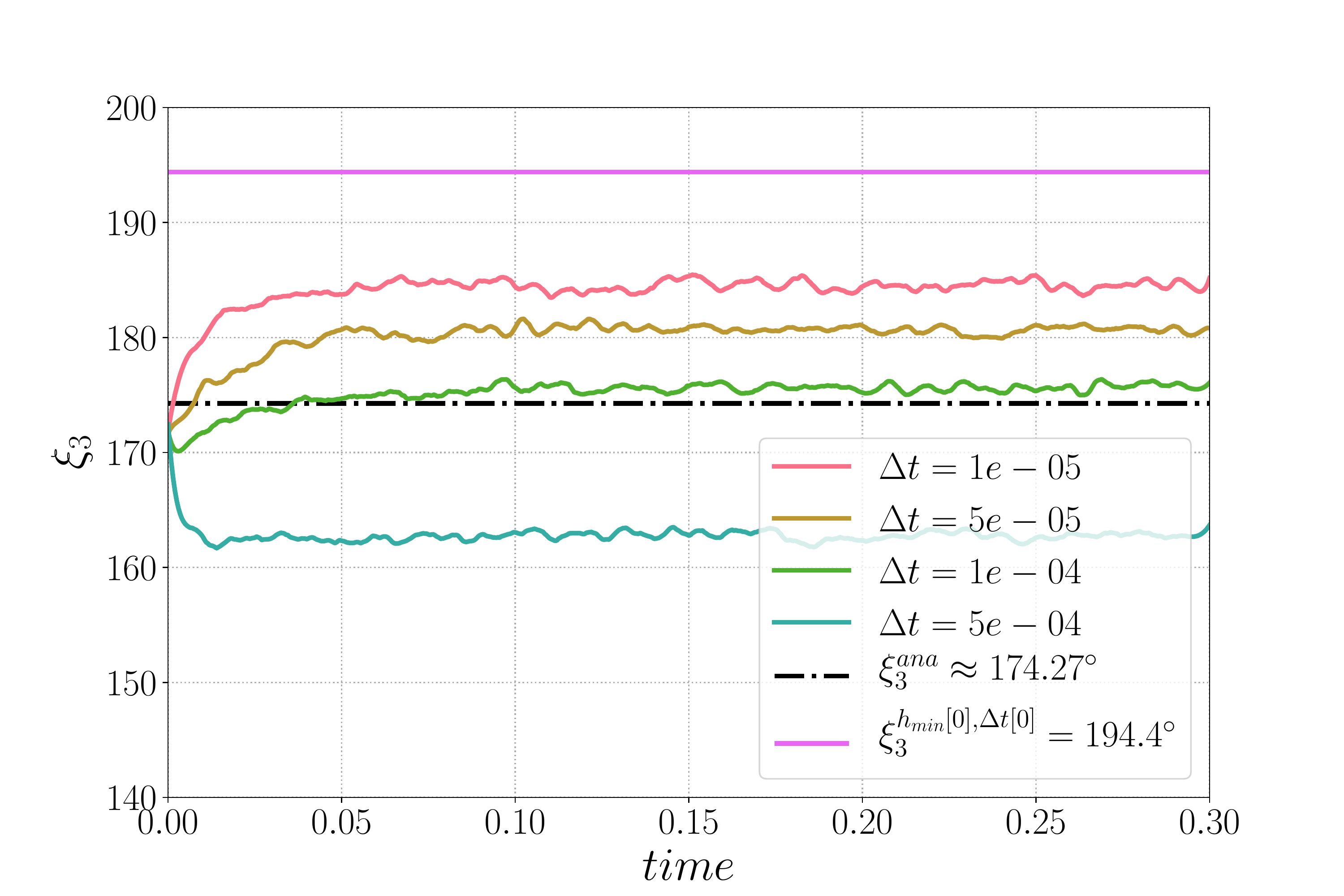}
  \end{subfigure}
  \begin{subfigure}[c]{0.495\textwidth}
    \centering
    \includegraphics[scale=0.25]{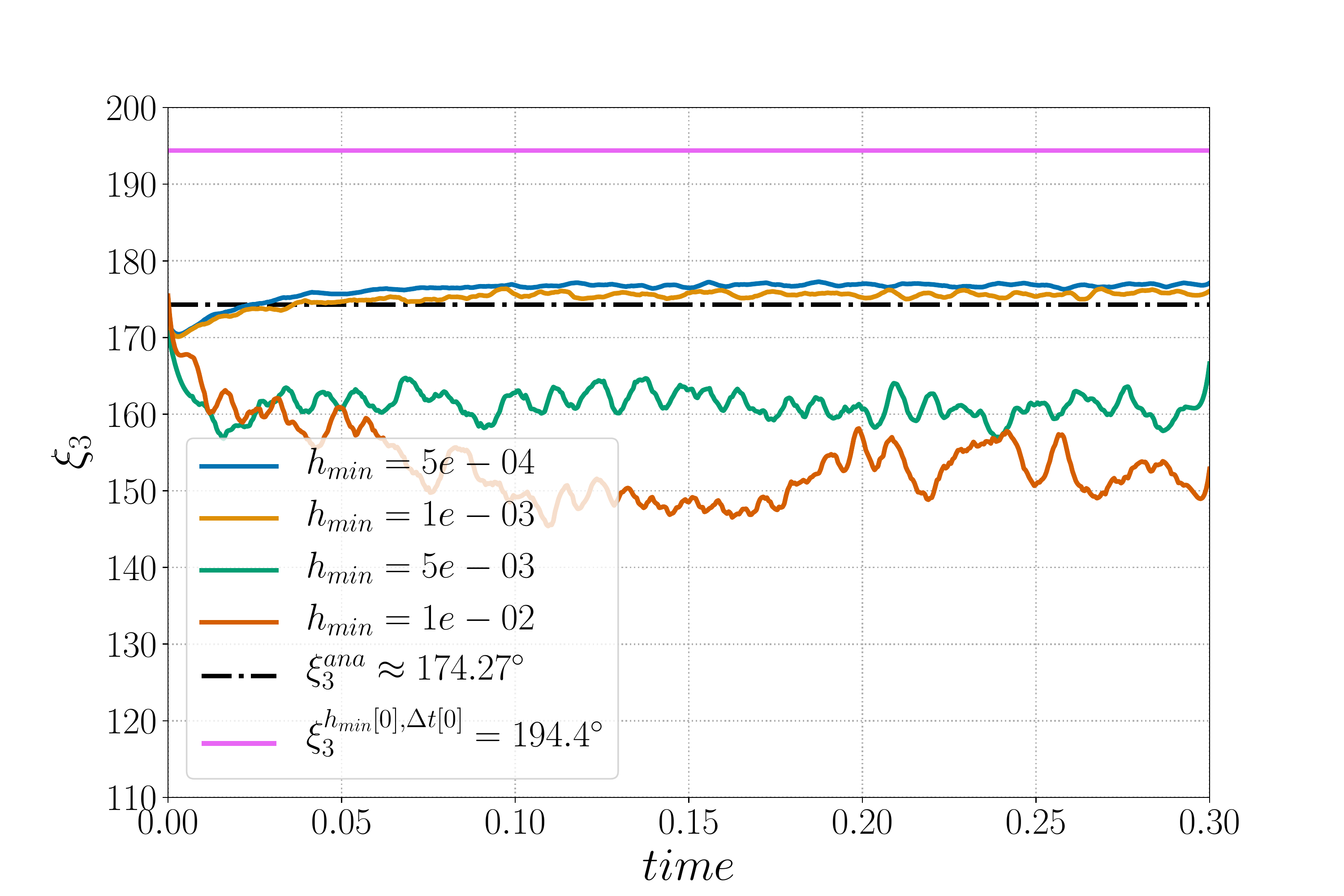} \\
  \end{subfigure}
  \begin{subfigure}[c]{0.495\textwidth}
    \centering
    \includegraphics[scale=0.25]{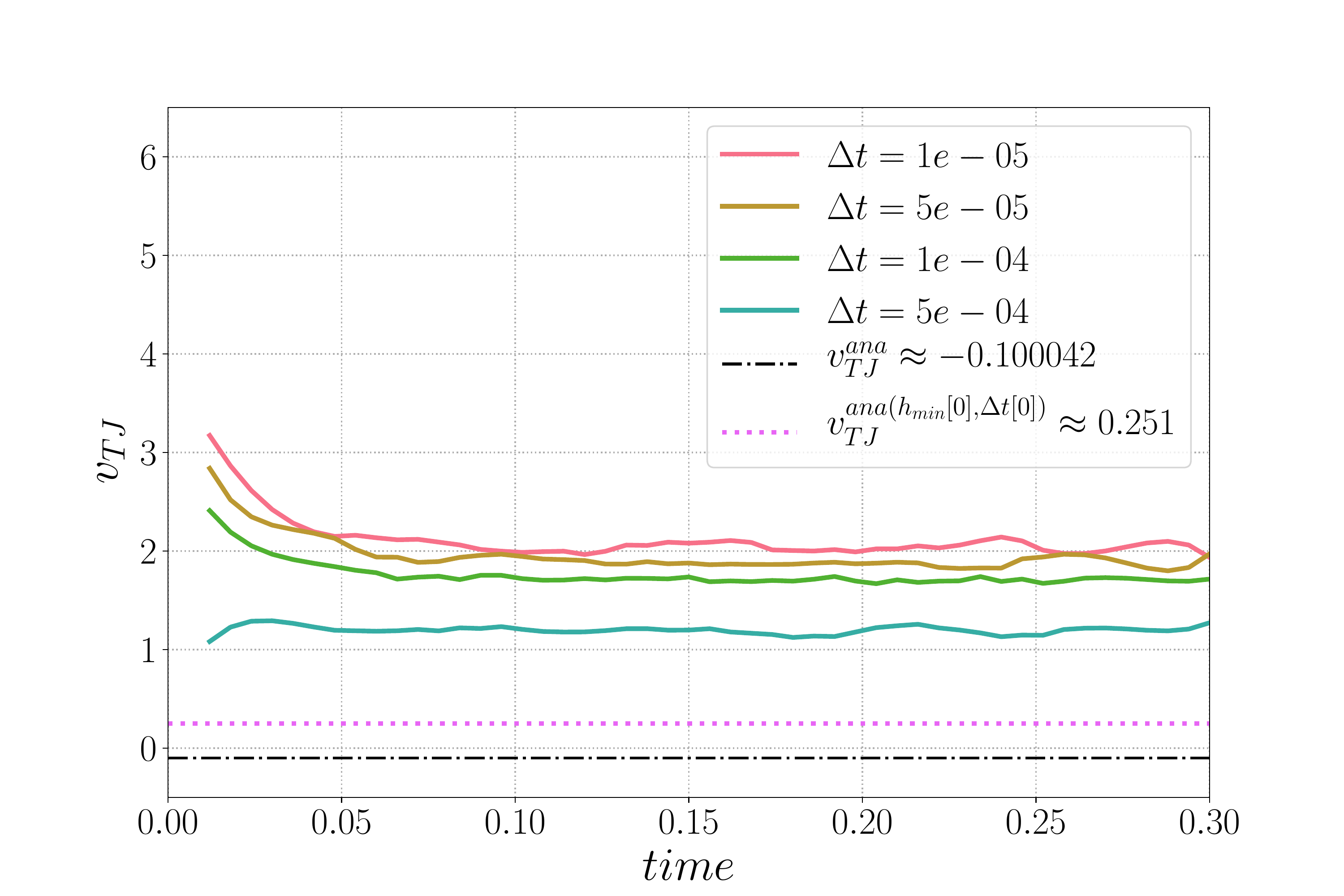}
  \end{subfigure}
  \begin{subfigure}[c]{0.495\textwidth}
    \centering
    \includegraphics[scale=0.25]{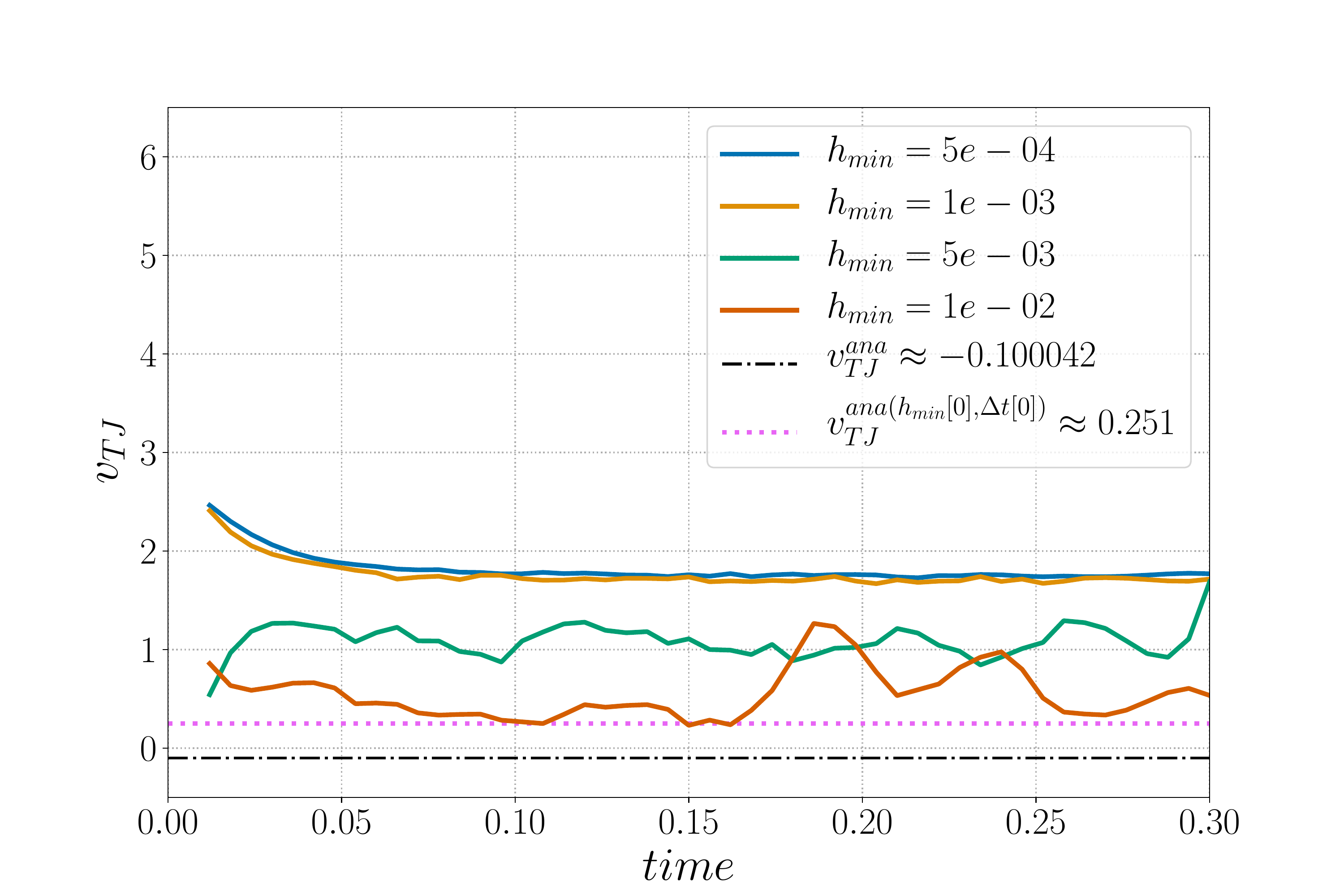}
  \end{subfigure}
  \caption{Sensibility analysis for the HetGrad formulation in (left) time for $h_{min}=1e-3$ and in (right) $h_{min}$ size for $\Delta t=1e-4$: (Top) Interfacial energy sensibility, (middle) Triple junction angle $\xi_3$ sensibility analysis, and (bottom) Triple junction velocity $v_{TJ}$ sensibility analysis.}
  \label{fig:TJHetGradAnalysis}
\end{figure}

\begin{figure}[h]
  \centering
  \begin{subfigure}[c]{0.495\textwidth}
    \centering
    \includegraphics[scale=0.25]{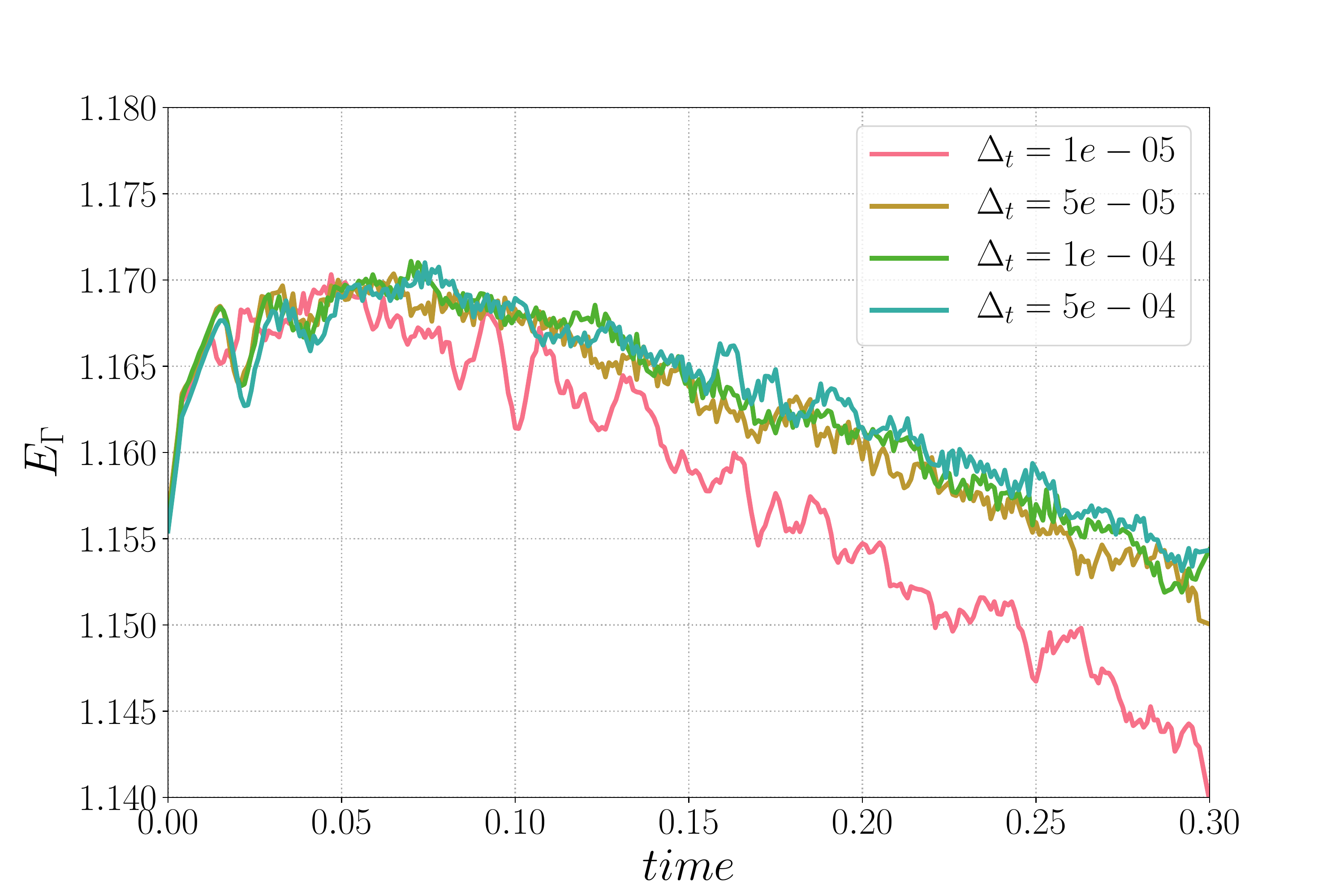}
  \end{subfigure}
  \begin{subfigure}[c]{0.495\textwidth}
    \centering
    \includegraphics[scale=0.25]{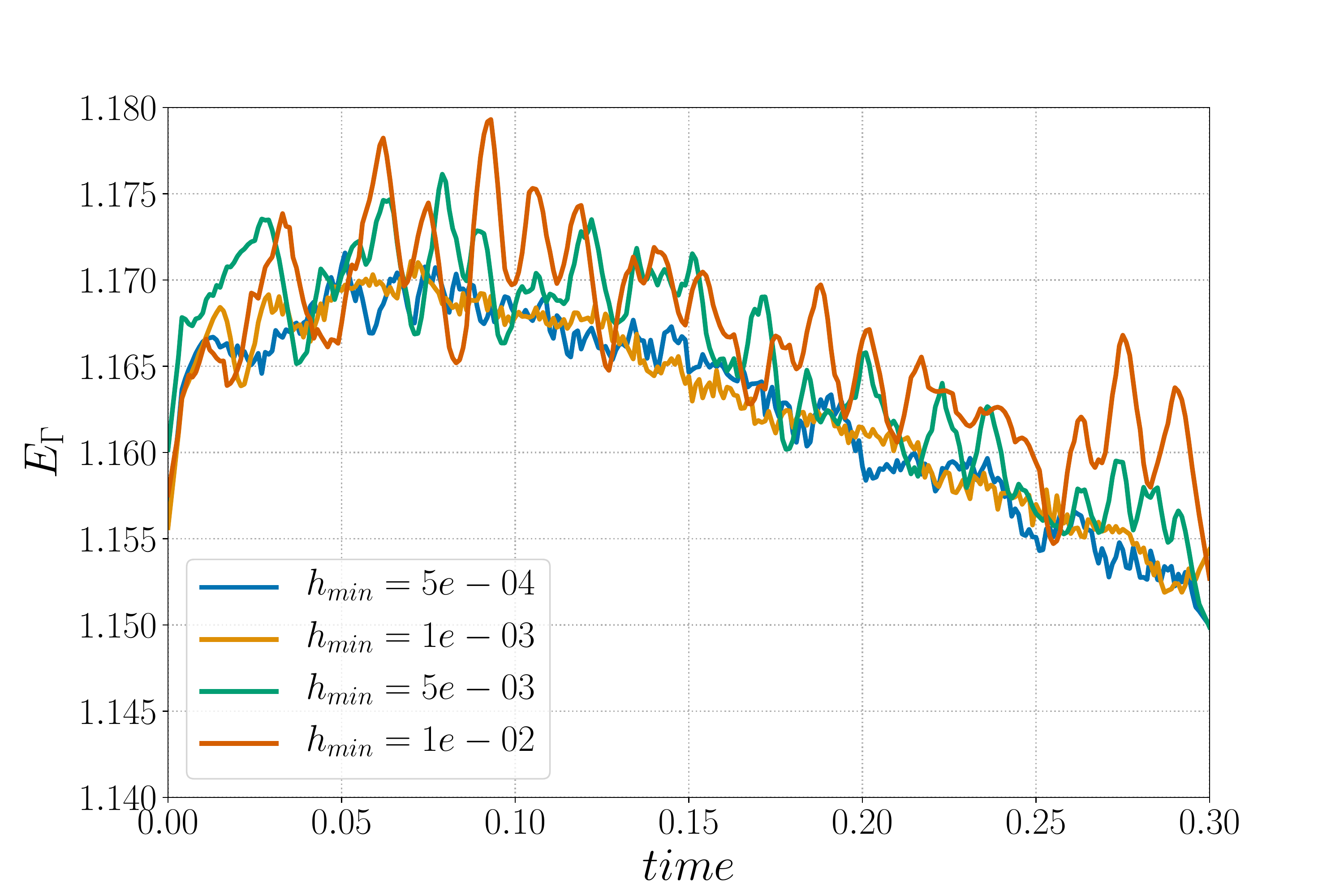} \\
  \end{subfigure}
  \begin{subfigure}[c]{0.495\textwidth}
    \centering
    \includegraphics[scale=0.25]{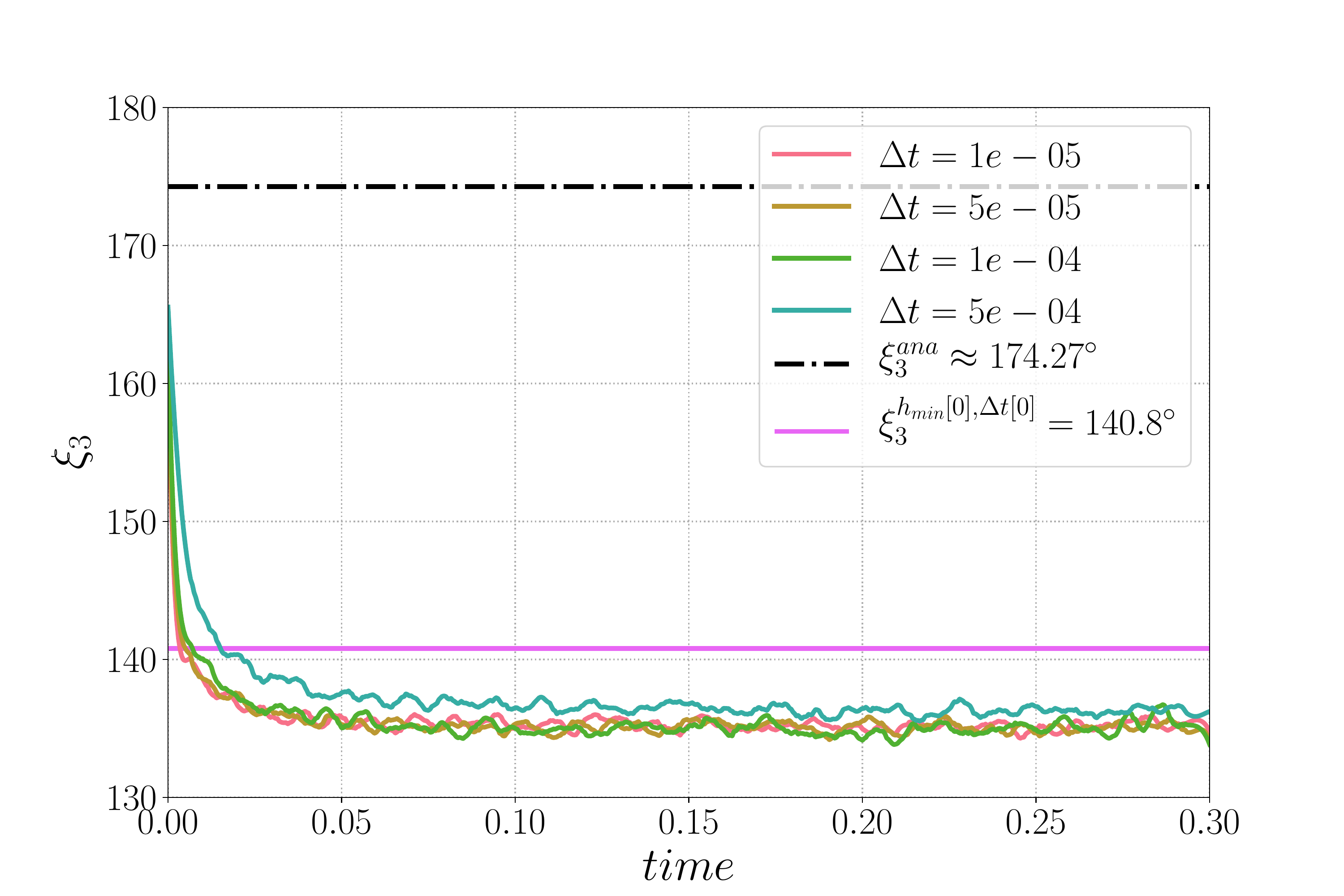}
  \end{subfigure}
  \begin{subfigure}[c]{0.495\textwidth}
    \centering
    \includegraphics[scale=0.25]{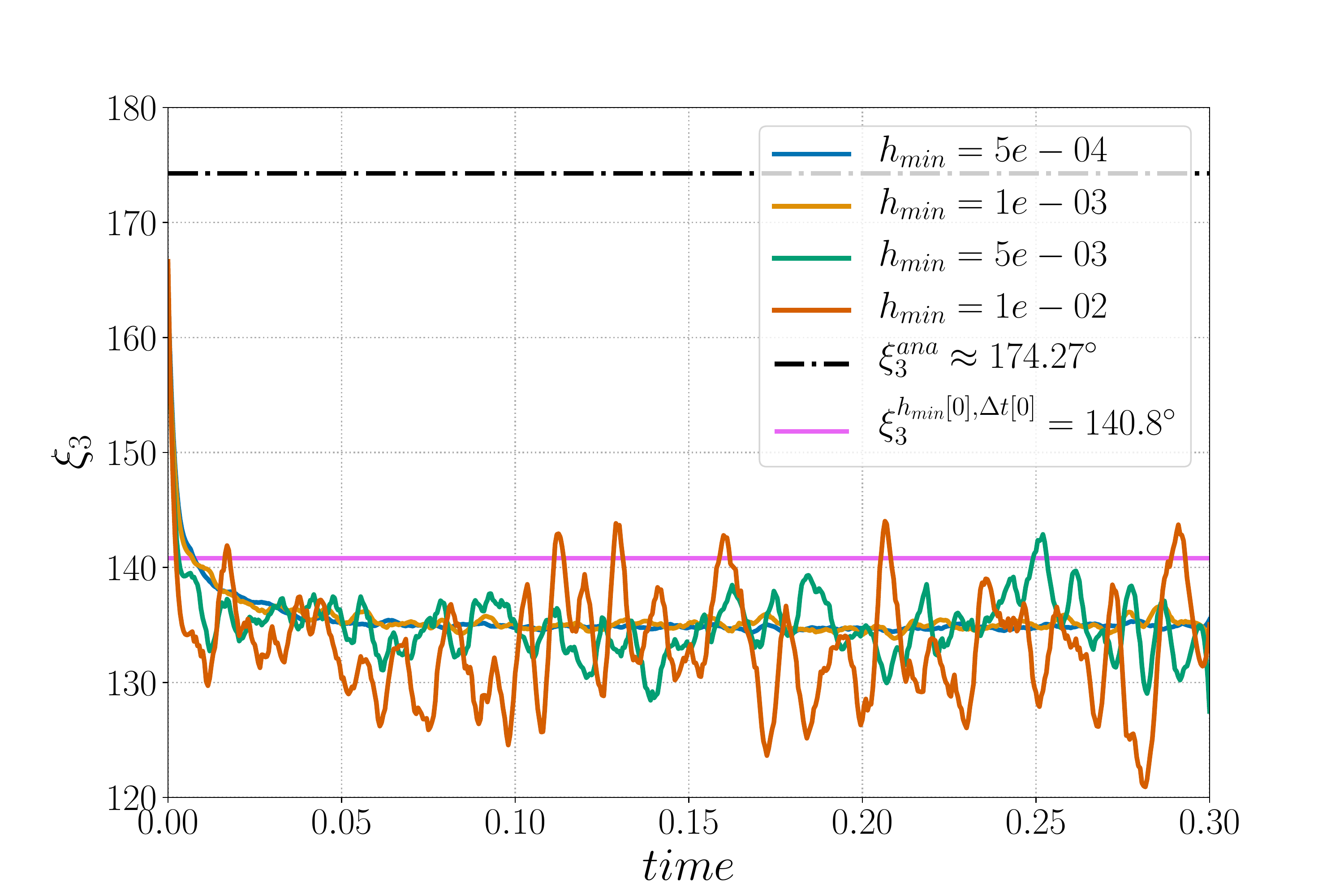} \\
  \end{subfigure}
  \begin{subfigure}[c]{0.495\textwidth}
    \centering
    \includegraphics[scale=0.25]{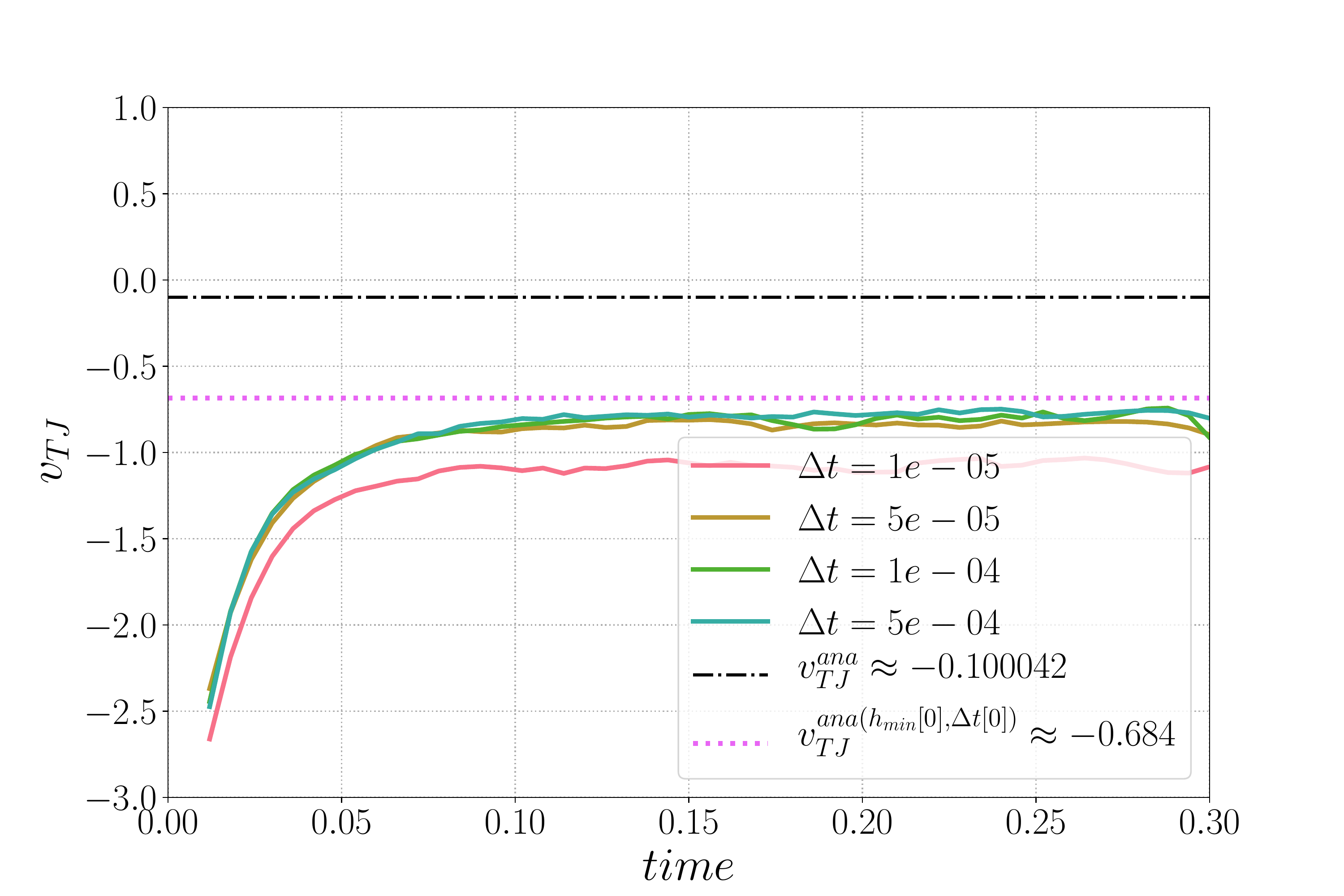}
  \end{subfigure}
  \begin{subfigure}[c]{0.495\textwidth}
    \centering
    \includegraphics[scale=0.25]{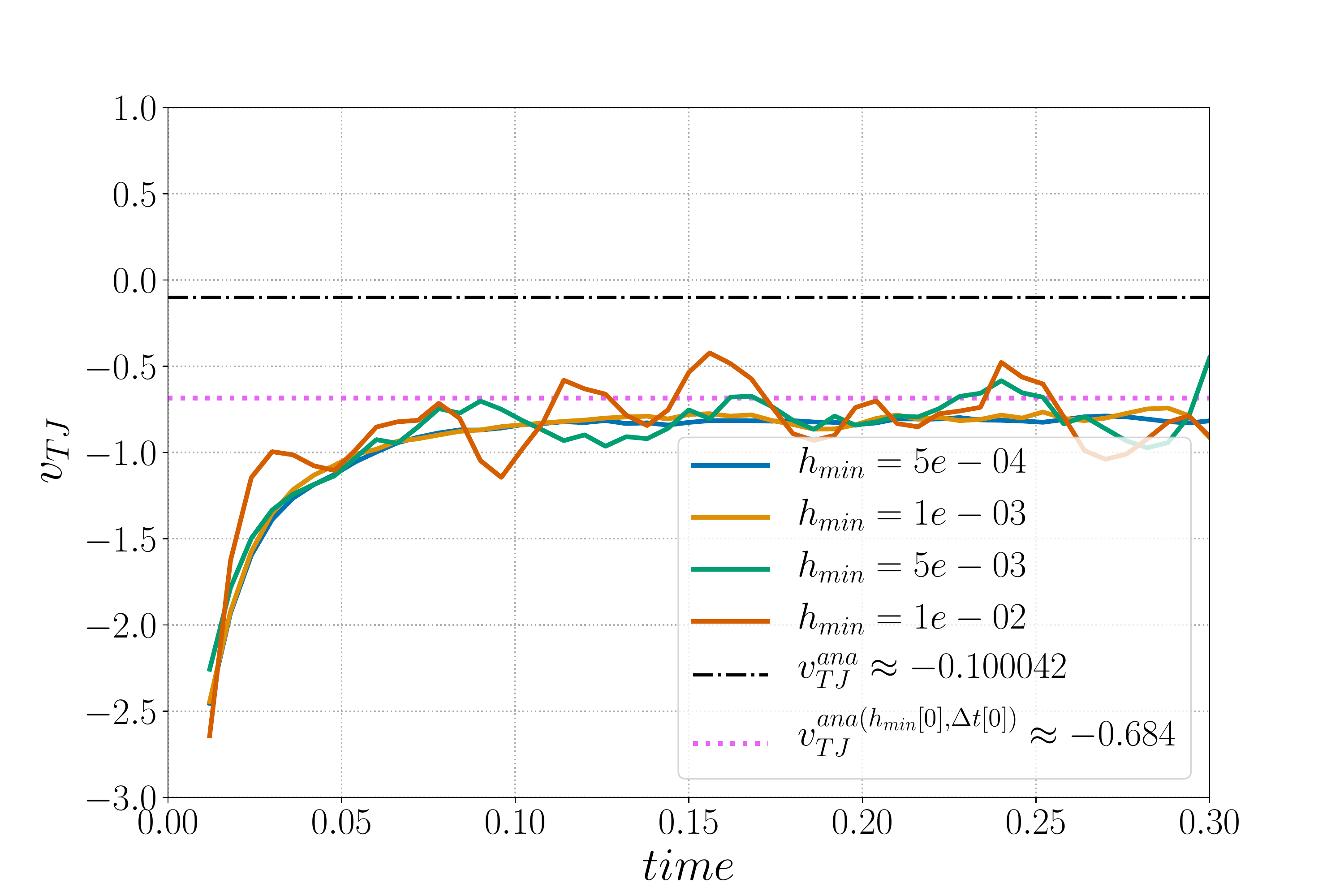}
  \end{subfigure}
  \caption{Sensibility analysis for the Aniso formulation in (left) time for $h_{min}=1e-3$ and in (right) $h_{min}$ size for $\Delta t=1e-4$: (Top) Interfacial energy sensibility, (middle) Triple junction angle $\xi_3$ sensibility analysis, and (bottom) Triple junction velocity $v_{TJ}$ sensibility analysis.}
  \label{fig:TJAnisoAnalysis}
\end{figure}

First, a sensibility analysis for the three formulations was carried out. The values of mesh size and time step used here are: $h_{max} = h_t = 1e-2$, $h_{min} = \{ 5e-4, \ 1e-3, \ 5e-3, \ 1e-2 \}$ and $\Delta t= \{1e-5, \ 5e-5, \ 1e-4, \ 5e-4 \}$. For all the cases, $\Phi_{min}$ and $\Phi_{max}$ are fixed respectively to $1e-2$ and $2e-2$. Fig.\ref{fig:TJdihedral} shows, for the different formulations, the triple junctions at $t=0.25$ using $h_{min}=5e-4$ and $\Delta t= 1e-5 $. One dihedral angle is depicted for different values of $\varepsilon$. In the following, $\xi_3^{h_{min}[k],\Delta t[k]}$ is used to define the converged value of the $\xi_3$ angle for the $k$-th value of the $h_{min}$ and $\Delta t$ datasets. Indeed, if the results described in figures~\ref{fig:TJHetAnalysis}, \ref{fig:TJHetGradAnalysis} and, \ref{fig:TJAnisoAnalysis} aim principally to compare the simulations with the quasi-steady state analytical values; it is also interesting to discuss the obtained converged value of $v_{TJ}$ as a function of the converged value of $\xi_3$ (i.e. if Eq.~\ref{eqn:c} is respected for these values).  

Fig.\ref{fig:TJHetAnalysis} illustrates the evolution of $E_{\Gamma}$, $\xi_3$ and $v_{TJ}$ using the Het formulation. Two stages appear in $E_{\Gamma}$, it increases initially before decreasing. The results illustrate the fact that the approach seems not converge, in time and space, towards the analytical solutions. However, in terms of the dihedral angle, the results converge towards $\xi_3^{h_{min}[0],\Delta t[0]}$ and the triple junction velocity converges toward the corresponding velocity $v_{TJ}^{ana(h_{min}[0],\Delta t [0])}$ (following Eq.~\ref{eqn:c}). The movement of the Het formulation is mostly influenced by the curvature of the interface, as exposed in section~\ref{sec:FrameworkDevelopment}, and one have to keep in mind that there is no additional terms that could influence the  movement of the interfaces. These results illustrate that the Het formulation, by considering heteregenous values of reduced mobility and the multiple junction treatment defined by Eq.\ref{eqn:normLS}, without rediscuting the capillarity driving pressure used in the kinetic equations is definitively not a good option when a convective/diffusive formulation is solved to model GG mechanism.        

\begin{figure}[h]
  \centering
  \includegraphics[scale=1.0]{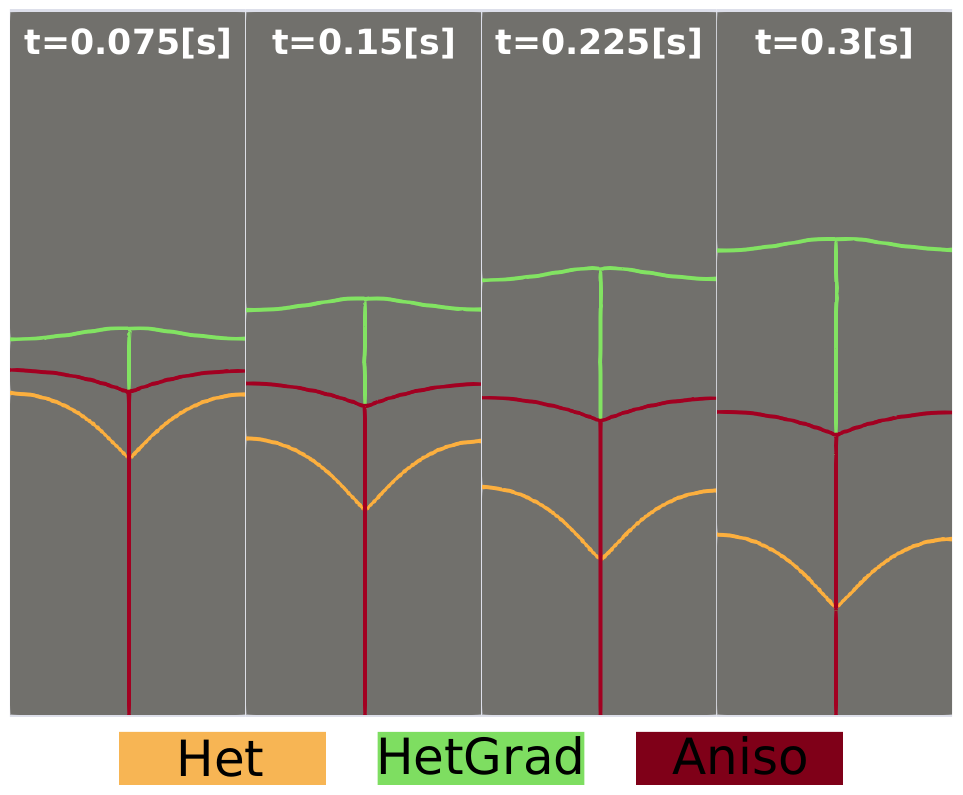}
  \caption{Evolution of the interfaces at different time steps of the three models and the initial microstructure colored in black.}
  \label{fig:TJInterfaces}
\end{figure}

The evolution of the HetGrad formulation is quite different, the interface evolves in the opposite direction (see figure~\ref{fig:TJInterfaces}) which explains that  $E_{\Gamma}$ increases during the simulation (see figure~\ref{fig:TJHetGradAnalysis}). An explanation of this evolution comes from the presence of the grain boundary energy gradient, $\nabla \gamma$, in the triple junction. The main purpose of this gradient is the correction of the triple junction dihedral angles and velocity. In figure~\ref{fig:TJHetGradAnalysis}, one can see that $\xi_3$ is closer to its analytical value and it also converges towards $\xi_3^{h_{min}[0],\Delta t[0]}$. Nevertheless, $\nabla \gamma$ also changes the kinetics of the interface because it is present along the interface and exerts a force that overcomes the effect of the curvature and generates a movement on the opposite direction. Regarding the velocity it does not converge towards the analytical value $v_{TJ}^{ana}$, nor the correlated value $v_{TJ}^{ana(h_{min}[0],\Delta t [0])}$.

The Aniso formulation has an additional term, the projection tensor $\mathbb{P}$, which takes into account the tangential changes of $\nabla \gamma$. Thanks to this term, the interface evolves in the right direction with a minimization of the boundary energy. From the evolution of $\xi_3$ and $v_{TJ}$, one can see that the simulation converges in time and space. Even if the values of $\xi_3$ do not fit precisely the analytical value, they converge towards $\xi_3^{h_{min}[0],\Delta t[0]}$. Moreover, the converged value of velocity is around $v_{TJ}^{ana(h_{min}[0],\Delta t [0])}$, meaning that the kinetics and topology of the triple junction are well correlated through Eq.~\ref{eqn:c}.

\begin{figure}[h]
  \centering
  \begin{subfigure}[c]{0.495\textwidth}
    \centering
    \includegraphics[scale=0.25]{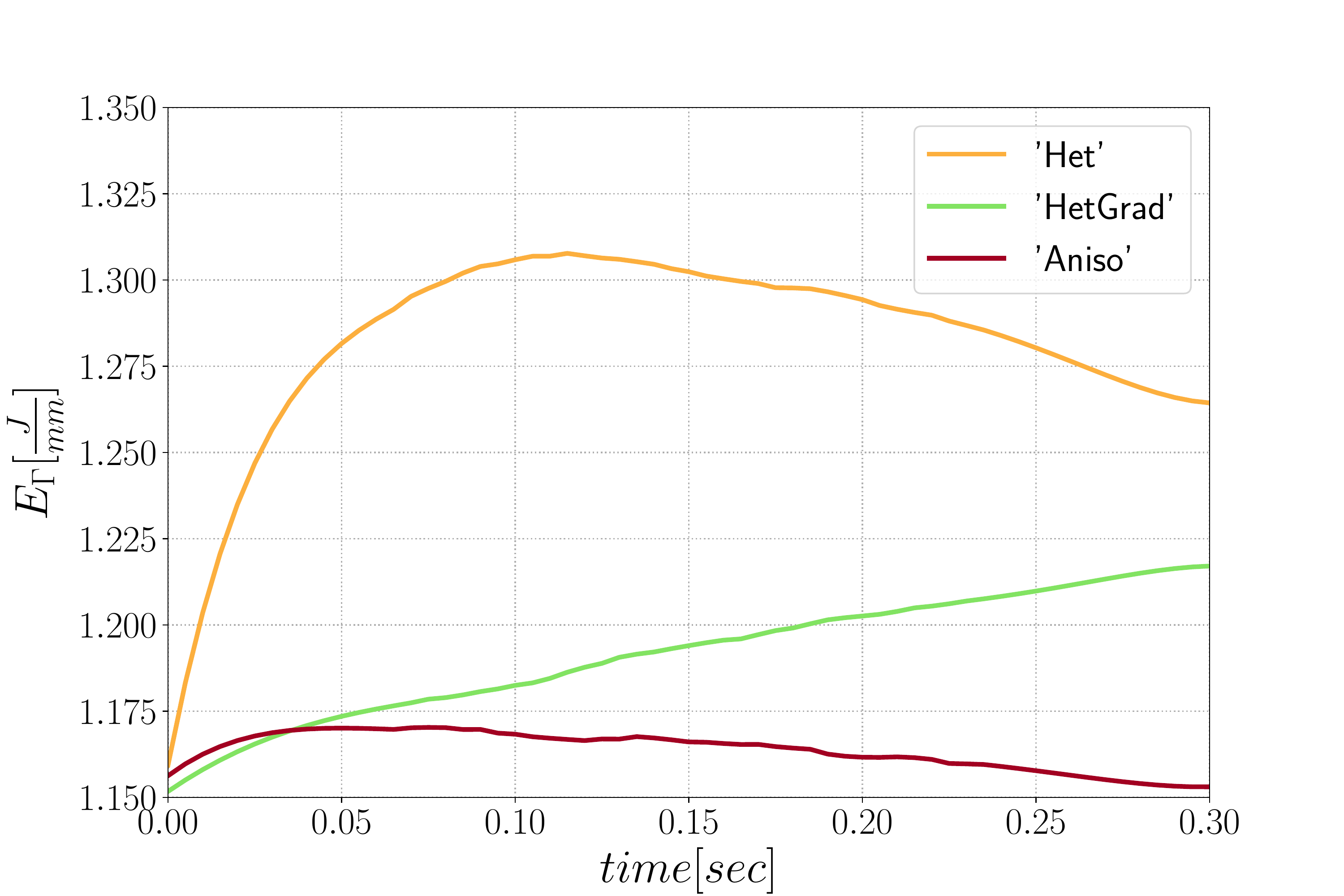}
    \caption{$E_{\Gamma}(t)$}
    \label{fig:TJInEnergy}
  \end{subfigure} \\
  \begin{subfigure}[c]{0.495\textwidth}
    \centering
    \includegraphics[scale=0.25]{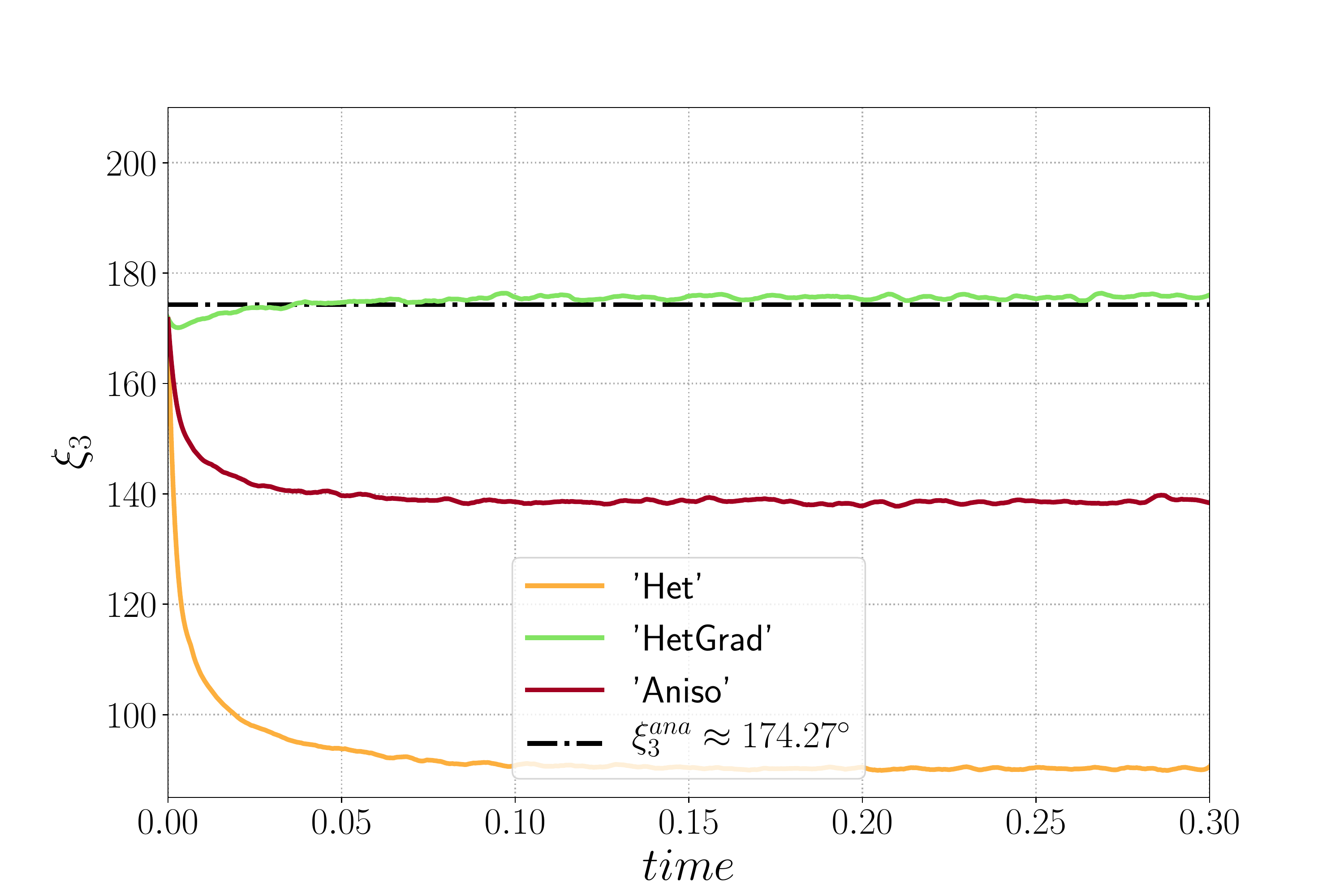}
    \caption{$\xi_3(t)$}
    \label{fig:TJxi3}
  \end{subfigure} 
  \begin{subfigure}[c]{0.495\textwidth}
    \centering
    \includegraphics[scale=0.25]{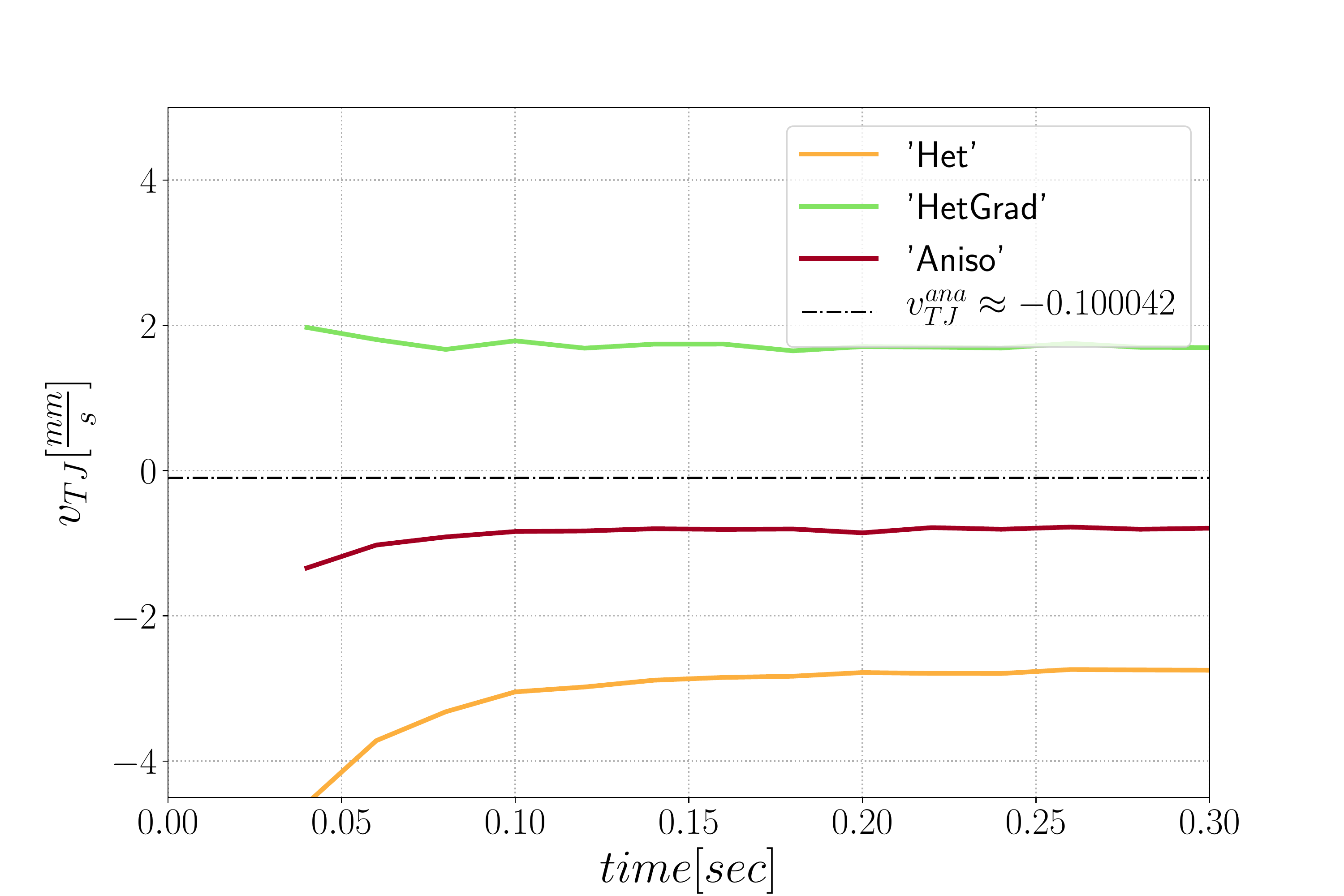}
    \caption{$v_{TJ}(t)$}
    \label{fig:TJcComp}
  \end{subfigure}
  \caption{Time series of $E_{\Gamma}$, $\xi_3$ and $v_{TJ}$ for the three formulations ($r=10$).}
\end{figure}

The evolution of the triple junction profile using the Het, HetGrad and Aniso formulations is illustrated in Fig. \ref{fig:TJInterfaces}. Both the Het and Aniso formulations produced the Grim Reaper profile while the profile produced by the HetGrad formulation evolves in the opposite direction. This is reflected in the values of the triple junction velocity shown in figures~\ref{fig:TJcComp}. From the comparison of the interfacial energy evolutions (Fig.\ref{fig:TJInEnergy}) and of the velocities (Fig.\ref{fig:TJcComp}),  one can see that the Aniso formulation has the best energetic behavior and a better approximation of the triple junction velocity. However, the best approximation of dihedral angles is obtain with the HetGrad formulation (Fig.\ref{fig:TJxi3}). 

\begin{figure}[h]
  \centering
  \begin{subfigure}{0.495\textwidth}
    \centering
    \includegraphics[scale=0.25]{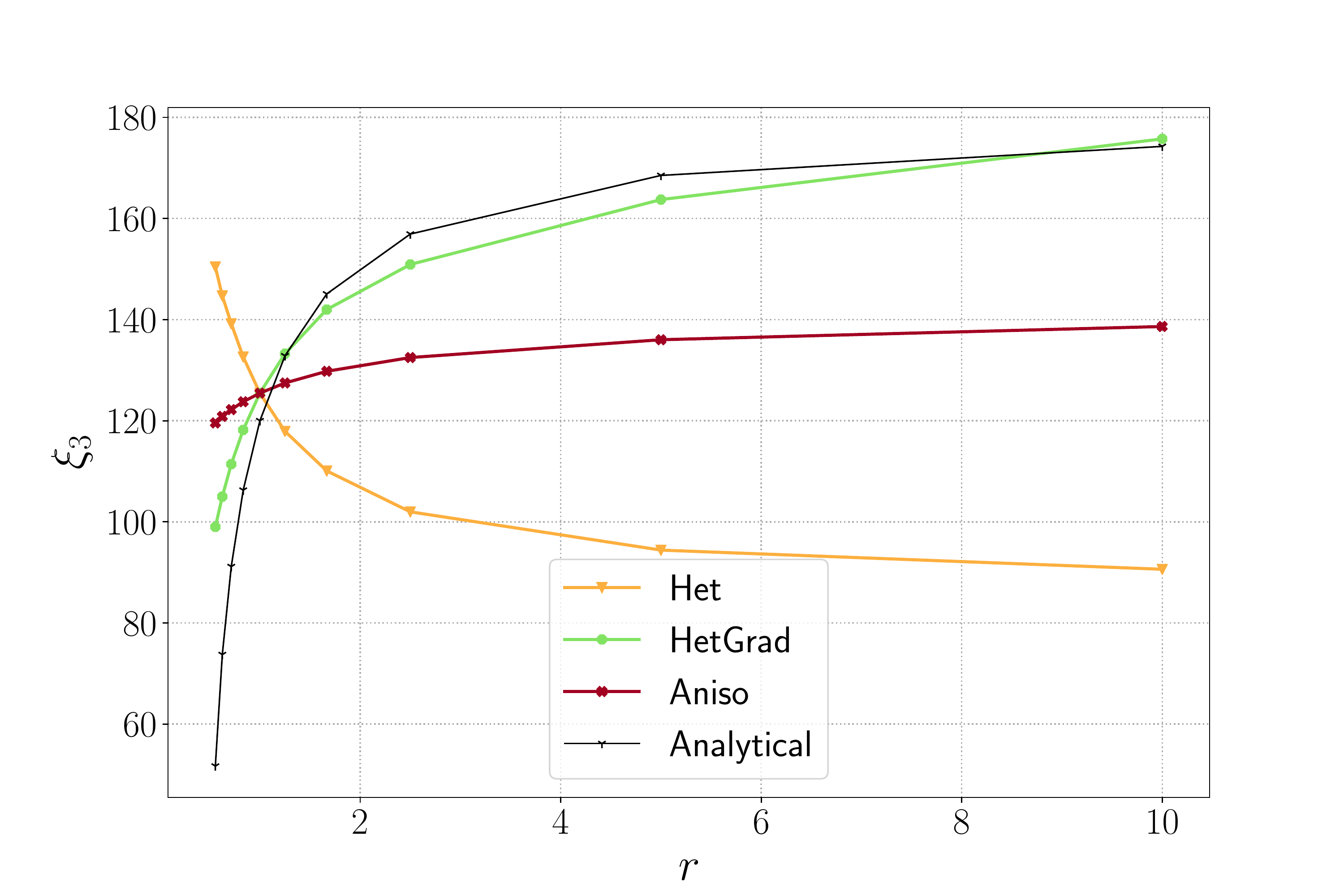}
    \caption{$\xi_3$}
    \label{fig:TJRerrorsXi3}
  \end{subfigure}
  \begin{subfigure}{0.495\textwidth}
    \centering
    \includegraphics[scale=0.25]{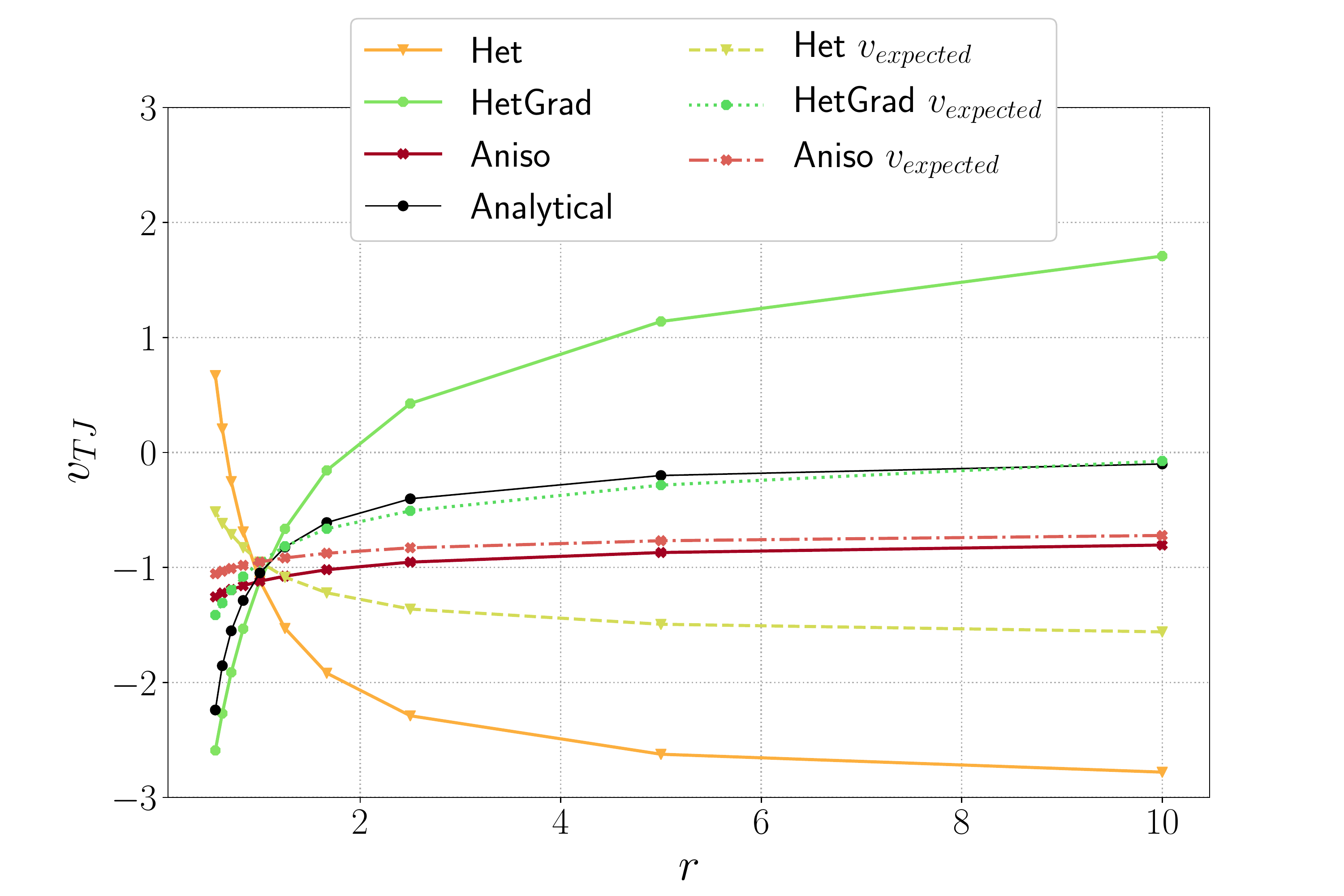}
    \caption{$v_{TJ}$} 
    \label{fig:TJRerrorsVtj}
  \end{subfigure}
  \caption{Variation of triple junction characteristics as a function of $r$ using the Het, HetGrad and Aniso formulation and the 1x3 domain. $h=1e-03$ and $\Delta t=1e-04$}\label{fig:TJRerrors}
\end{figure}

The level of anisotropy defined here is high ($r=10$), this order of value have been also discussed in the literature \citep{Fausty2018, eiken2020discussion, miyoshi2020accuracy} and remains necessary to discuss realistic polycrystal aggregates (coherent twin energy, for example). In figure~\ref{fig:TJRerrors}, the effect of the anisotropy level ($r$ value) on the top dihedral angle and the triple junction velocity is illustrated. We have carried simulations using $h=1e-3$, $\Delta t=1e-4$ and $r\in\{ 0.55, \ 0.625, \ 0.714, \ 0.833, \ 1.0, \ 1.25, \ 1.66, \ 2.5, \ 5, \ 10 \}$, which are equivalent to $\gamma_{bot}\in\{ 1.8, \ 1.6, \ 1.4, \ 1.2, \ 1.0, \ 0.8, \ 0.6, \ 0.4, \ 0.2, \ 0.1 \}$. These results allow us to conclude that the Het methodology is not adapted whatever the $r$ value. Interestingly, the HetGrad formulation seems very good for $\xi_3$ and 
$v_{TJ}$ for $r<1.5$, but the migration direction ends up being reversed for higher r values while keeping an excellent profile for the equilibrium angles. Finally, if the angle respect is slightly worse for the Aniso formulation, the respect of the triple junction speed is much better as soon as $r>1$. In fig.~\ref{fig:TJRerrorsVtj}, the three additional dashed lines represent the expected velocity for the $\xi_3$ values obtained after reaching equilibrium illustrated in fig.~\ref{fig:TJRerrorsXi3} using equation \ref{eqn:c}. One can see that the Het and HetGrad formulations correlate $\xi_3$ and $v_{TJ}$ for $r<1$. On the other hand, the Aniso formulation correlates $\xi_3$ and $v_{TJ}$ for every $r$ value. A good correlation could be advantageous if one wants to do more realistic simulations where correct kinetics and topology of the microstructure are of significant importance.

\subsection{Effect of the boundary conditions}

\begin{figure}[h]
  \centering
  \includegraphics[scale=0.9]{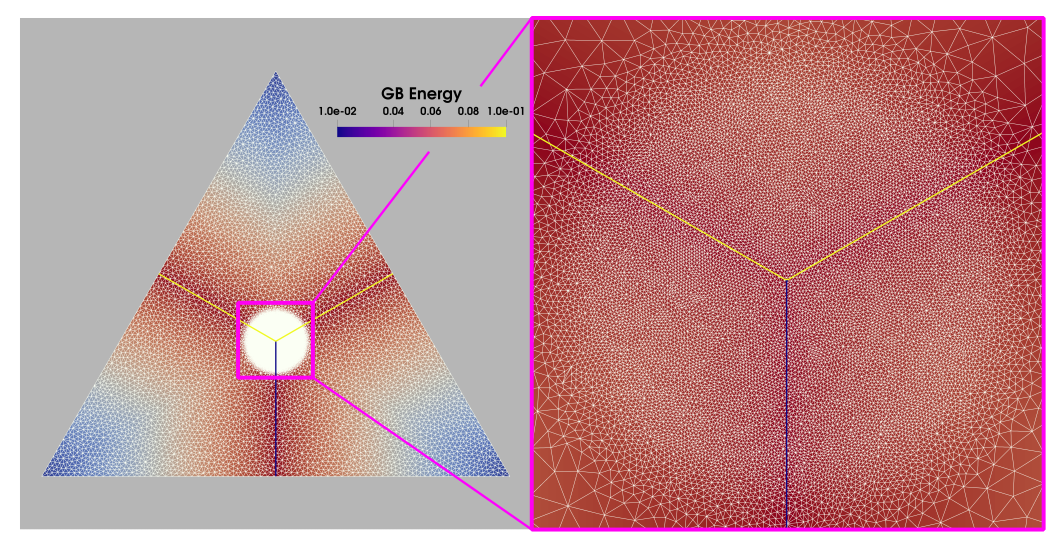}
  \caption{Initial configuration of the triangular case with a refined isotropic mesh around the triple junction and a coarse mesh outside the triple junction with $r=10$.}
  \label{fig:TJTRIANGLE}
\end{figure}

In \cite{Fausty2018}, the authors proposed the HetGrad formulation and performed several simulations for different values of $r$. The authors compared the dihedral angles against the analytical Grim Reaper values (see Eq.~\ref{eqn:xi3}) and found a very good estimation of the dihedral angles. A triangular domain was used with an initial triple junction equilibrium at 120° and Dirichlet boundary conditions (fixing the GB in the border domain). In other words, a final configuration respecting the Young's equilibrium is attended without the possibility to describe the transient state with an analytical solution. In order to study the Aniso formulation behavior, the same case is presented here. An isotropic mesh is used with a local adaptation around the triple junction where the mesh is refined in a circle of radius $\varepsilon=0.05$ allowing the simulation to be more computationally efficient in terms of CPU time and memory storage. Fig.~\ref{fig:TJTRIANGLE} illustrates the mesh around the triple junction, one can see the change of the mesh size close to the triple junction. 

\begin{figure}[h]
  \centering

    \includegraphics[scale=0.25]{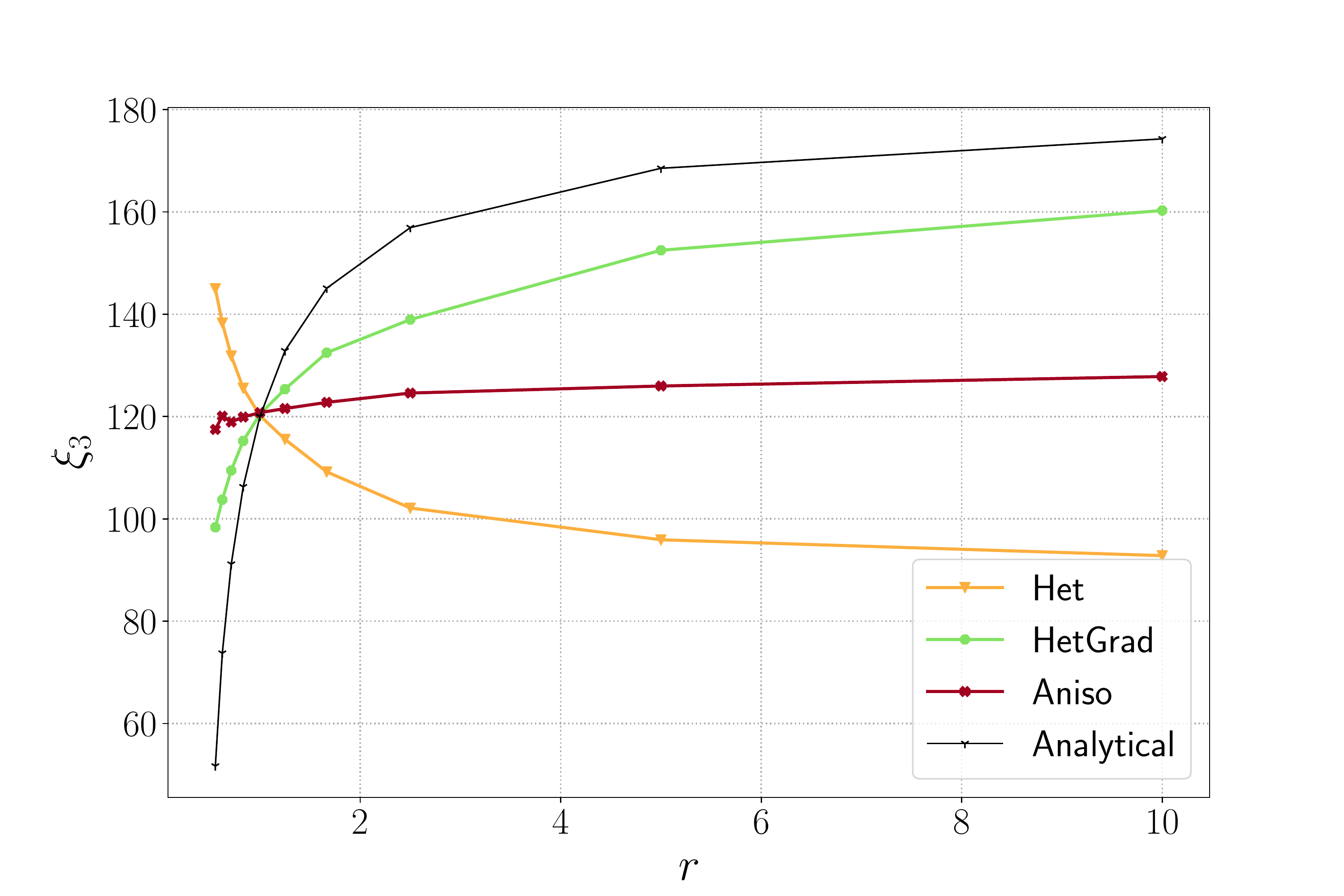}

  \caption{Triangular case with Dirichlet boundary conditions : Variation of the triple junction top dihedral angle $\xi_3$ as a function of $r$ using the Het, HetGrad and Aniso formulations.}\label{fig:TJTriangleBC}
\end{figure}

Multiple simulations were carried in order to study the effect of $r$. The constant parameters are the GB mobility $\mu=1$, the GB energy of the top interfaces $\gamma_{top}=0.1$, the mesh size at the triple junction $h_{TJ}=0.001$, the mesh size outside the triple junction $h=0.01$ and the time step $\Delta t =1e-4$. As for the case presented before, the GB energy of the bottom interface is changed to obtain $r\in\{ 0.55, \ 0.625, \ 0.714, \ 0.833, \ 1.0, \ 1.25, \ 1.66, \ 2.5, \ 5, \ 10 \}$ ($\gamma_{bot}\in\{ 0.18, \ 0.16, \ 0.14, \ 0.12, \ 0.1, \ 0.08, \ 0.06, \ 0.04, \ 0.02, \ 0.01 \}$).  In fig.~\ref{fig:TJTriangleBC} one can see the same tendencies as in fig.~\ref{fig:TJRerrors}, being the HetGrad formulation the best option in terms of dihedral angles prediction.

\begin{figure}[h]
  \centering
  \includegraphics[scale=0.9]{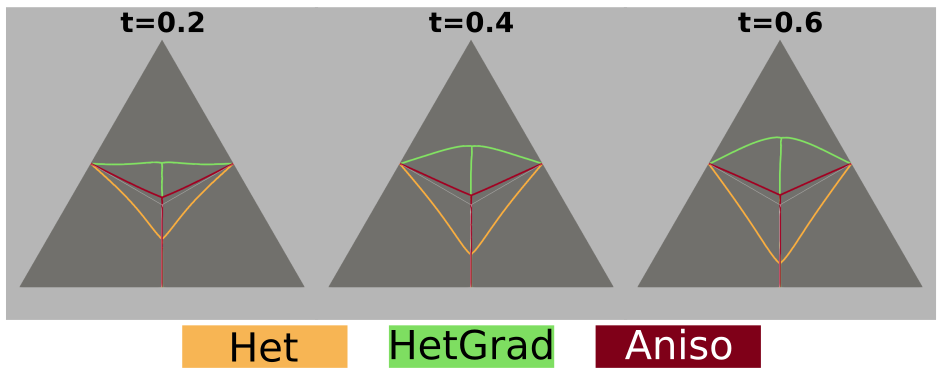}
  \caption{Interface evolution using the Het, Hetgrad and Aniso formulation of the triple junction with $r=10$.}
  \label{fig:TJTRIANGLE_Int_r10}
\end{figure}

\begin{figure}[h]
  \centering
  \includegraphics[scale=0.9]{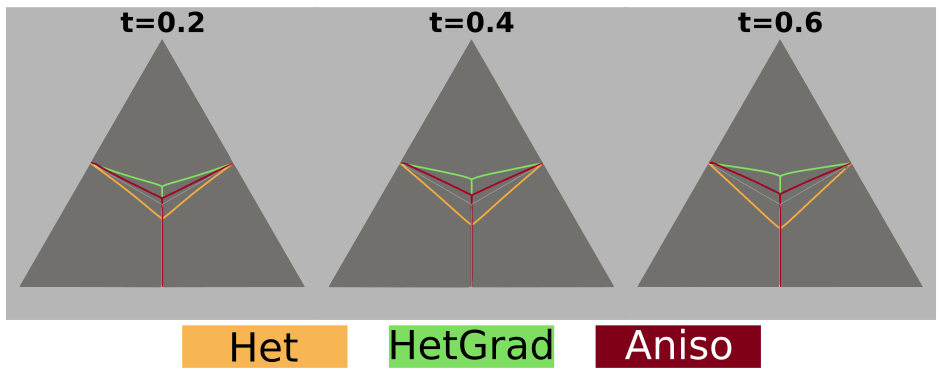}
  \caption{Interface evolution using the Het, Hetgrad and Aniso formulation of the triple junction with $r=1.66$.}
  \label{fig:TJTRIANGLE_Int_r1-6}
\end{figure}

Fig.~\ref{fig:TJTRIANGLE_Int_r10} shows the interface evolution with $r=10$, the evolution is similar for the Grim Reaper example in fig.~\ref{fig:TJInterfaces}. The Het and Aniso formulations exhibit a Grim Reaper-like profile while the HetGrad formulation evolves in the upward direction. This may seem wrong, however, for this particular geometry an upward movement is expected for $r>1$ in order to match the analytical angles and as the initial angles are fixed to 120°. Thus, one can say that the interface obtained with the Het formulation evolves in the wrong direction. On the other hand, the movement obtained by the HetGrad formulation exaggerates the expected displacements and the interface is highly curved. Another illustration of the interface movement is shown in Fig.~\ref{fig:TJTRIANGLE_Int_r1-6} for $r=1.66$, the HetGrad and Aniso formulation have a correct evolution of the interface and the dihedral angle is closer to the analytical value as exposed in Fig.~\ref{fig:TJTriangleBC}.

\begin{figure}[h]
  \centering
  \begin{subfigure}{0.495\textwidth}
    \centering
    \includegraphics[scale=0.25]{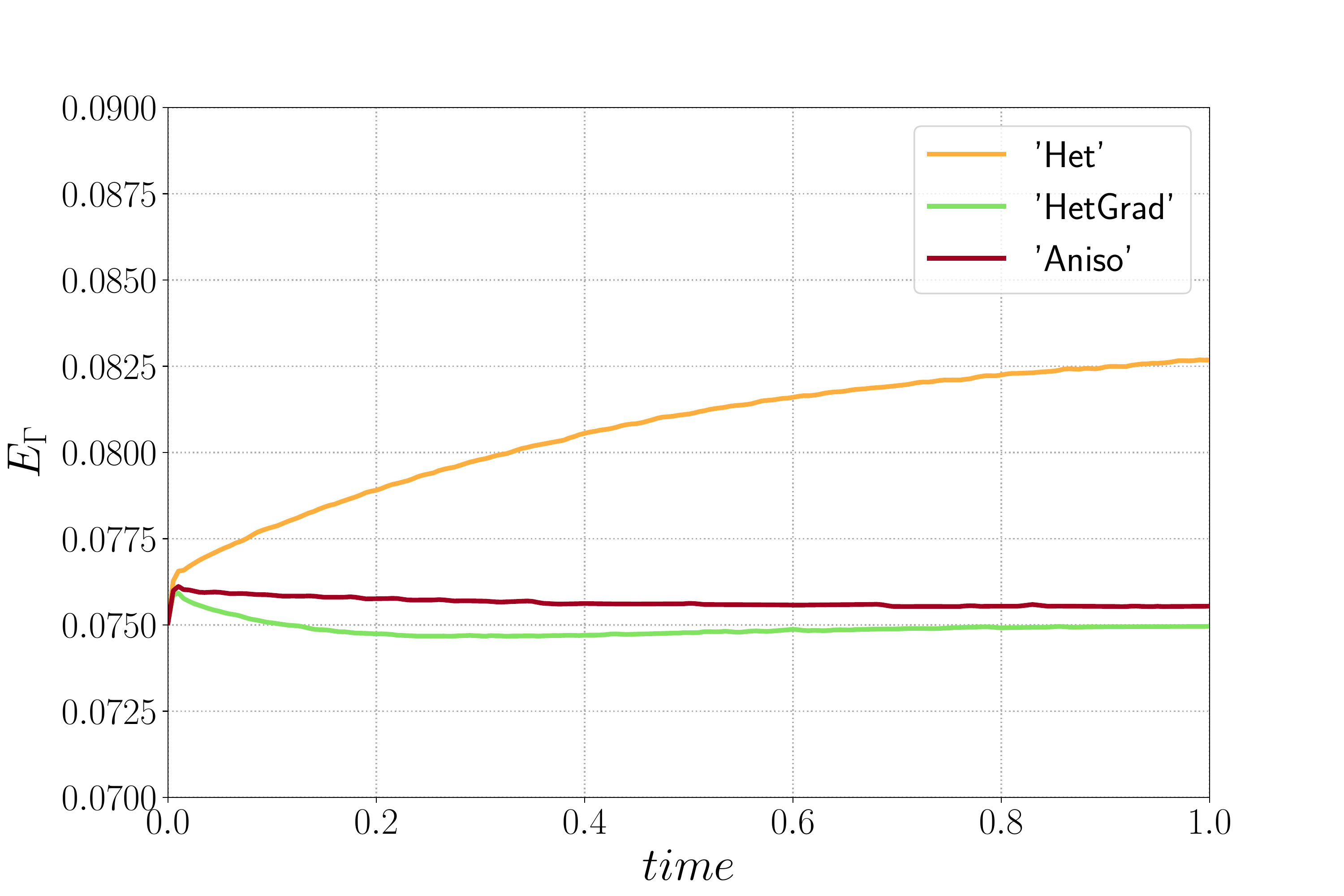}
    \caption{$r=1.66$}
  \end{subfigure}
  \begin{subfigure}{0.495\textwidth}
    \centering
    \includegraphics[scale=0.25]{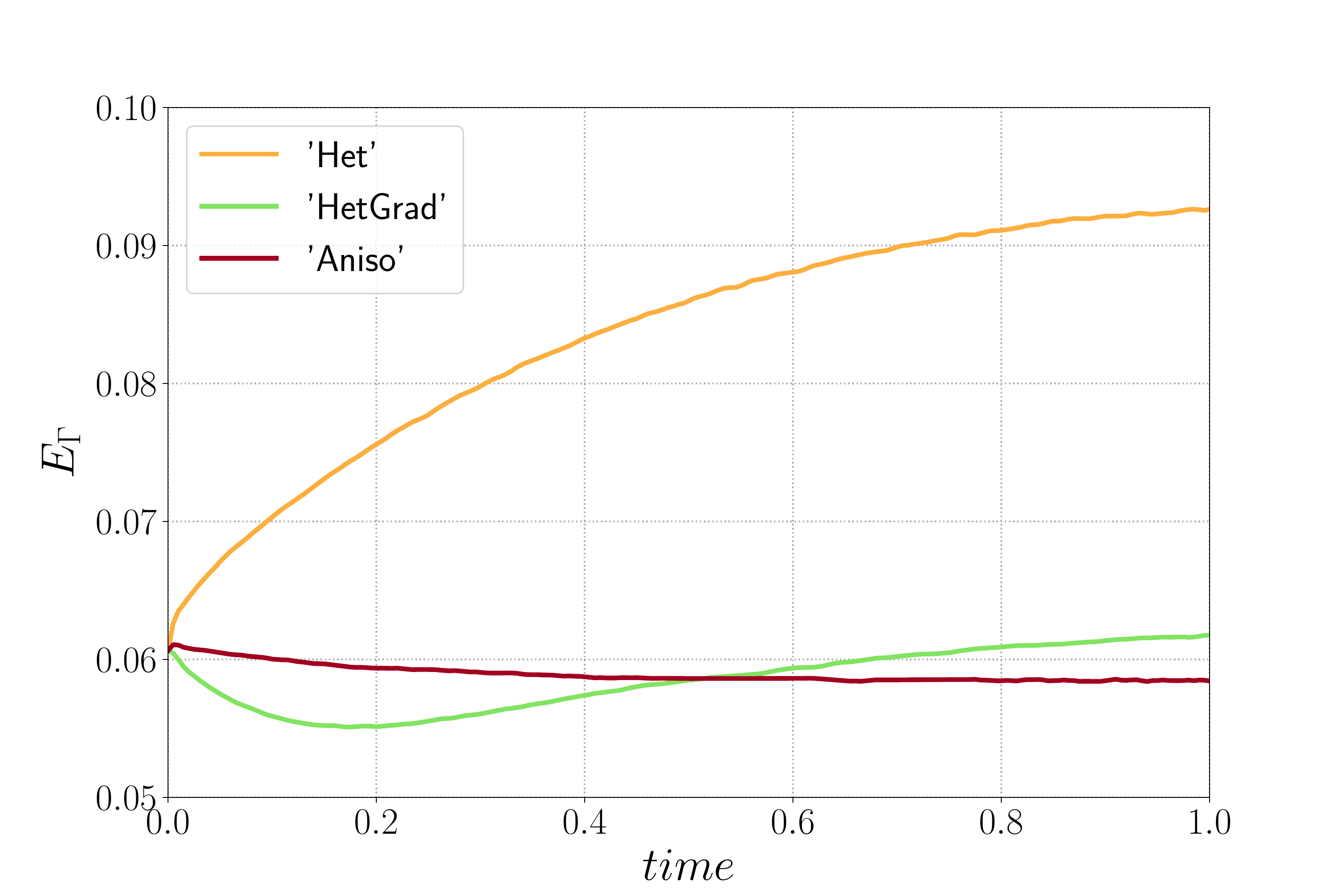}
    \caption{$r=10$}
  \end{subfigure}
  \caption{Variation of the interfacial energy $E_{\Gamma}$ using two different values of $r$}\label{fig:TJTriangleIntenergy}
\end{figure}

In fig.\ref{fig:TJTriangleIntenergy} one can see the evolution of $E_{\Gamma}$. The trends are similar to the previous test case. The HetGrad and Aniso formulations have a better energetic behavior and the Aniso formulation remains the best option for high anisotropy levels.

\subsection{Conclusion}
These results highlight that the Aniso formulation seems to be the most physically acceptable approach regarding the velocity of the triple junction and the interfacial energy evolution. Additionally it also represents the dihedral angles correctly for a wide range of anisotropy levels. Nevertheless, these results must be reinforced with large scale simulations of polycrystals which is the subject of the next section.

\section{Effect of the texture and heterogeneous GB properties during GG simulations for a polycrystalline microstructure}
\label{sec:PX}

\begin{figure}[h]
  \centering
  \includegraphics[scale=0.25]{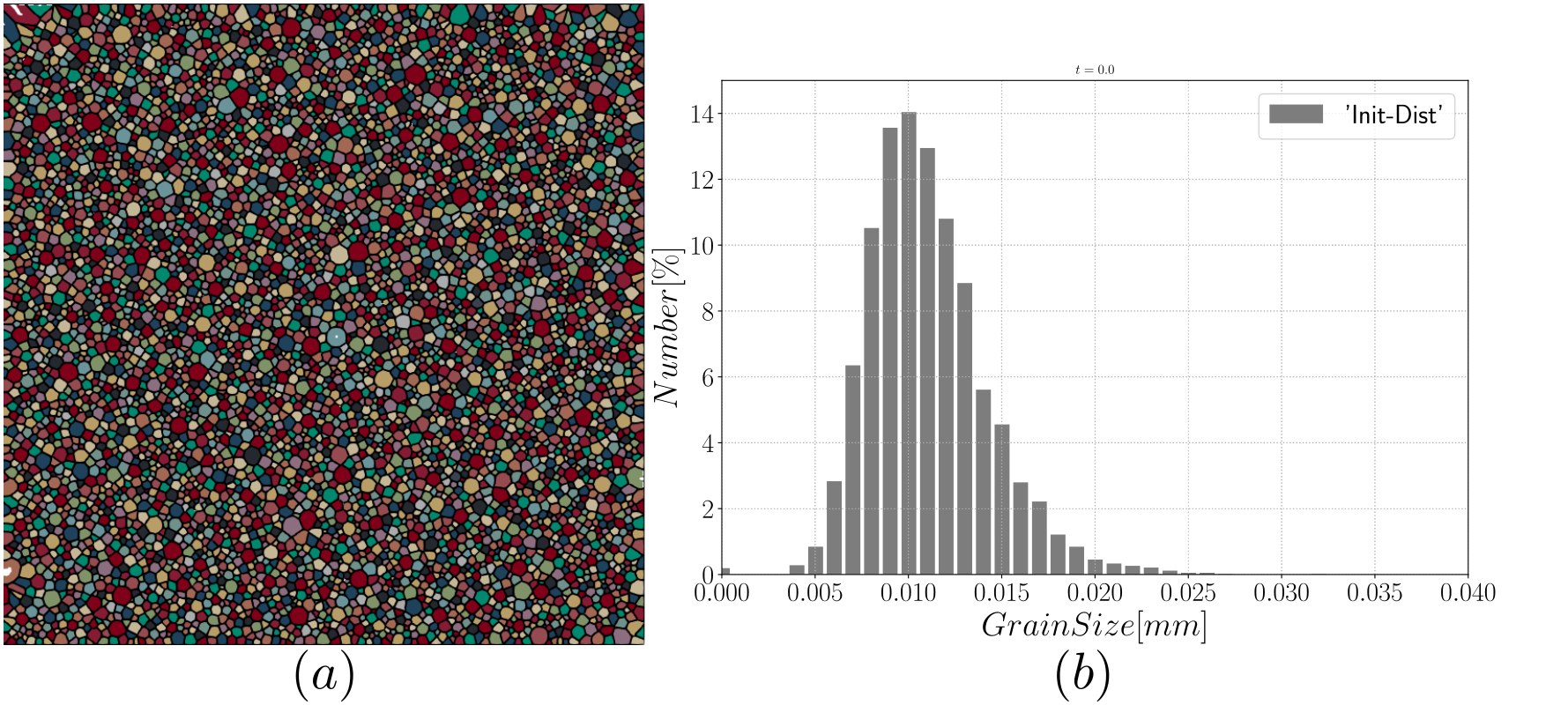}
  \caption{Initial microstructure (a) with 5000 grains and the grain size distribution in number (b). }
  \label{fig:5000G}
\end{figure}

In this section we study a representative GB network in $2D$. Figure \ref{fig:5000G} exhibits the initial characteristics of the microstructure, it consists of a square domain with length $L=1.6mm$ and 5000 grains generated using a Laguerre-Voronoi tessellation \citep{Hitti2012} based on an optimized sphere packing algorithm \citep{hitti2013optimized} with a log-normal distribution for the mean grain size weighted in number. The grain size $R$ of each grain is defined as $\sqrt{S/\pi}$ with $S$ its surface (i.e. defined as the radius of the equivalent circular grain of same surface). Anisotropic remeshing is used following Eq.\ref{eqn:Mesh} with a refinement close to the interface, the mesh size in the tangential direction (as well as far from the interface) is fixed at $h_t=5 \mu m$ and at $h_n=1 \mu m$ in the normal direction. The time step is fixed at $\Delta t = 10 s$. This section is mainly devoted to study the heterogeneity of both GB energy and mobility using the four introduced grain growth formulations. Finally, the same study is performed using a different texture. 

\subsection{Effect of the heterogeneity}
\label{ssec:Random}

Here we use a misorientation dependent GB energy and mobility defined respectively with a Read-Shockley (RS) function \citep{ReadShockley} and a Sigmoidal (S) function proposed by Humphreys in \citep{humphreys1997unified}:
\begin{align}
  \left\{
  \begin{array}{l}
    \gamma(\theta) = \gamma_{max} \frac{\theta}{\theta_0} \left( 1 - ln \left( \frac{\theta}{\theta_0} \right) \right), \ \theta < \theta_0 \\
    \gamma_{max}, \ \theta \ge \theta_0
\end{array}
\label{eqn:Gamma}
\right .
\end{align}

\begin{align}
  \mu(\theta)=\mu_{max} \left( 1 - exp \left( -5 \left(\frac{ \theta}{\theta_0} \right)^4 \right) \right),  \label{eqn:Mob}
\end{align}
where $\theta$ is the disorientation, $\mu_{max}$ and $\gamma_{max}$ are the maximal GB mobility and energy, respectively. $\theta_0=30 \degree $ is the disorientation defining the transition from a low angle grain boundary (LAGB) to a high angle grain boundary (HAGB). $\theta_0$ is normally considered to be between $15-20 \degree  $ but here this parameter is exaggerated to exacerbate the heterogeneity of the system. The maximal values for the GB properties are $\mu_{max}=1.379 mm^4 J^{-1} s^{-1} $, and $\gamma_{max}=6$e$-7 J mm^{-2} $ and are typical for a stainless steel \citep{CRUZFABIANO2014305}. 

\begin{figure}[h]
  \centering
  \begin{subfigure}{0.495\textwidth}
    \centering
    \includegraphics[scale=0.25]{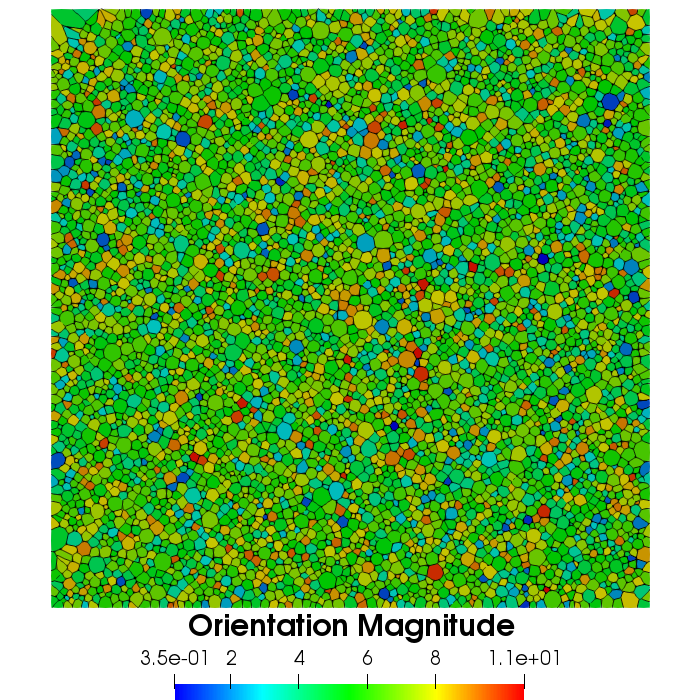}
    \caption{Orientation}
  \end{subfigure}
  \begin{subfigure}{0.495\textwidth}
    \centering
    \includegraphics[scale=0.25]{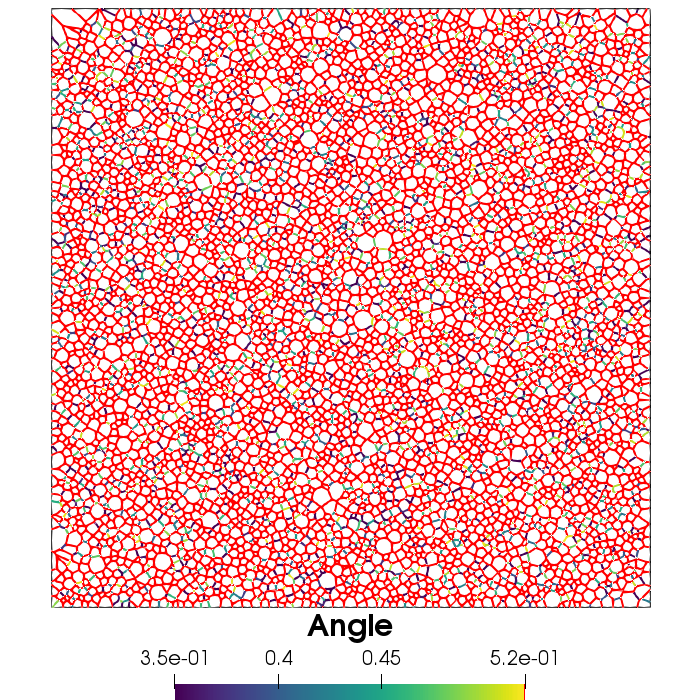}
    \caption{Disorientation}
  \end{subfigure}
  \caption{Initial crystallographic characteristics.}\label{fig:PX5000XOrien}
\end{figure}

\begin{figure}[h]
  \centering
  \begin{subfigure}[c]{0.495\textwidth}
    \centering
    \includegraphics[scale=0.25]{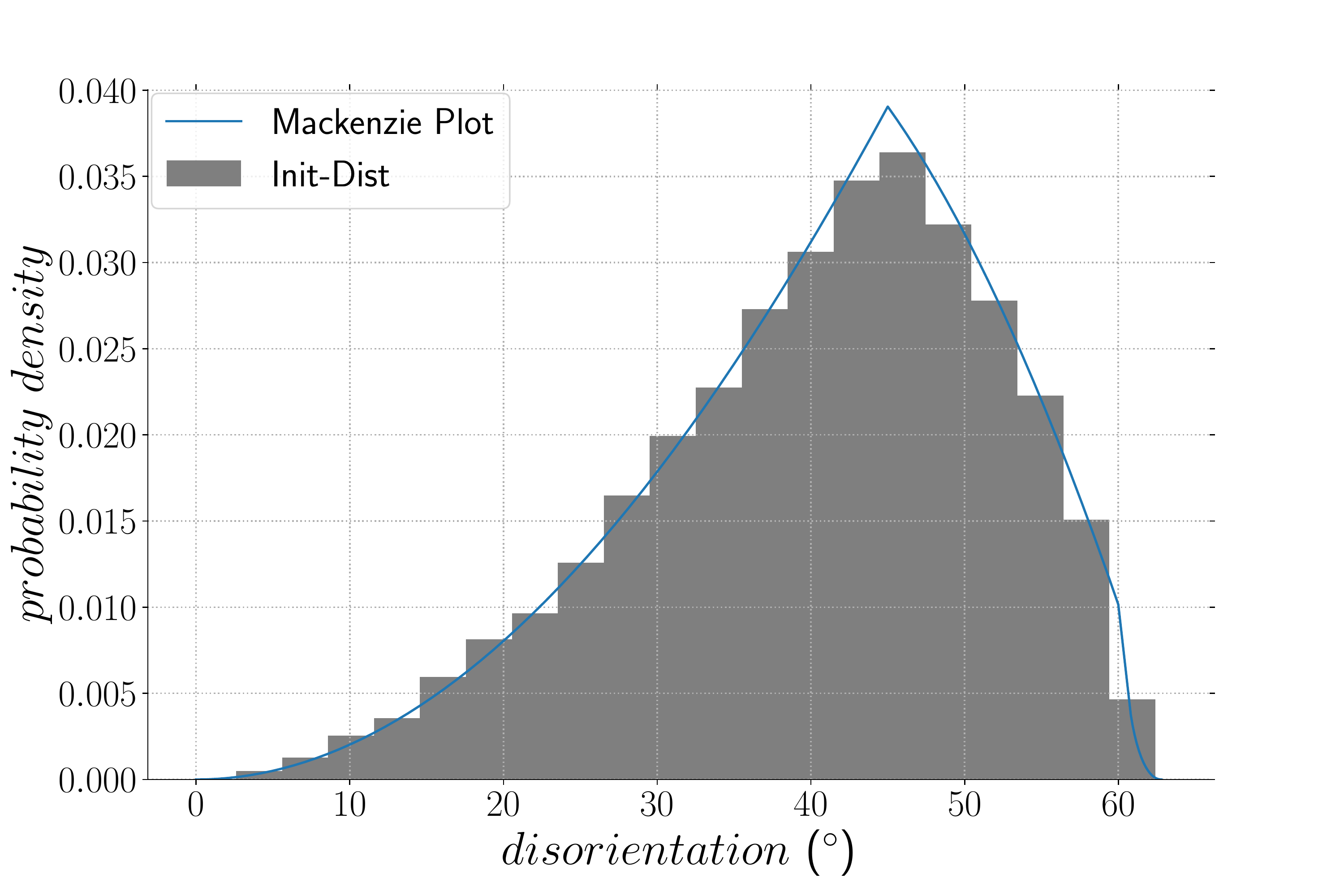}
    \caption{Initial disorientation distribution function}
  \end{subfigure} 
  \begin{subfigure}[c]{0.494\textwidth}
    \centering
    \includegraphics[scale=0.25]{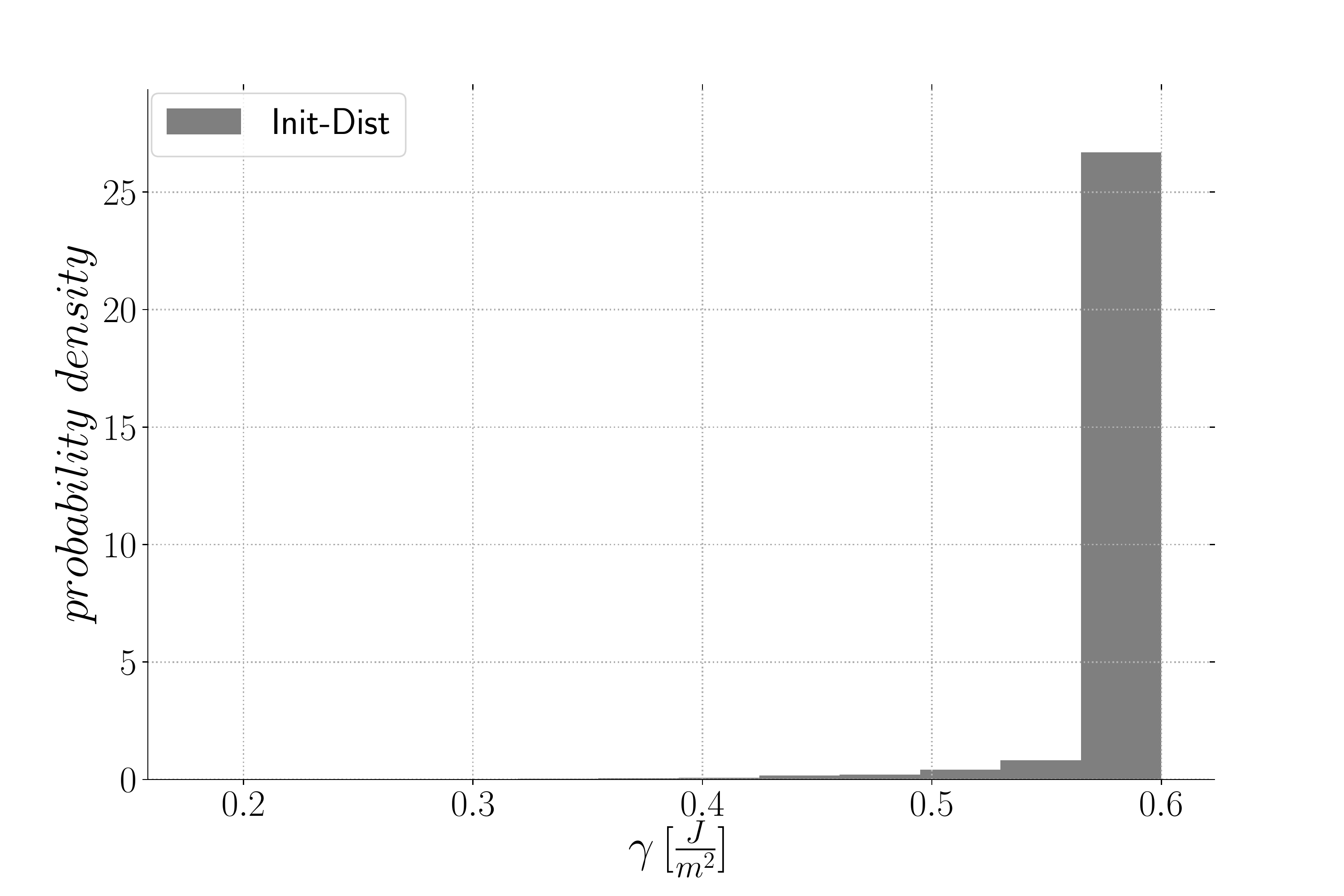}
    \caption{Initial grain boundary energy distribution}
  \end{subfigure}
  \caption{Initial GB characteristics.}
  \label{fig:PX5000CrysChar}
\end{figure}

Figure \ref{fig:PX5000XOrien} shows the orientation field using the vector magnitude $O_G = \sqrt{\varphi_1^2 + \phi^2 + \varphi_2^2 }$ where $(\varphi_1, \phi,\varphi_2)$ are the three Euler angles. The Euler angles defining the crystallographic orientations generated in this case are generated randomly, leading to a Mackenzie-like disorientation distribution function \citep{mackenzie1958second}. As the Read-Shockley model is used to define $\gamma$, the GB energy is concentrated at high values as illustrated in figure~\ref{fig:PX5000CrysChar}. 

\subsubsection{Heterogeneous grain boundary energy}
\label{sssec:PX1000GRS}

\begin{figure}[h]
  \centering
  \begin{subfigure}{0.48\textwidth}
    \centering
    \includegraphics[scale=0.25]{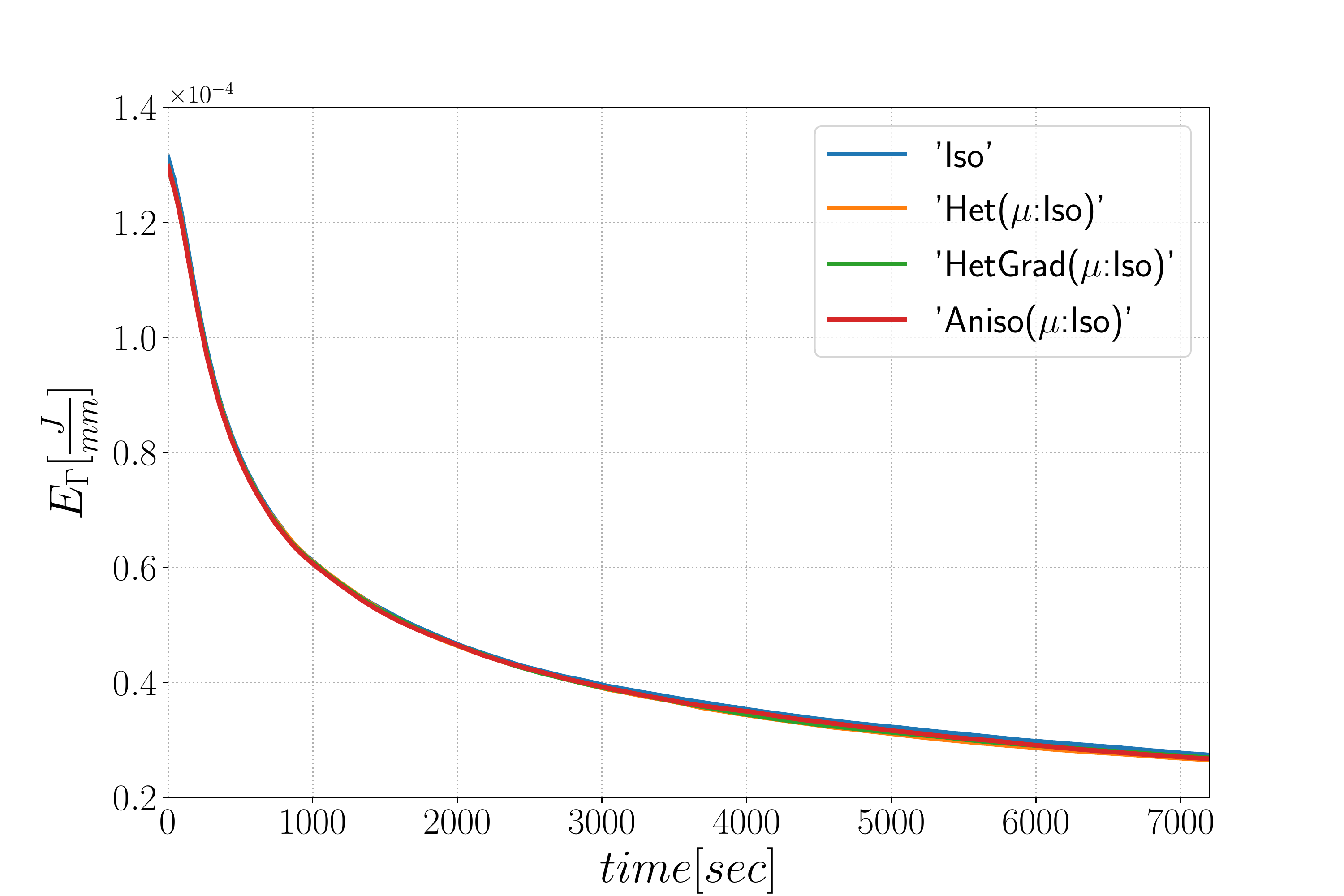}
    \caption{$E_{\Gamma}=f(t)$}
  \end{subfigure} 
  \begin{subfigure}{0.48\textwidth}
    \centering
    \includegraphics[scale=0.25]{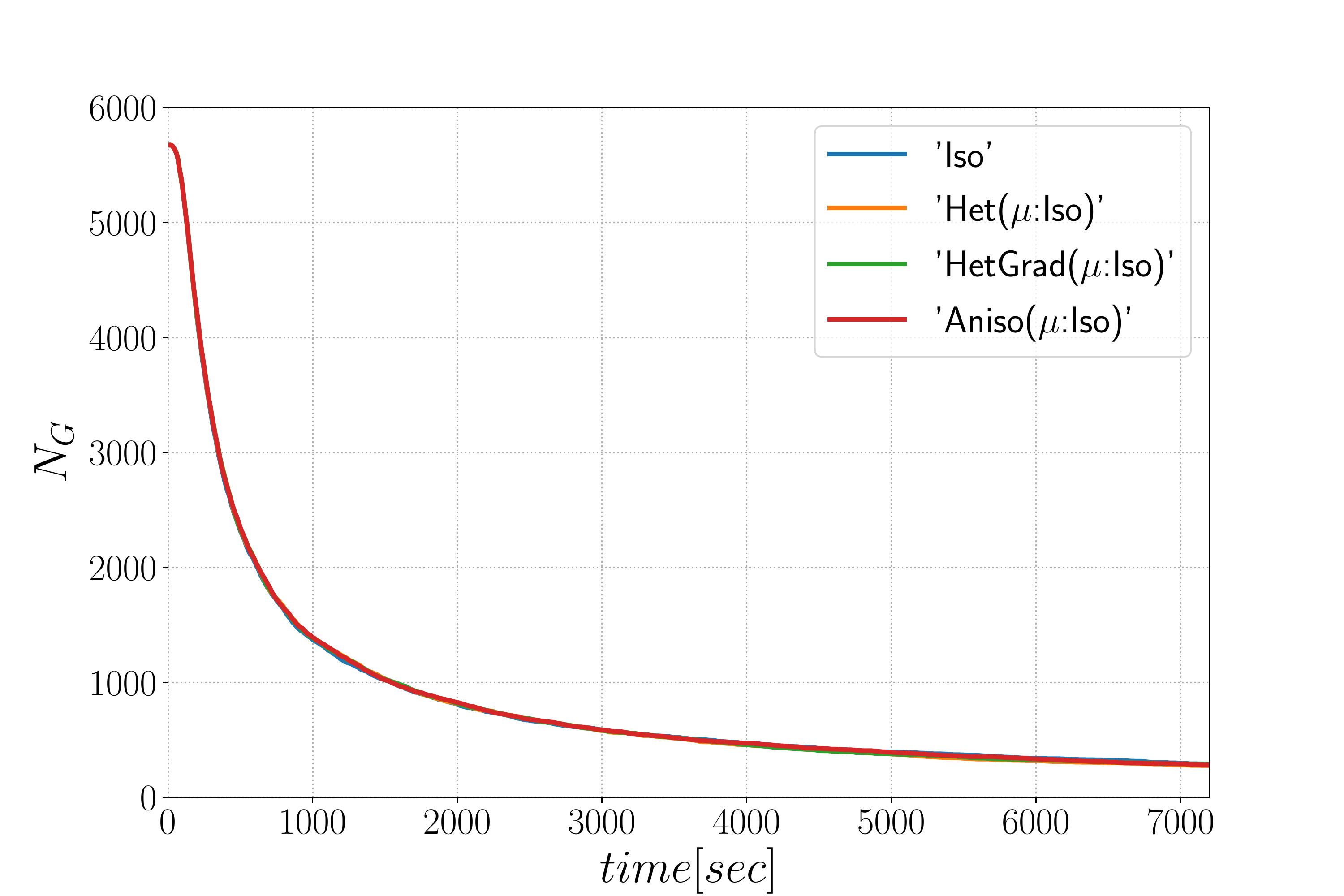}
    \caption{$N_g=f(t)$}
  \end{subfigure} \\
  \begin{subfigure}{0.48\textwidth}
    \centering
    \includegraphics[scale=0.25]{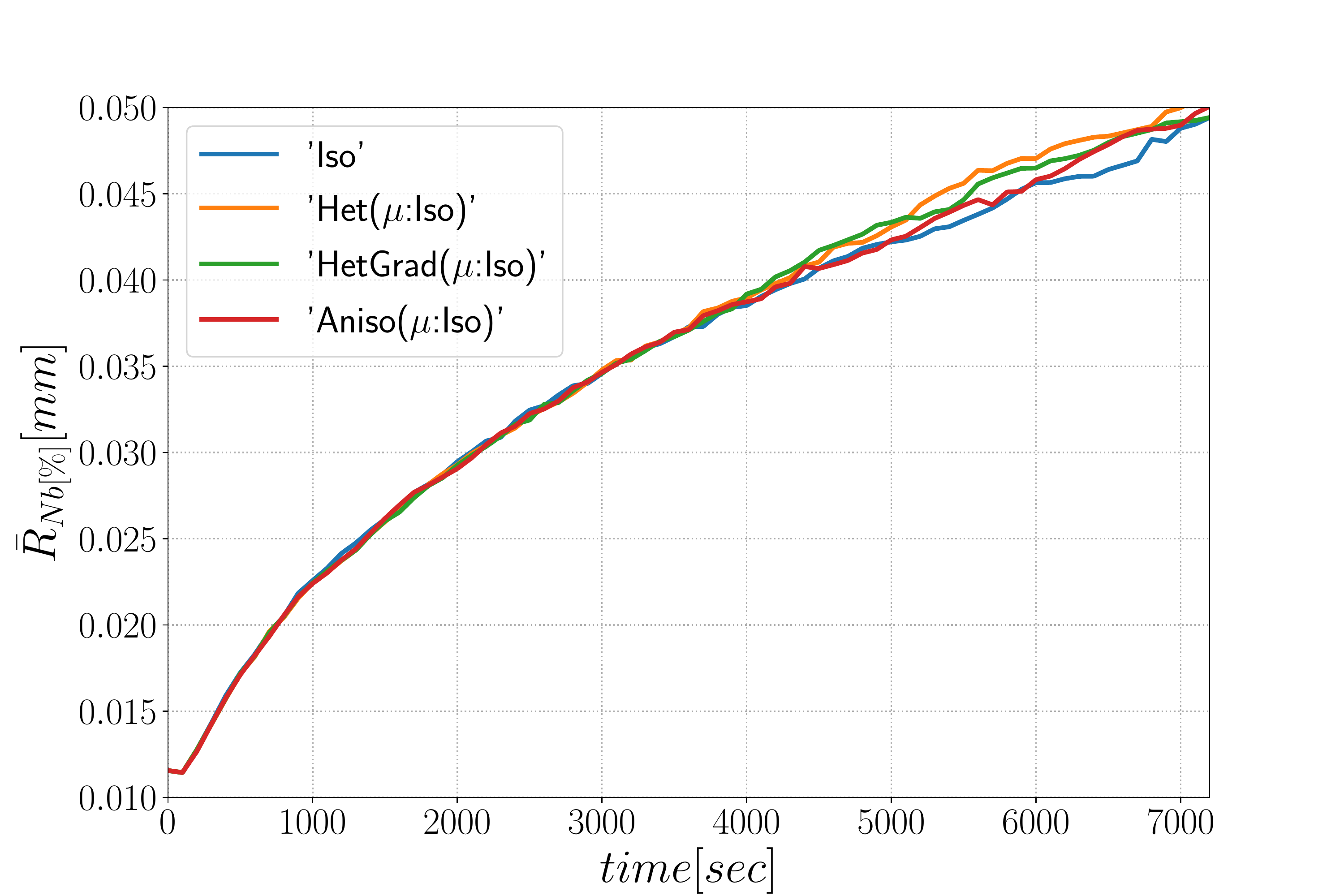}
    \caption{$\bar{R}_{Nb[\%]}=f(t)$}
  \end{subfigure}
  \begin{subfigure}{0.48\textwidth}
    \centering
    \includegraphics[scale=0.25]{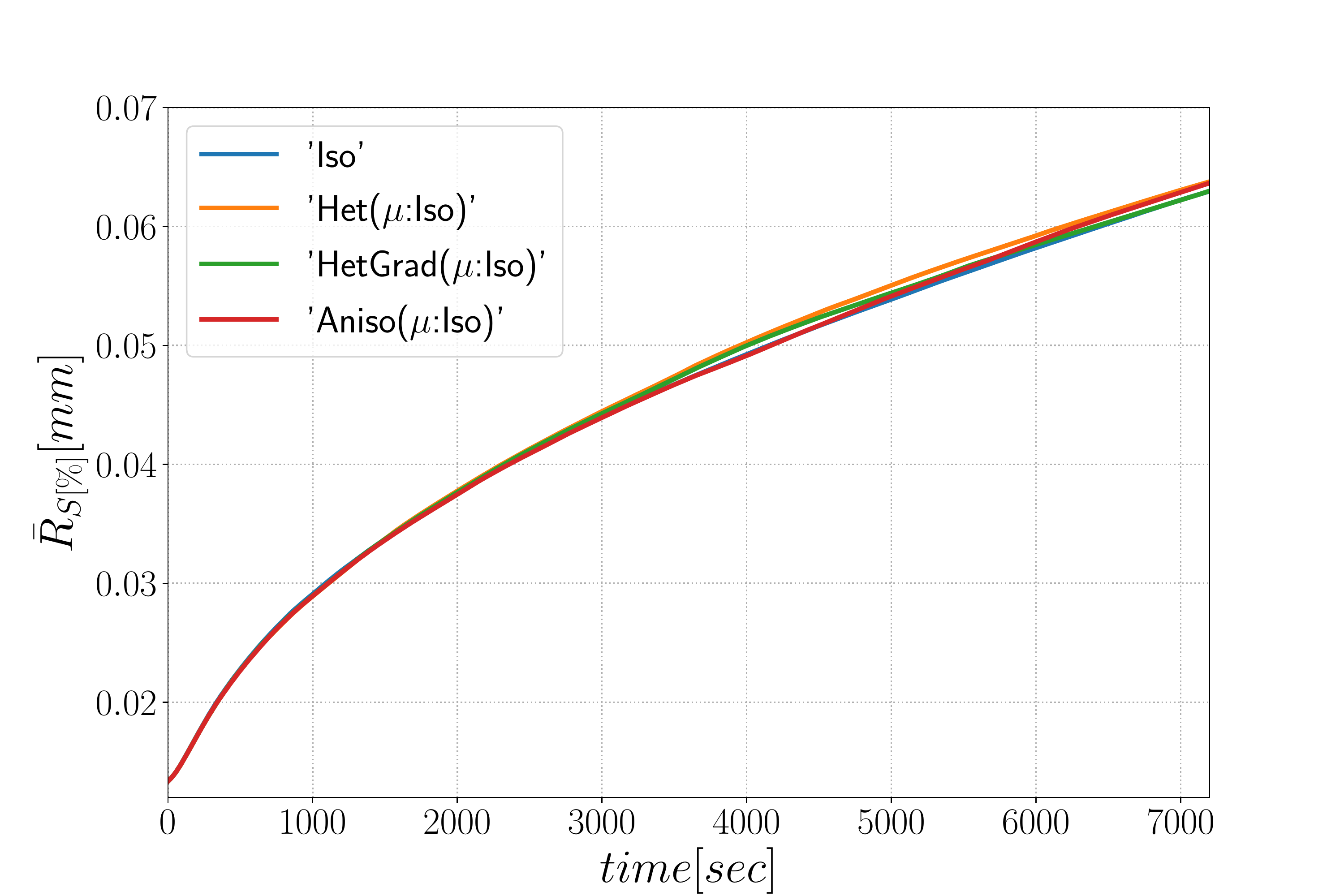}
    \caption{$\bar{R}_{S[\%]}=f(t)$}
  \end{subfigure}
  \caption{Time evolution for the different formulations: (a) the total GB energy, (b) the number of grains, (c) the mean radius weighted in number and (d) the mean radius weighted in surface.}\label{fig:PX5000MuIsoMeanV}
\end{figure}

\begin{figure}[h]
  \centering
  \includegraphics[scale=0.3]{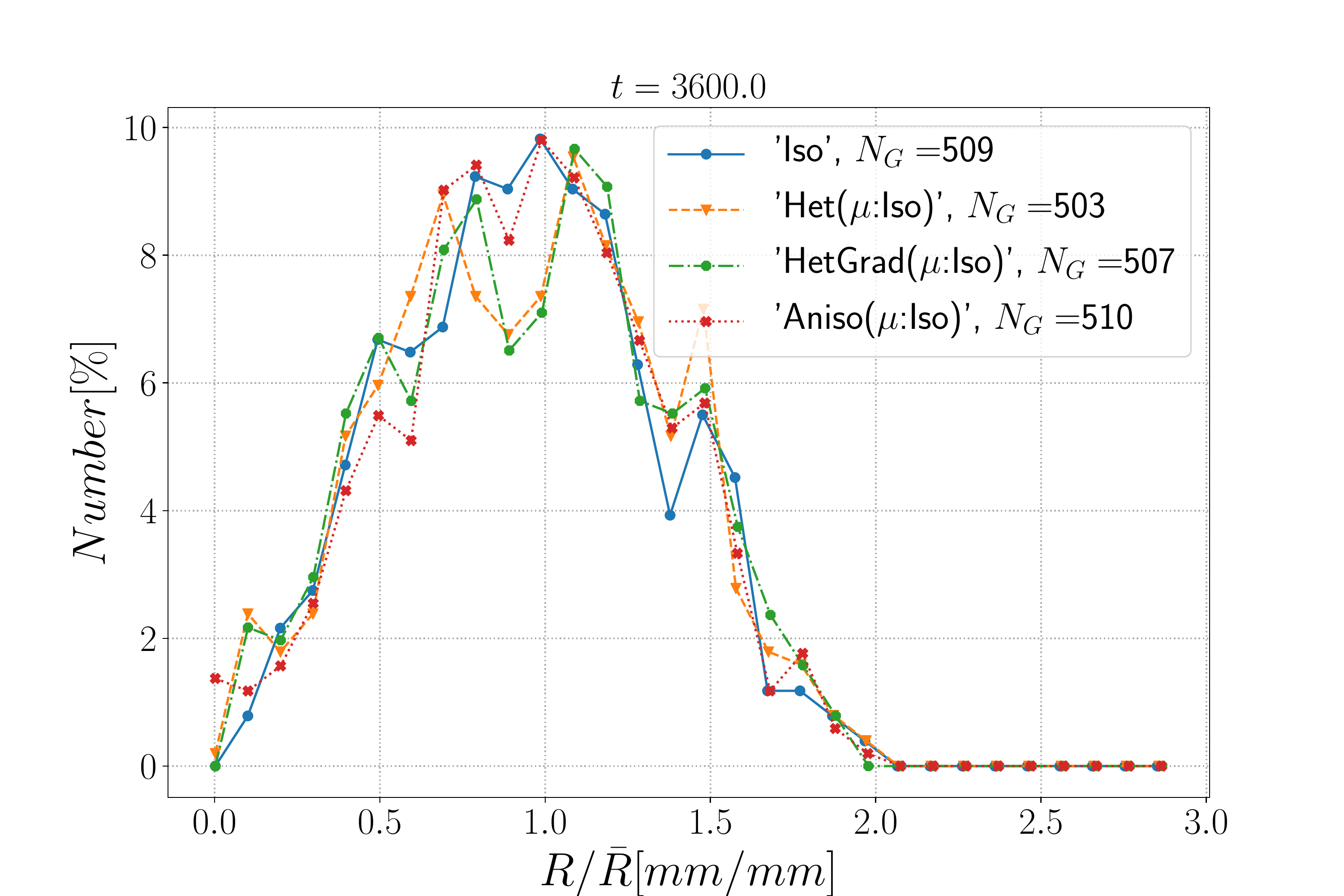}
  \caption{Grain radius distribution weighted in number at $t=1h$.}
  \label{fig:PX5000GSD1h}
\end{figure}

\begin{figure}[h]
  \centering
  \begin{subfigure}{0.48\textwidth}
    \centering
    \includegraphics[scale=0.25]{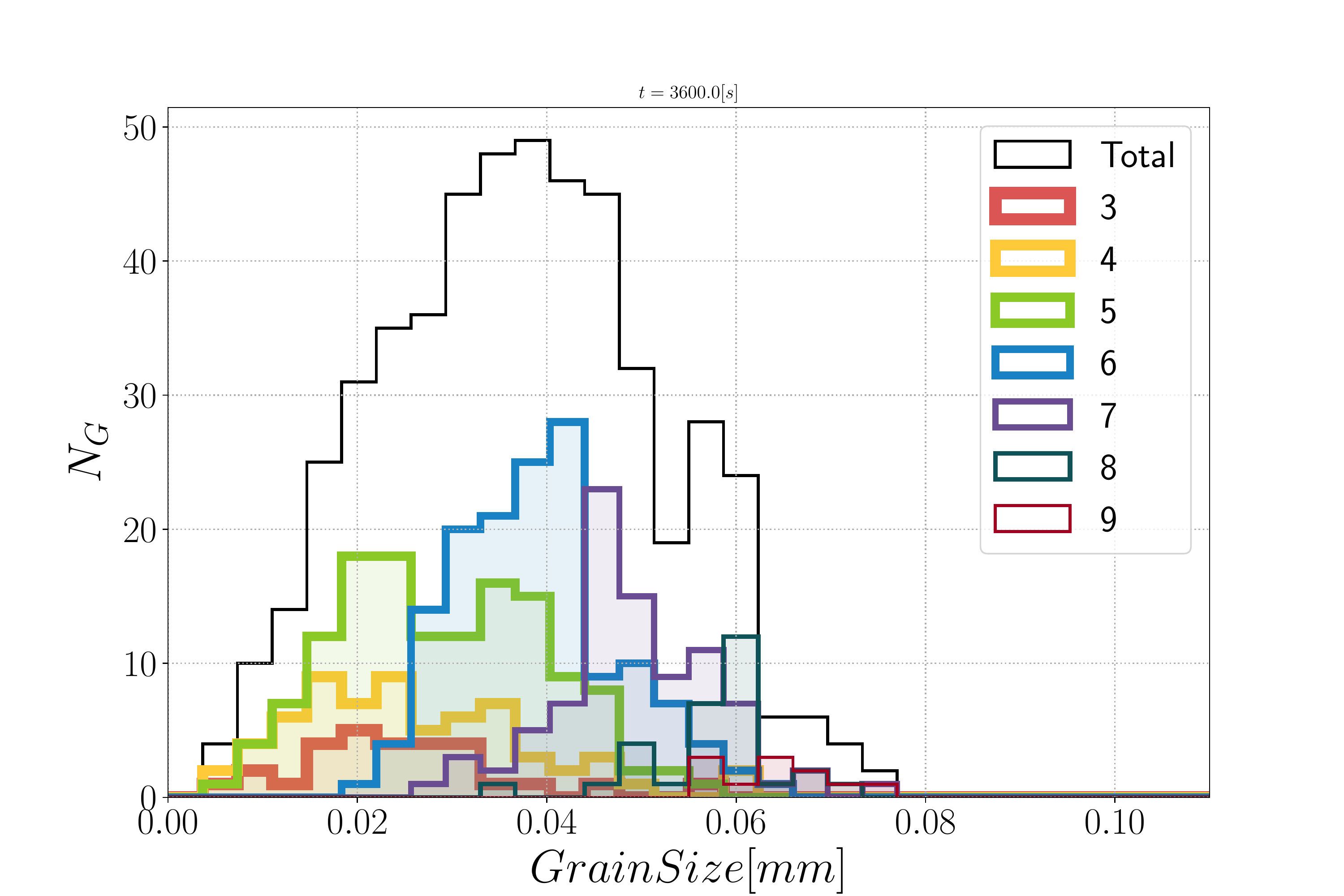}
    \caption{Iso}
  \end{subfigure} 
  \begin{subfigure}{0.48\textwidth}
    \centering
    \includegraphics[scale=0.25]{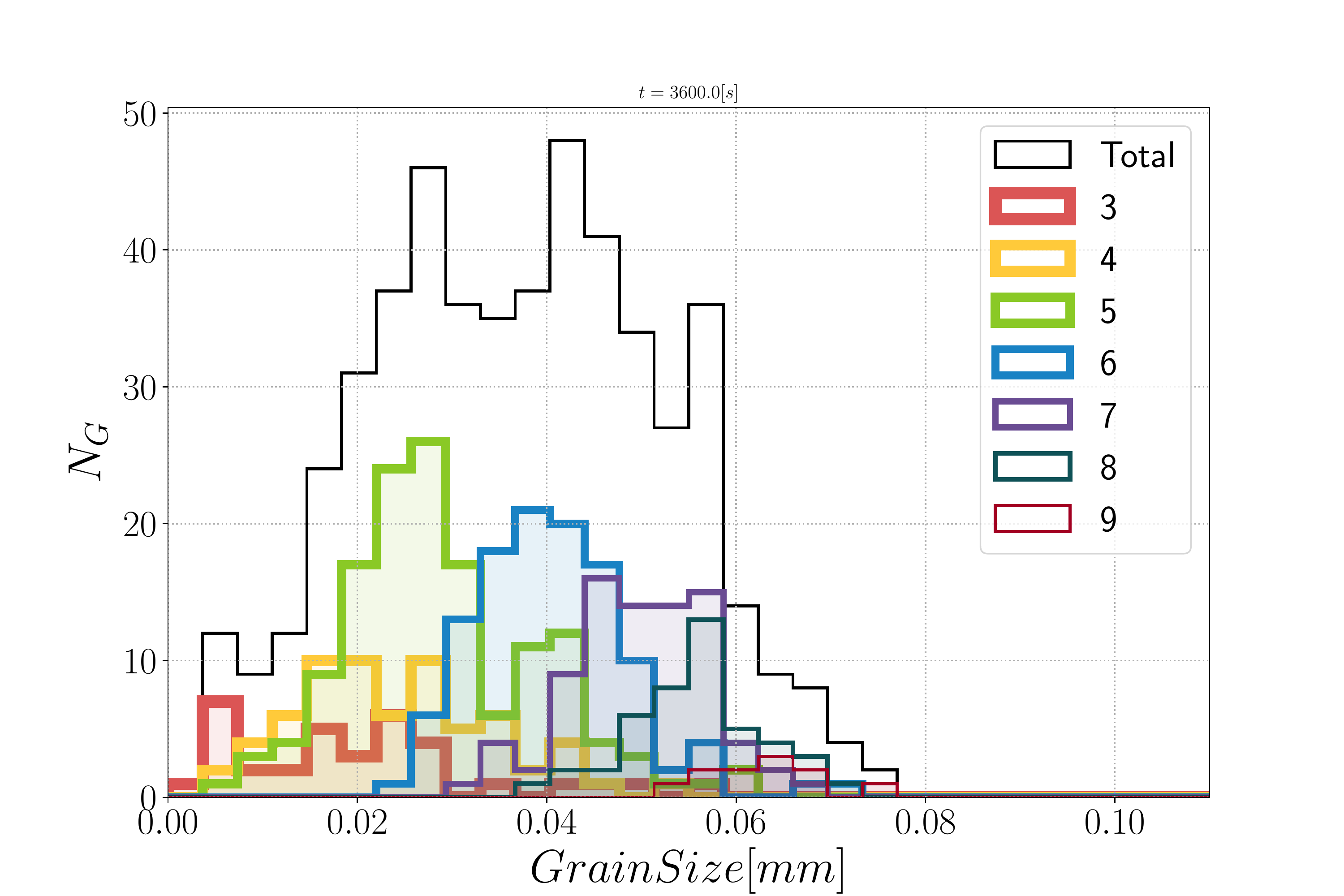}
    \caption{Het($\mu$:Iso)}
  \end{subfigure} \\
  \begin{subfigure}{0.48\textwidth}
    \centering
    \includegraphics[scale=0.25]{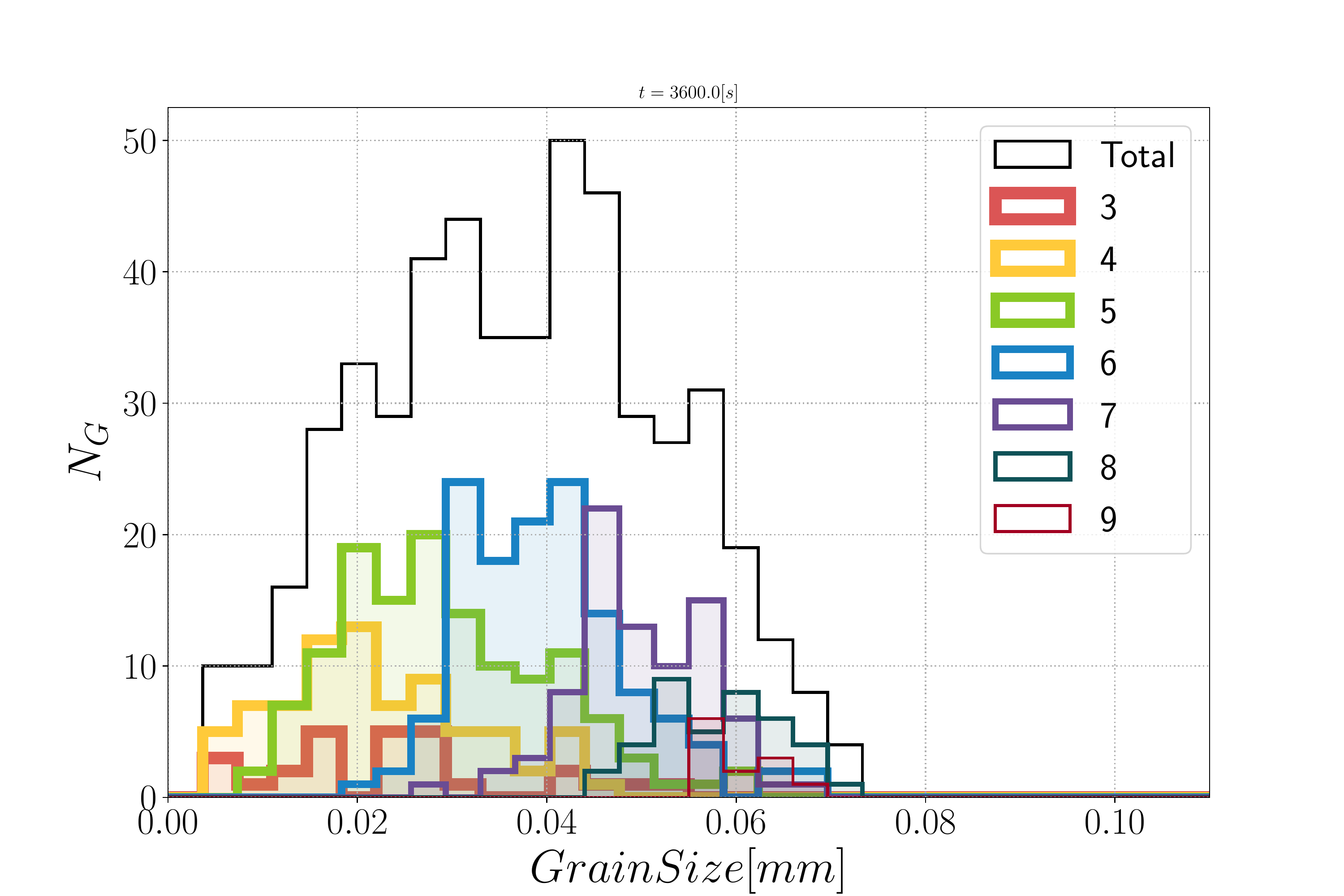}
    \caption{HetGrad($\mu$:Iso)}
  \end{subfigure}
  \begin{subfigure}{0.48\textwidth}
    \centering
    \includegraphics[scale=0.25]{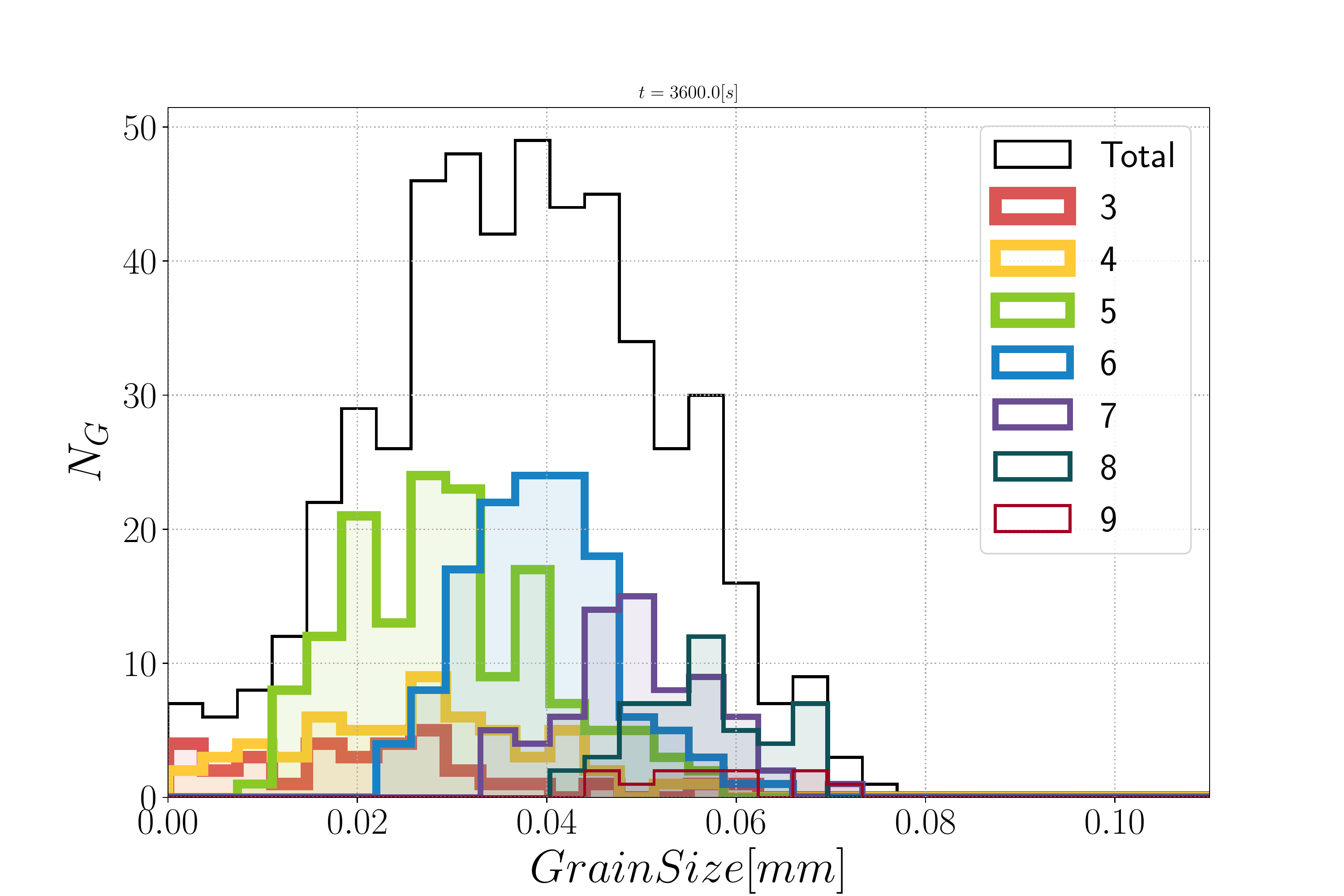}
    \caption{Aniso($\mu$:Iso)}
  \end{subfigure}
  \caption{Grain size distribution and contributions from every group of grains of same coordination number from 3 to 9 at $t=1h$.}\label{fig:PX5000MuRadCooHist}
\end{figure}

In this section GB energy is defined using Eq.~\ref{eqn:Gamma} and GB mobility is assumed isotropic. Hence, the Het, HetGrad, and Aniso formulations are presented as "Het($\mu$:Iso)", "HetGrad($\mu$:Iso)" and "Aniso($\mu$:Iso)". The results are summarized from figures~\ref{fig:PX5000MuIsoMeanV} to \ref{fig:PX5000MuRadCooHist}. First, it is noticeable that all the formulations have a similar evolution concerning the total grain boundary energy $E_{\Gamma}$, the number of grains $N_g$ and the mean grain size weighted in number $\bar{R}_{Nb[\%]}$ or in surface $\bar{R}_{S[\%]}$. Additionally, if the grain size distribution weighted by number is normalized (figure~\ref{fig:PX5000GSD1h}), one can recognize that all the formulations have similar distributions and the minima have similar values with respect to the mean radius. Similar results for the ``Iso'' and ``HetGrad'' formulations with heterogeneous GB energy defined by the Read-Shockley model were already reported \citep{Fausty2020}. 

The slow evolution of the mean values has been reported as a consequence of little local-heterogeneity produced by a Mackenzie type disorientation distribution function (DDF) and/or a low value of $\theta_0$ \citep{holm2001misorientation, chang2014effect, gruber2009misorientation, elsey2013simulations}. If the DDF starts as a Mackenzie distribution, the value of GB mobility and energy is focused at higher values, thus the microstructure cannot easily find a path to minimize its energy faster and the DDF changes slightly from its initial Mackenzie form. In other words the initial configuration is almost isotropic. Slight differences can be observed after $t=1h$ for the different formulations, this may due to the low final number of grains ($N_G \approx 500$).

Regarding the morphology of the microstructures at $t=1h$, the grains are equiaxed. If we divide the total group of grains in classes divided by the number of neighbors (defined as the coordination number in the following), $n$, an interesting analysis regarding the morphology of grains could be done. In figure~\ref{fig:PX5000MuRadCooHist}, the contribution of every class is depicted and at $t=0s$ most of the grains verify $n=5$. After one hour, one can directly appreciate that the classes with $n=6$ is the main class using the four formulations. This agree with theoretical predictions of grain boundary motion with isotropic GB energy which promotes triple junctions with dihedral angles near $120^{\circ}$ \citep{cahn1991stability}. This aspect illustrates again the limited impact of the considered anisotropy in this configuration.

\subsubsection{Heterogeneous grain boundary energy and mobility}
\label{sssec:PX1000GRSMuS}

\begin{figure}[h]
  \centering
  \includegraphics[scale=0.25]{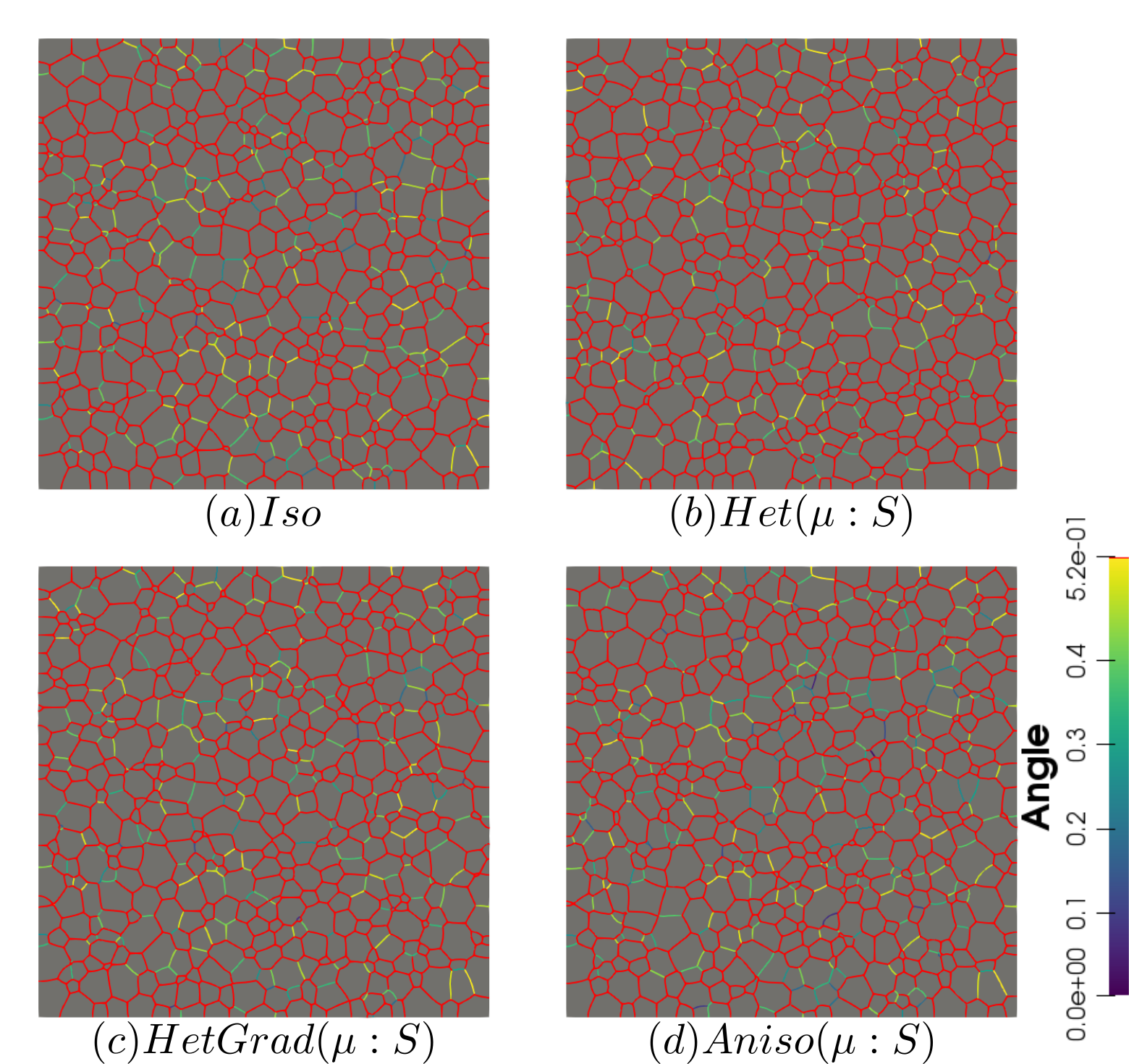}
  \caption{Disorientation of the boundaries using the four formulations with heterogeneous grain boundary mobility at $t=1h$, boundaries with a disorientation higher than $30\degree$ are colored in red.}\label{fig:PX5000MuSIntMiso}
\end{figure}

\begin{figure}[h]
  \centering
  \begin{subfigure}{0.48\textwidth}
    \centering
    \includegraphics[scale=0.25]{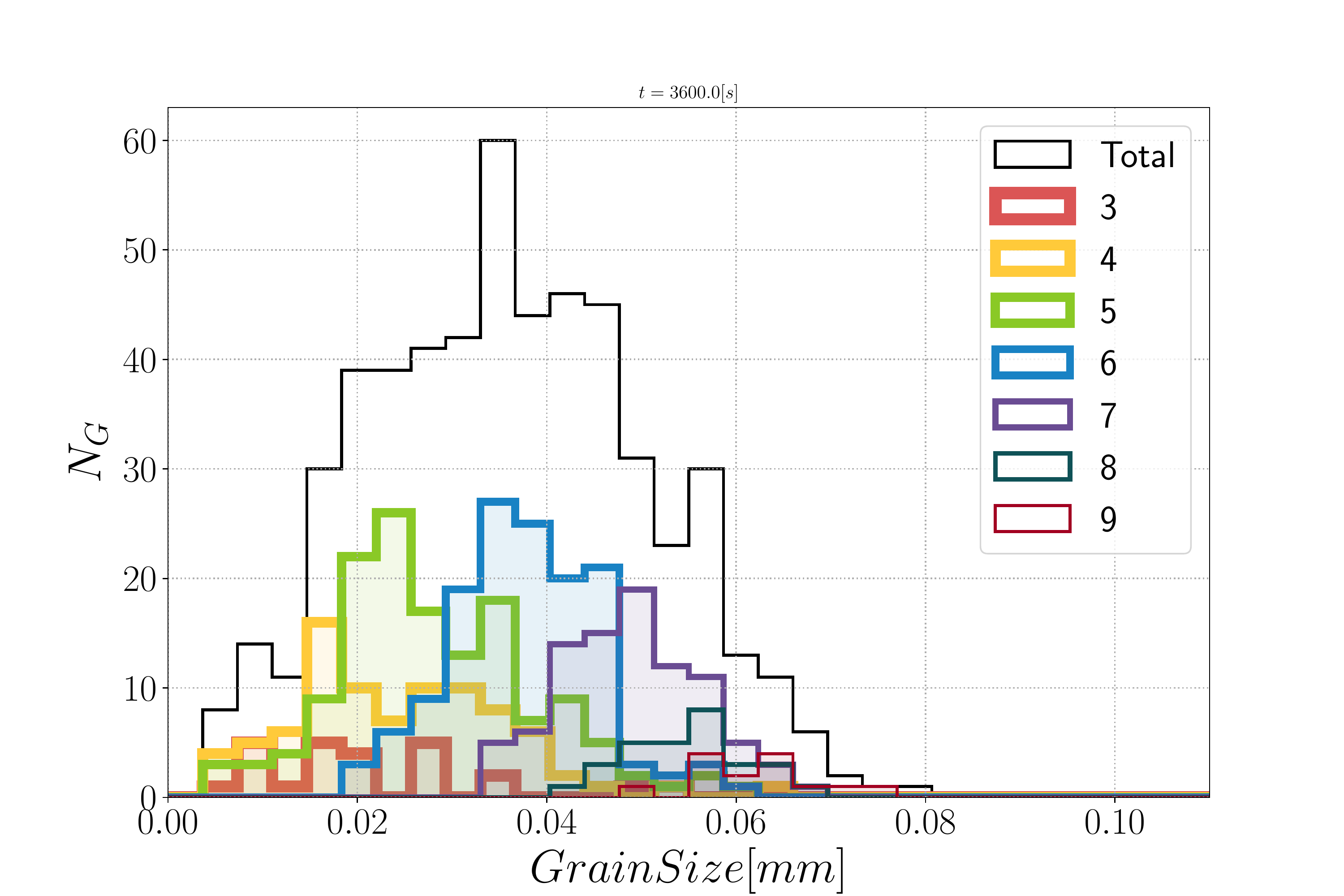}
    \caption{HetGrad($\mu$:S)}
  \end{subfigure}
  \begin{subfigure}{0.48\textwidth}
    \centering
    \includegraphics[scale=0.25]{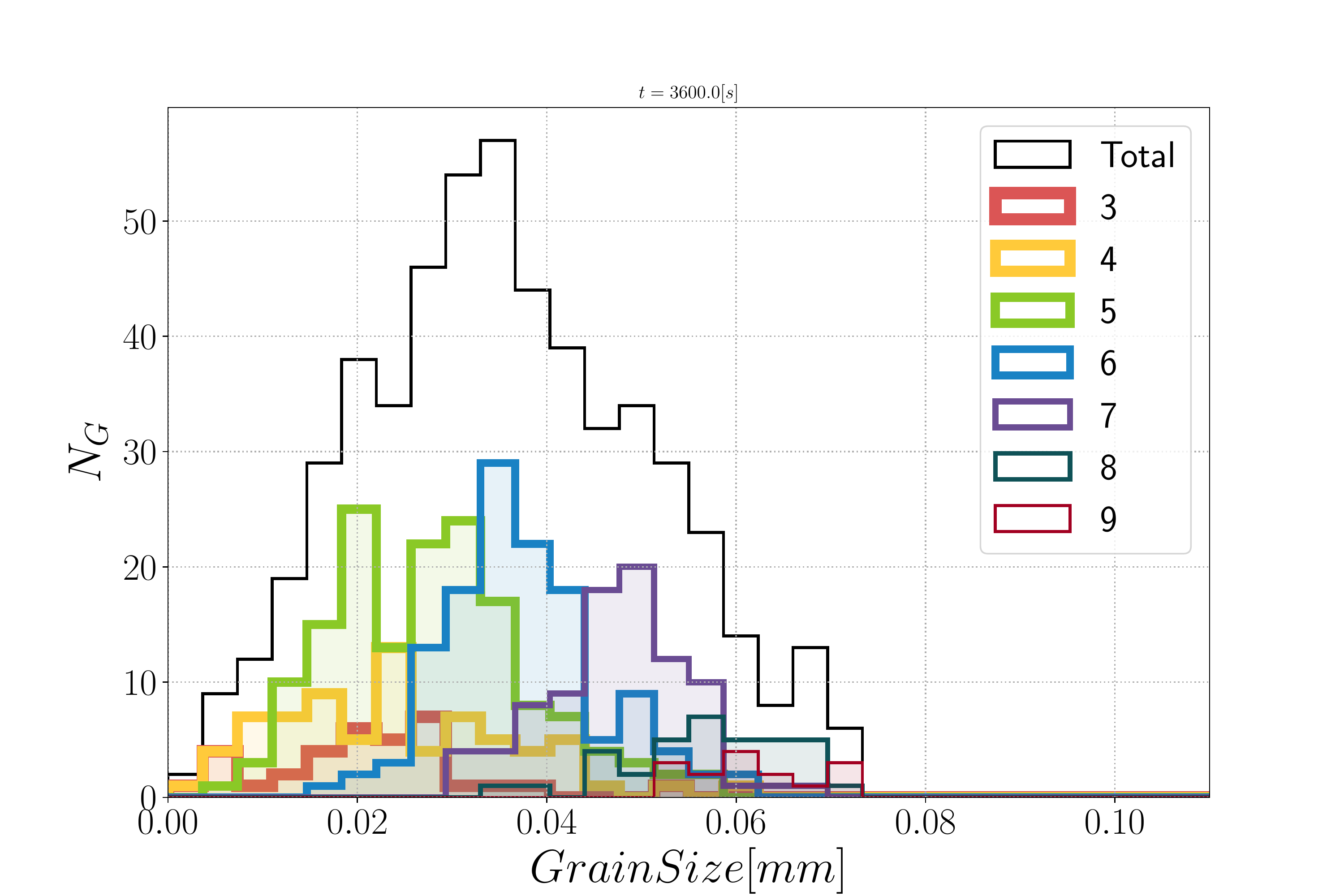}
    \caption{Aniso($\mu$:S)}
  \end{subfigure}
  \caption{Grain size distribution and contributions from every group of grains of same coordination number from 3 to 9 at $t=1h$.}\label{fig:PX5000MuSRadCooHist}
\end{figure}

In this section both GB energy and mobility are heterogeneous, respectively defined with Eqs.~\ref{eqn:Gamma} and \ref{eqn:Mob}, for that reason the names introduced above are replaced by "Het($\mu$:S)", "HetGrad($\mu$:S)" and "Aniso($\mu$:S)". In order to compare the results presented above, the same initial microstructure and crystallographic orientations are used. The mean values evolution and distributions remain similar among the four formulations and keep similar values as presented before. The heterogeneous GB mobility may affect the morphology of the microstructure due to a retarding effect from boundaries with disorientation lower than $\theta_0$. There is similarity between the four microstructures shown in figure~\ref{fig:PX5000MuSIntMiso} showing mostly equiaxed grains. Two important aspects of these microstructures are that the microstructure obtained by the ``Het'' formulation is the most dissimilar with a lower number of boundaries with disorientation inferior to $\theta_0$. Second, the presence of low angle boundaries ($\theta < 30 \degree$) looks higher using the Anisotropic formulation. Nevertheless, this is not reflected in the interfacial energy evolution nor the DDF (see Fig.\ref{fig:PX5000MuSRadCooHist}). 

\begin{figure}[h]
  \centering
  \begin{subfigure}[c]{0.48\textwidth}
    \centering
    \includegraphics[scale=0.25]{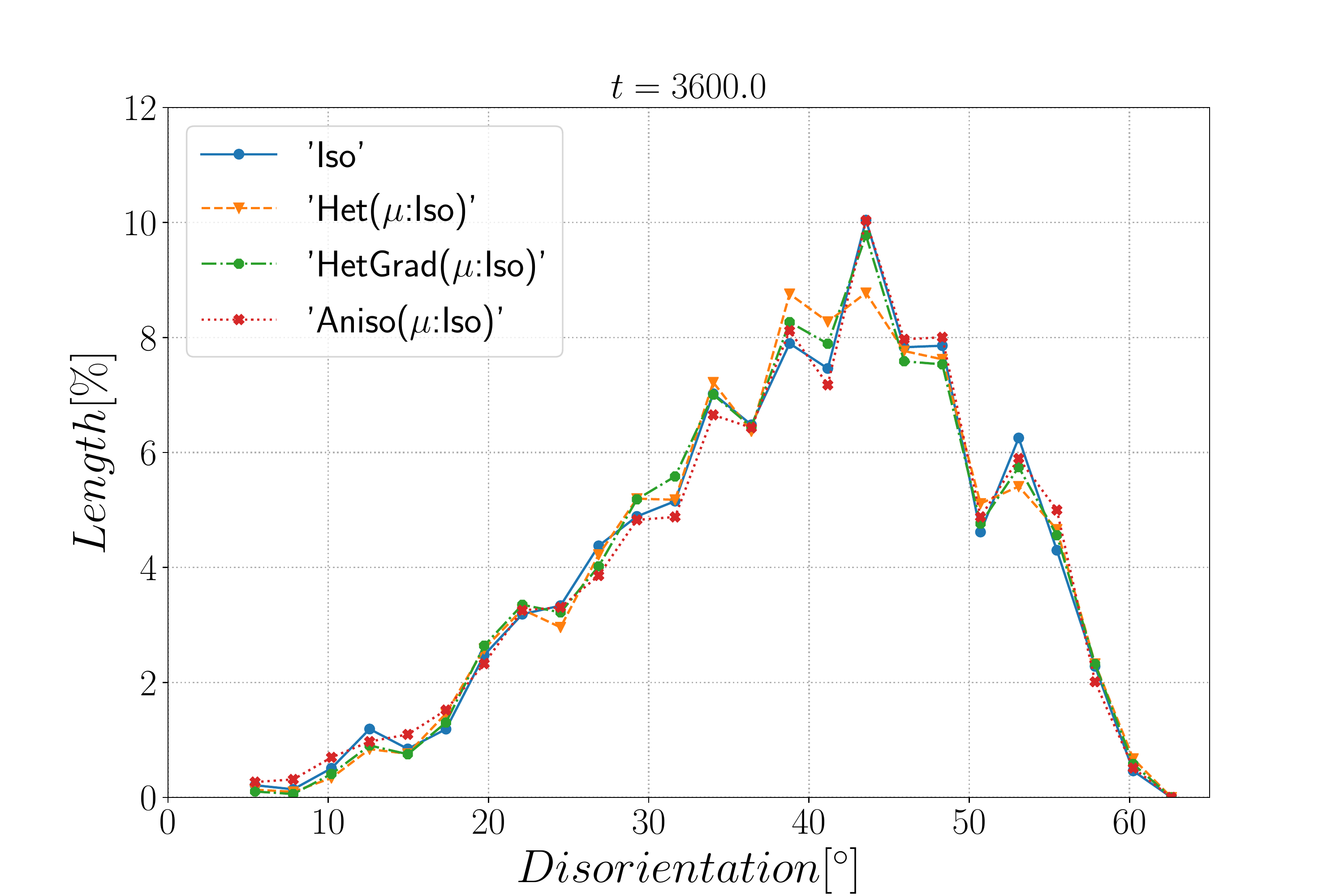}
    \caption{$\mu$ isotropic}
  \end{subfigure}
  \begin{subfigure}[c]{0.48\textwidth}
    \centering
    \includegraphics[scale=0.25]{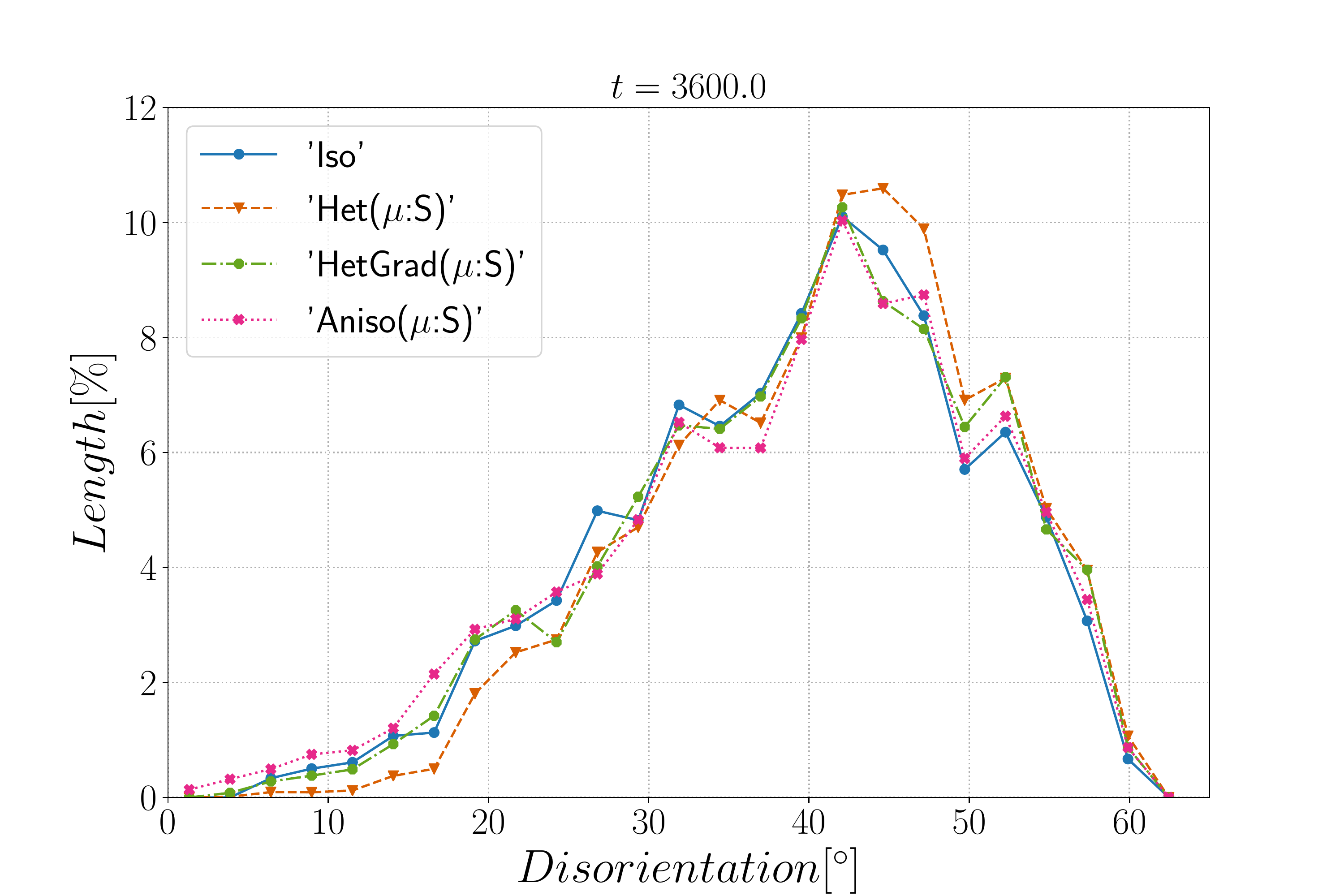}
    \caption{$\mu$ sigmoidal}
  \end{subfigure}
  \caption{Disorientation distribution function at $t=1h$ using an isotropic (a) and a heterogeneous (b) mobility.}
  \label{fig:PX5000DDGBF}
\end{figure}

Finally, figure~\ref{fig:PX5000DDGBF} shows the disorientation distribution function using both an isotropic and heterogeneous mobility at $t=1 h$. As said before, the initial Mackenzie-type distribution evolves slowly, a slow preference of low angles boundaries is found. Using heterogeneous mobility affect slightly the DDF, one can see that the Anisotropic formulation (Aniso($\mu$:S)) exacerbates low values of disorientation reflected in higher values in the distribution at $0 < \theta < 10^{\circ}$. Due to the Mackenzie like DDF, the GB energy distribution is concentrated around $\gamma_{max}$ leading to microstructures with triple junctions angles around $120 \degree$ (see figure~\ref{fig:PX5000MuSIntMiso}). These results are in accordance with prior works \citep{cahn1991stability, holm1993microstructural, chang2014effect}. 

At this point one can see that for an untextured polycrystal with an initial Mackenzie-like DDF, the evolution of the GB network and of the GB energy and mobility fields are similar to an Isotropic case. That is the fundamental reason explaining the weak differences among the results of the different formulations. The results exhibit similar evolution of mean values, distributions and grain morphologies. In order to study the behavior of the different formulations for a wider spectrum of GB properties, the next section is devoted to study the effect of a strong texture using the four formulations with isotropic and heterogeneous mobility.

\subsection{Effect of a strong texture}
\label{ssec:PX1000FT}

\begin{figure}[h]
  \centering
  \begin{subfigure}[c]{0.48\textwidth}
    \centering
    \includegraphics[scale=0.25]{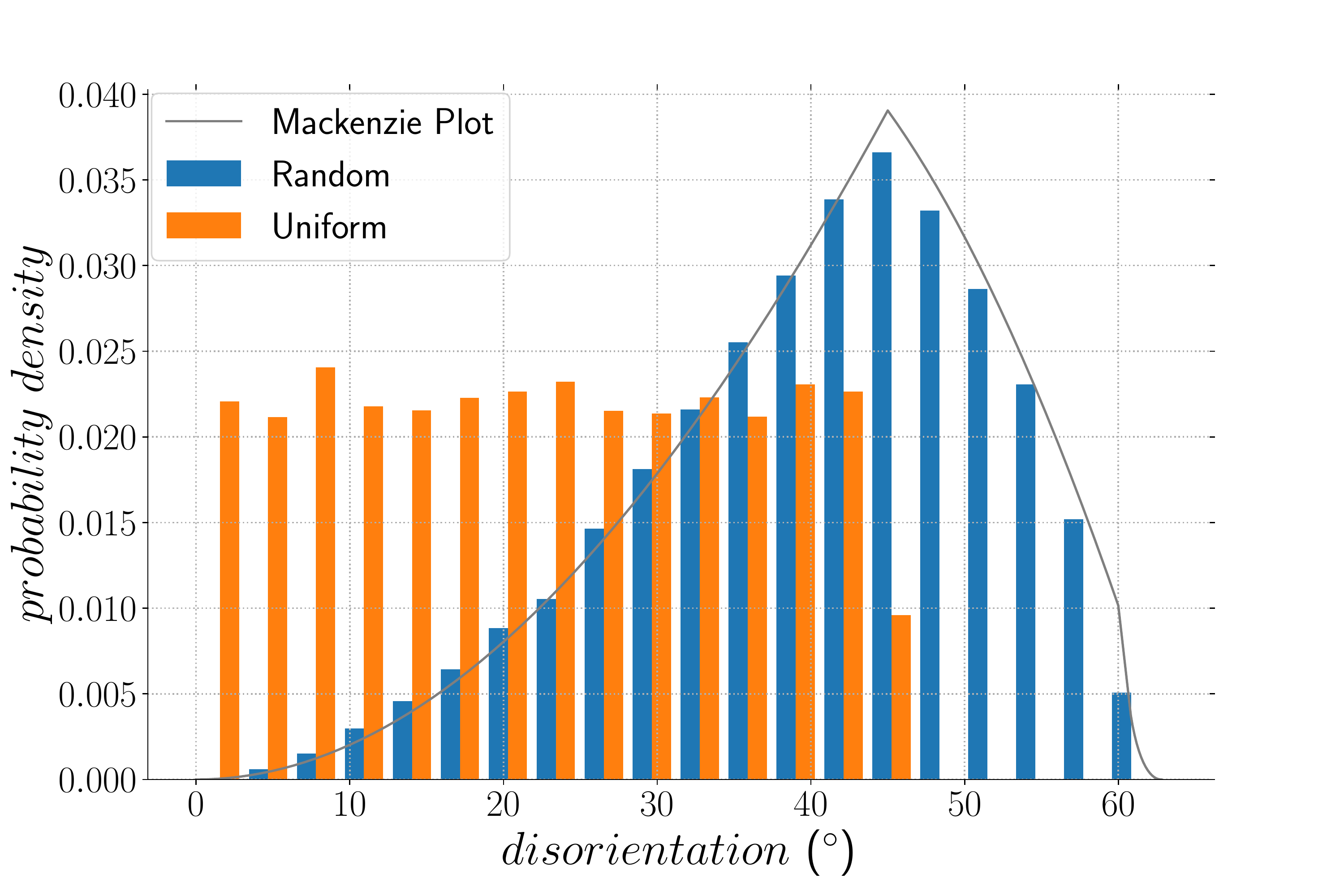}
  \end{subfigure}
  \begin{subfigure}[c]{0.48\textwidth}
    \centering
    \includegraphics[scale=0.25]{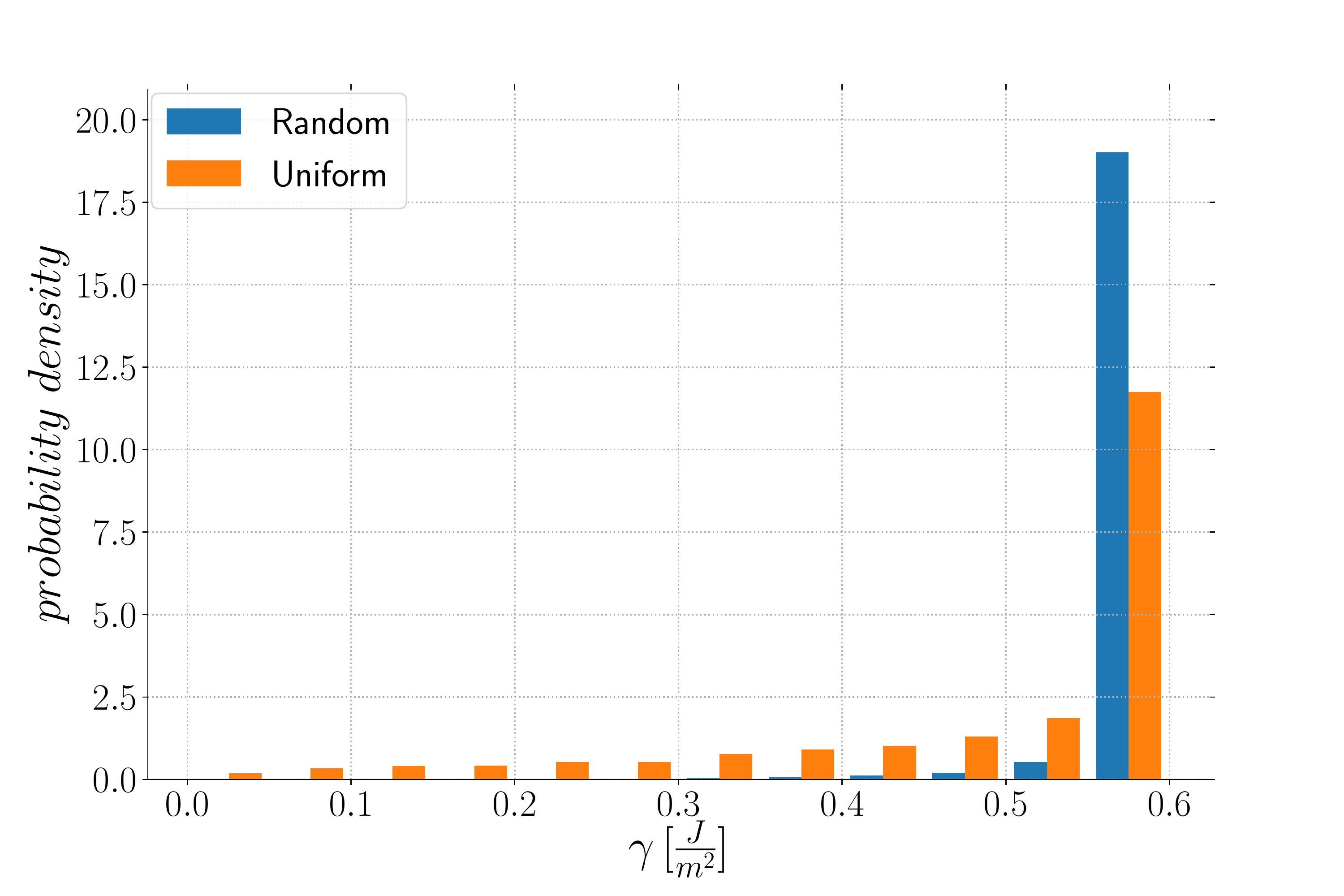}
  \end{subfigure}
  \caption{Comparison of GB distribution properties, (a) DDf and (b) GBED.}
  \label{fig:PX5000DDFFTComp}
\end{figure}

Here the crystallographic orientations are defined differently: one Euler angle is generated randomly with a uniform distribution function and the two others are constants. As a result, the final disorientation distribution is more uniform as seen in figure~\ref{fig:PX5000DDFFTComp}. Properties are defined using equations~\ref{eqn:Gamma} and \ref{eqn:Mob}, and the transition disorientation angle is set to $30 \degree$ as previously. The main effect of the wider resulting GB energy distribution (GBED) is the increase of local anisotropy at triple junctions as illustrated in Fig.\ref{fig:PX5000DDFFTComp} comparatively to the previous test case (Mackenzie-like DDF). 

\subsubsection{Heterogeneous grain boundary energy}
\label{sssec:PX1000GRSMuIsoFT}

\begin{figure}[h]
  \centering
  \begin{subfigure}{0.48\textwidth}
    \centering
    \includegraphics[scale=0.25]{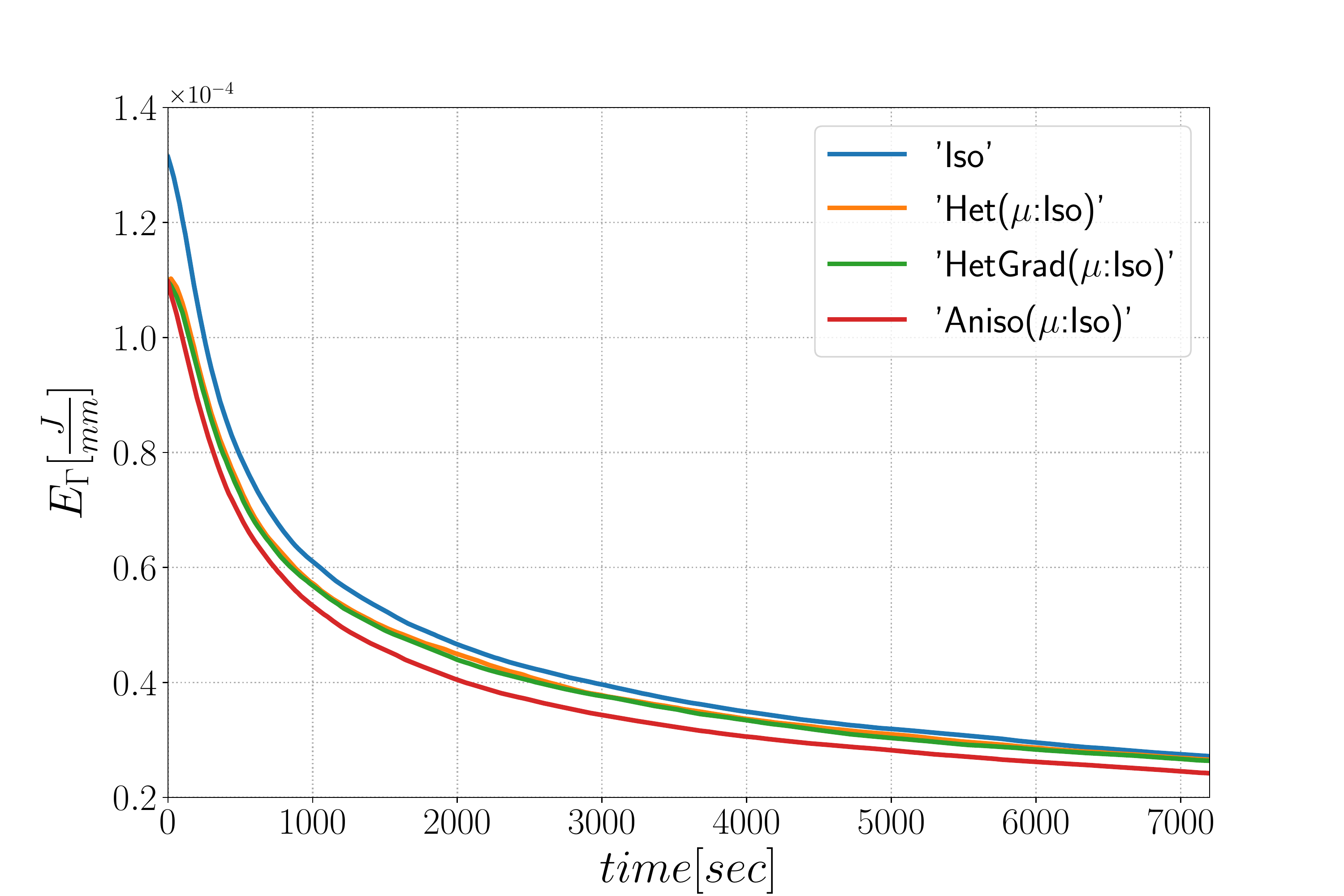}
    \caption{$E_{\Gamma}=f(t)$}
  \end{subfigure} 
  \begin{subfigure}{0.48\textwidth}
    \centering
    \includegraphics[scale=0.25]{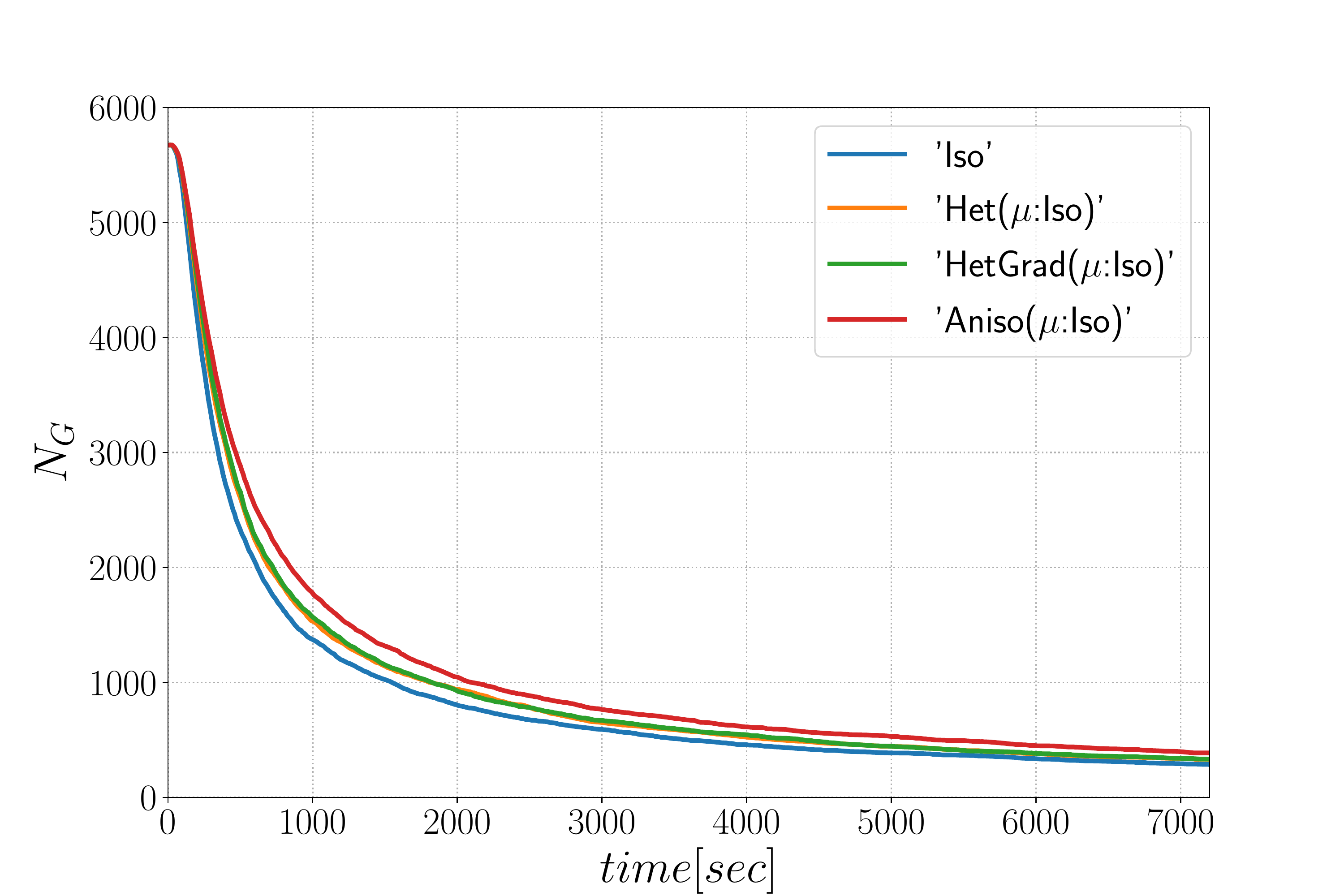}
    \caption{$N_g=f(t)$}
  \end{subfigure} \\
  \begin{subfigure}{0.48\textwidth}
    \centering
    \includegraphics[scale=0.25]{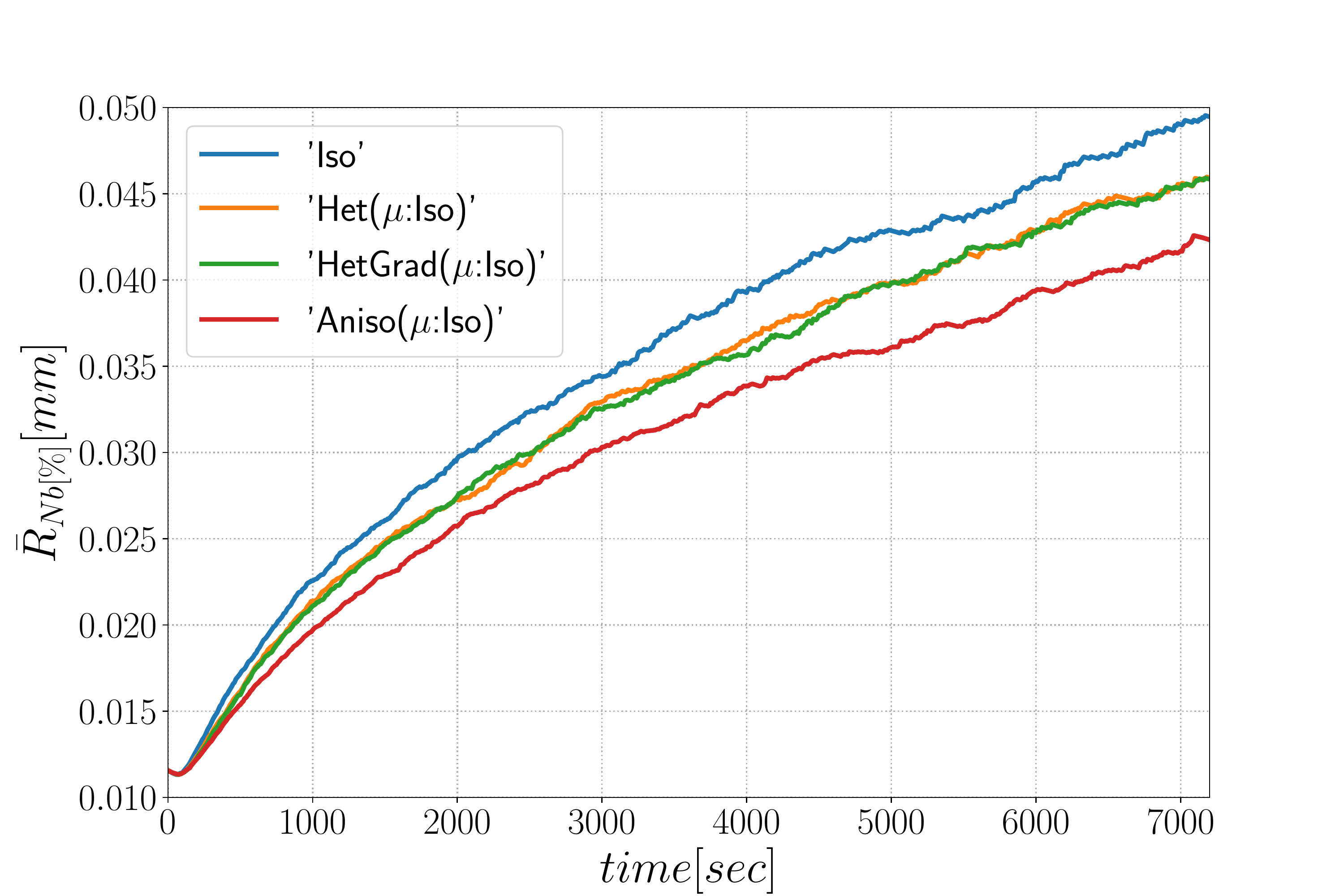}
    \caption{$\bar{R}_{Nb[\%]}=f(t)$}
  \end{subfigure}
  \begin{subfigure}{0.48\textwidth}
    \centering
    \includegraphics[scale=0.25]{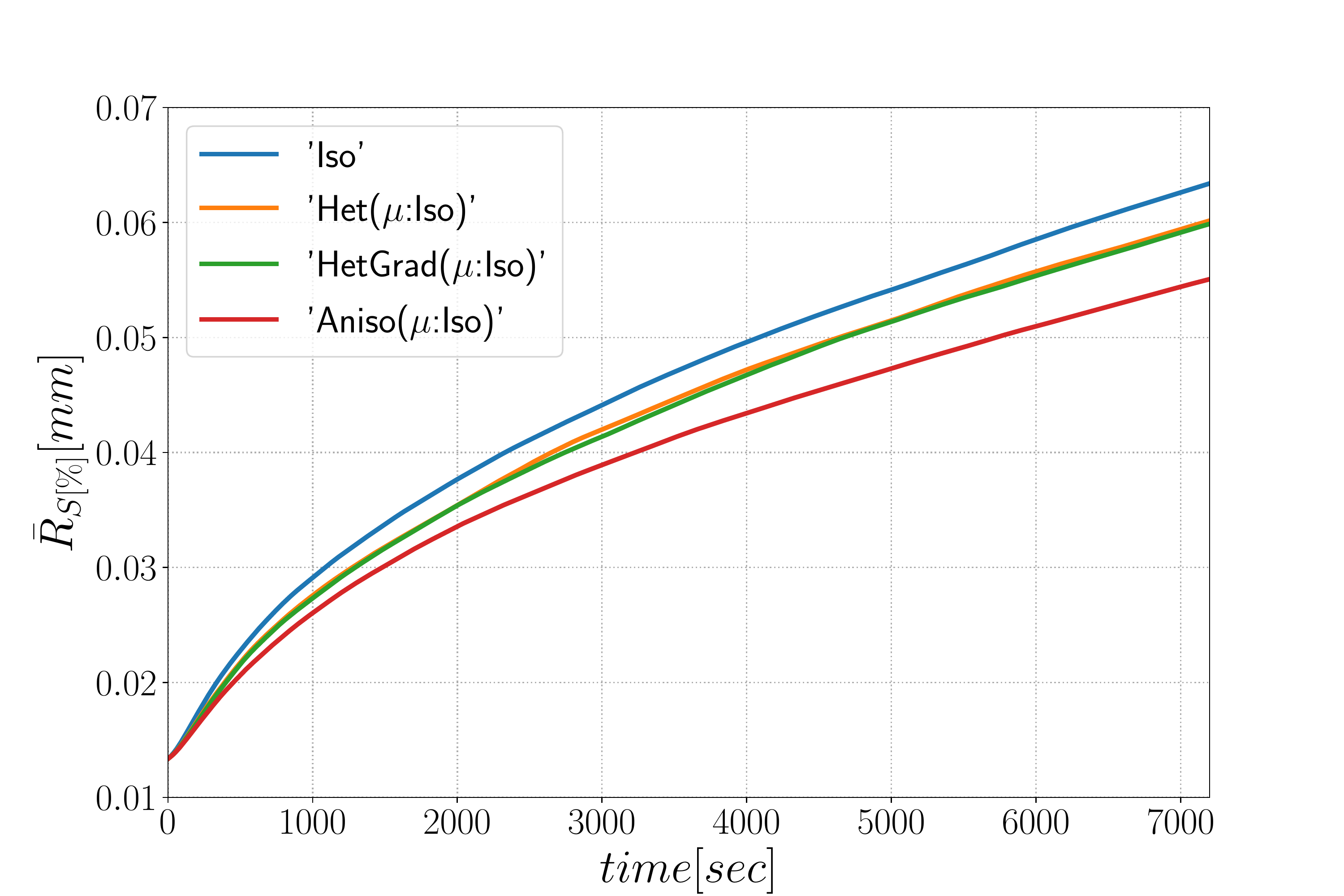}
    \caption{$\bar{R}_{S[\%]}=f(t)$}
  \end{subfigure}
  \caption{Time evolution for the different formulations: (a) the total GB energy, (b) the number of grains, (c) the mean radius weighted in number and (d) the mean radius weighted in surface.}\label{fig:PX5000MuIsoFTMeanV}
\end{figure}

The results described in Fig.\ref{fig:PX5000MuIsoFTMeanV} illustrate that the Iso formulation predicts the fastest evolution. Additionally one can see that the interfacial energy is better minimized using the Aniso formulation. From these results one can infer that the Isotropic formulation seems not adapted in this context. For a wider range of anisotropy level as the one used in this test case, particular coordination number with $n=4,5$ may be more present \citep{cahn1991stability, holm1993microstructural}. However, the Iso formulation promotes equiaxed grains ($n=6$). Once again this tendency discredits the Isotropic approach for highly heterogeneous interfaces. 

\begin{figure}[h]
  \centering
  \begin{subfigure}{0.48\textwidth}
    \centering
    \includegraphics[scale=0.25]{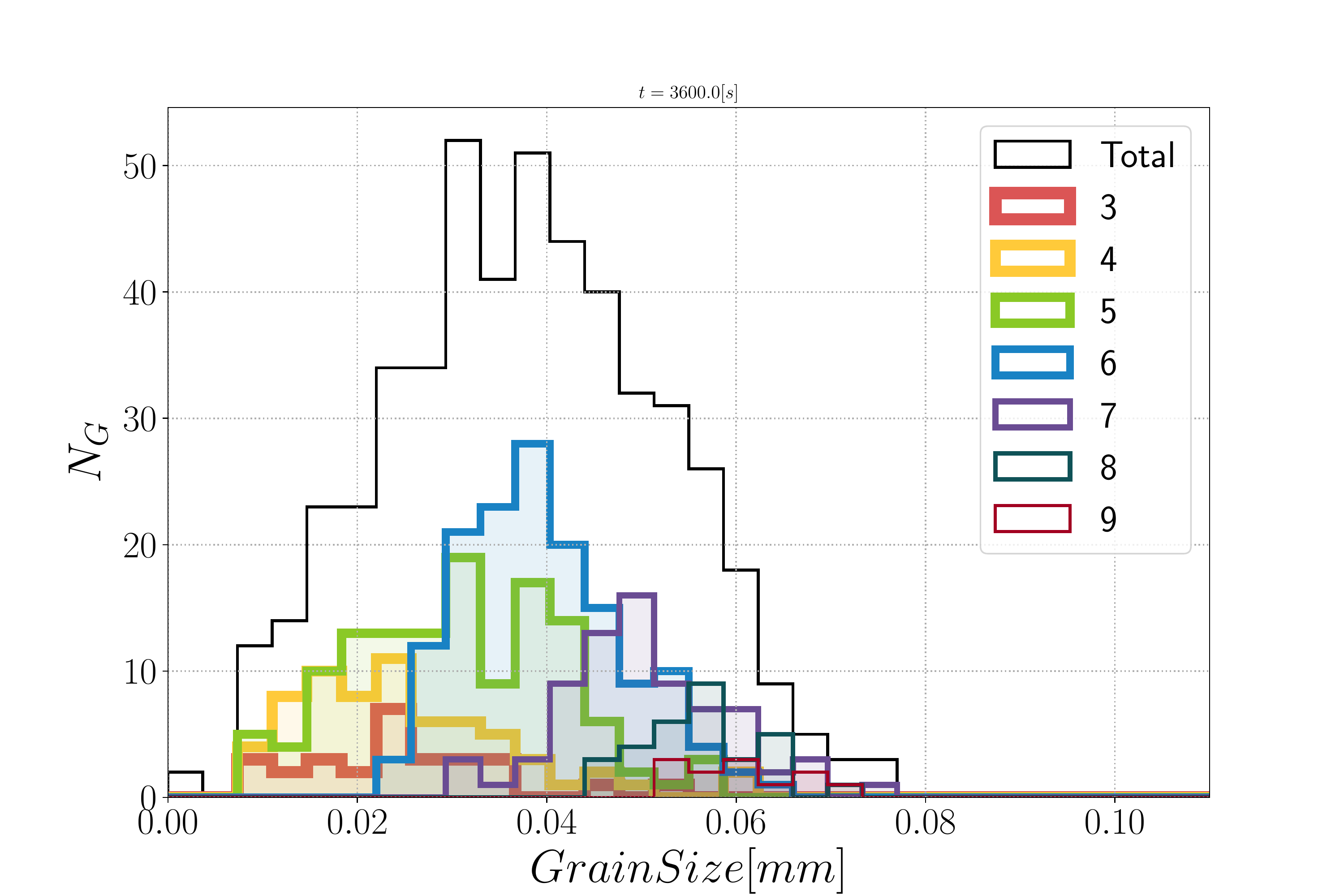}
    \caption{Iso}
  \end{subfigure} 
  \begin{subfigure}{0.48\textwidth}
    \centering
    \includegraphics[scale=0.25]{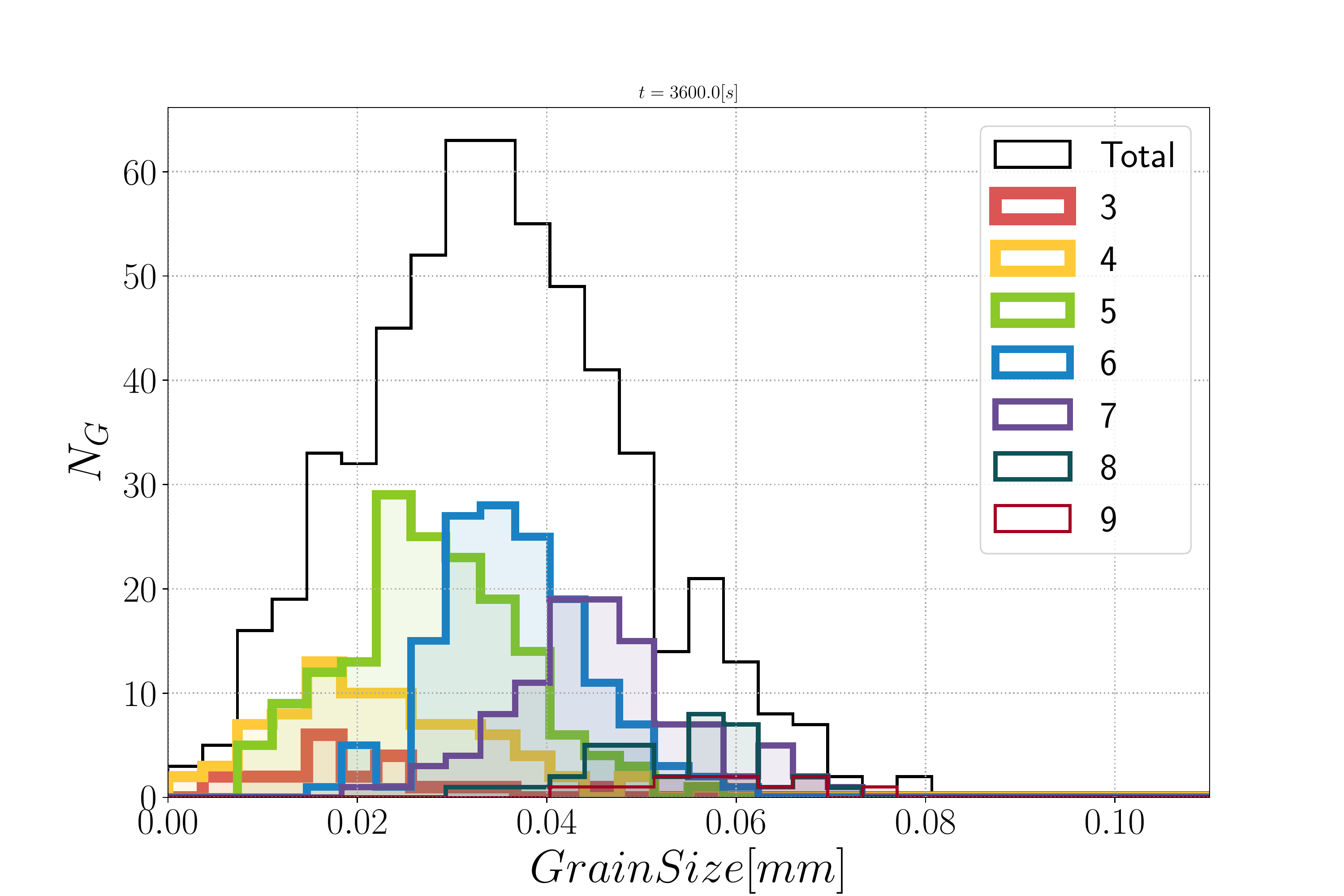}
    \caption{Het($\mu$:Iso) }
  \end{subfigure} \\
  \begin{subfigure}{0.48\textwidth}
    \centering
    \includegraphics[scale=0.25]{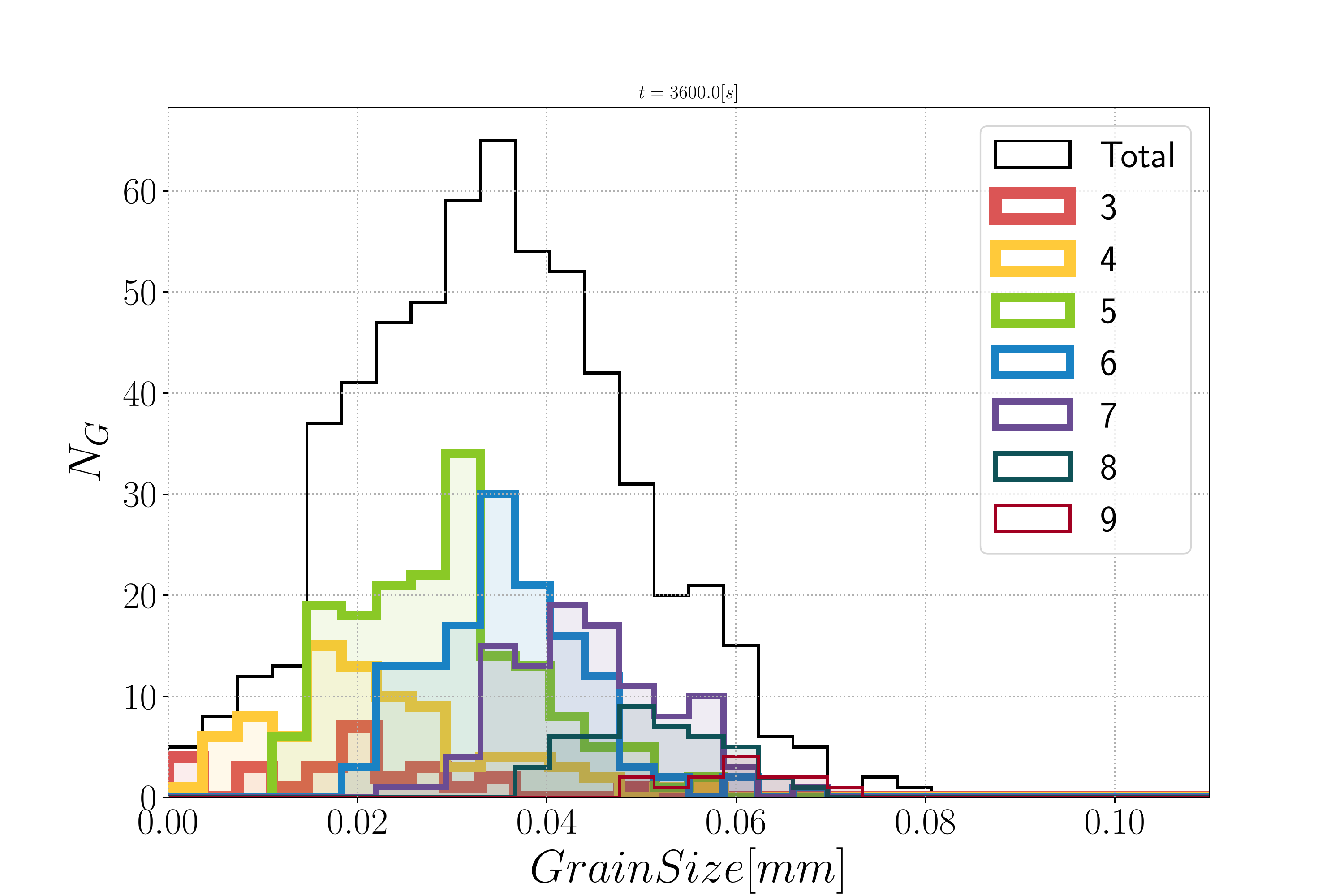}
    \caption{HetGrad($\mu$:Iso) }
  \end{subfigure}
  \begin{subfigure}{0.48\textwidth}
    \centering
    \includegraphics[scale=0.25]{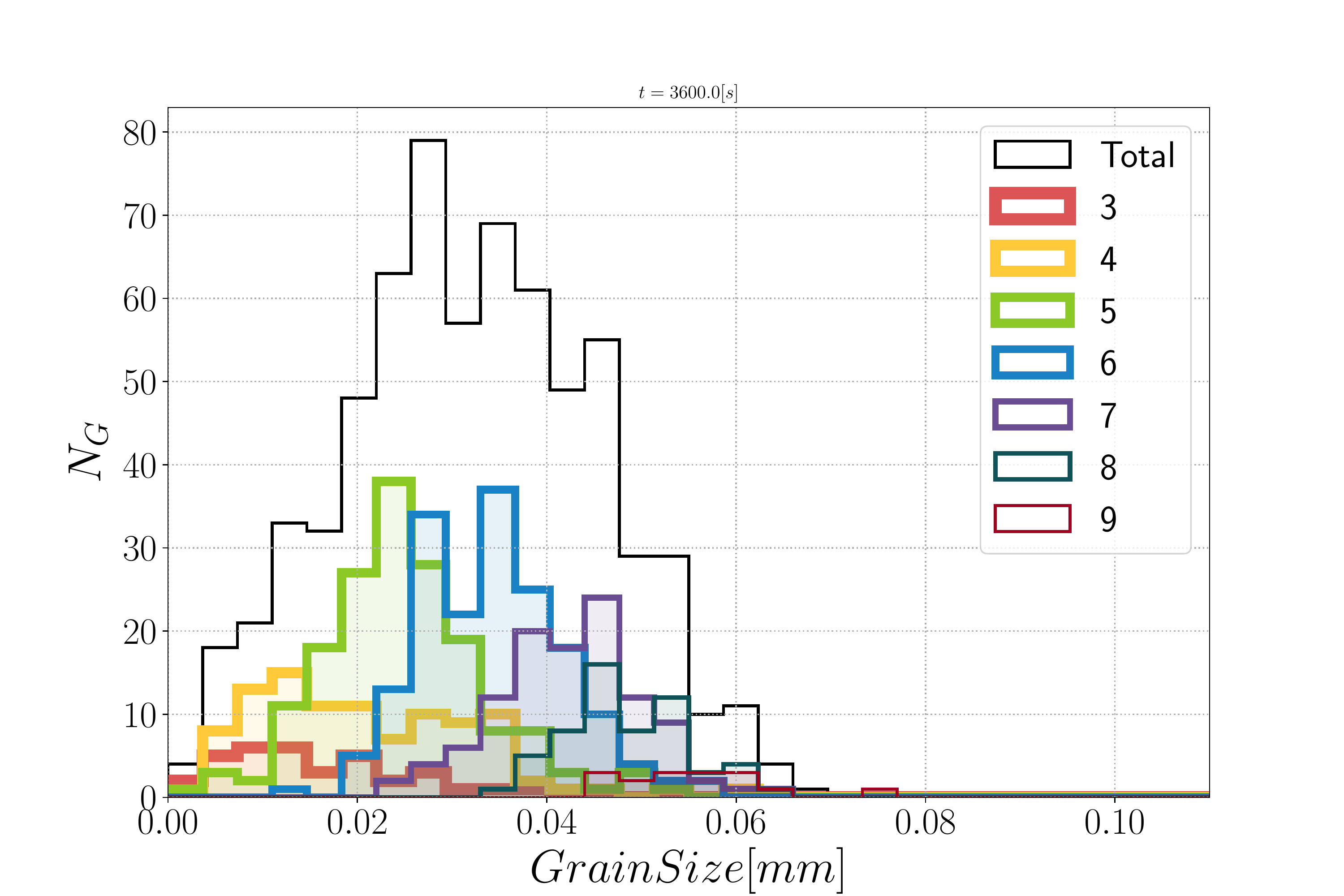}
    \caption{Aniso($\mu$:Iso) }
  \end{subfigure}
  \caption{Grain size distribution and contributions from every group of grains of same coordination number from 3 to 9 at $t=1h$.}\label{fig:PX1000MuIsoFTBoundary}
\end{figure}

\begin{figure}[h]
  \centering
  \includegraphics[scale=0.25]{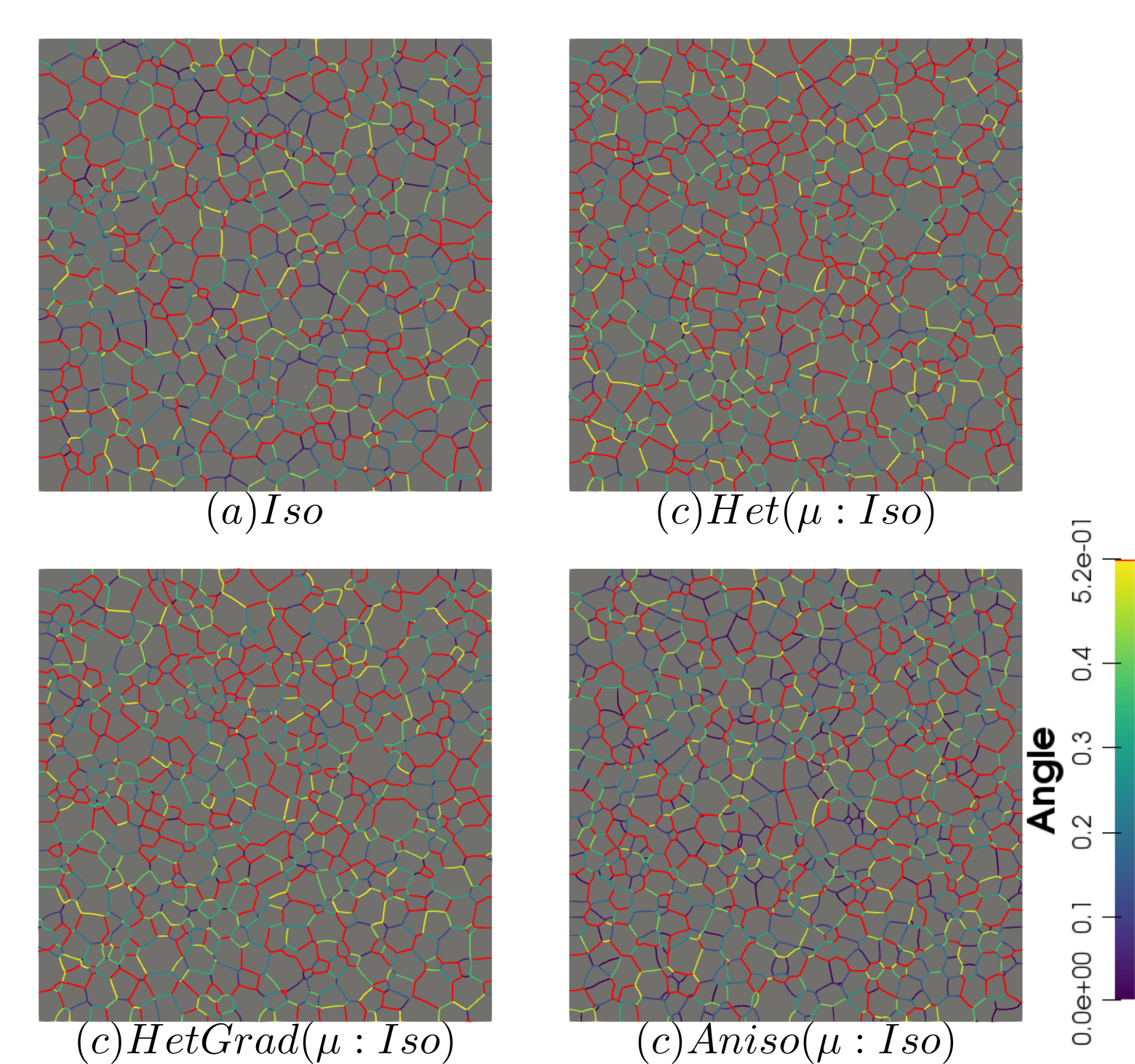}
  \caption{Disorientation of the boundaries using the four formulations with homogeneous grain boundary mobility at $t=1h$. Boundaries with a disorientation higher than $30\degree$ are colored in red.}\label{fig:PX1000MuIsoFTBoundaryDis}
\end{figure}

\begin{figure}[h]
  \centering
  \includegraphics[scale=0.25]{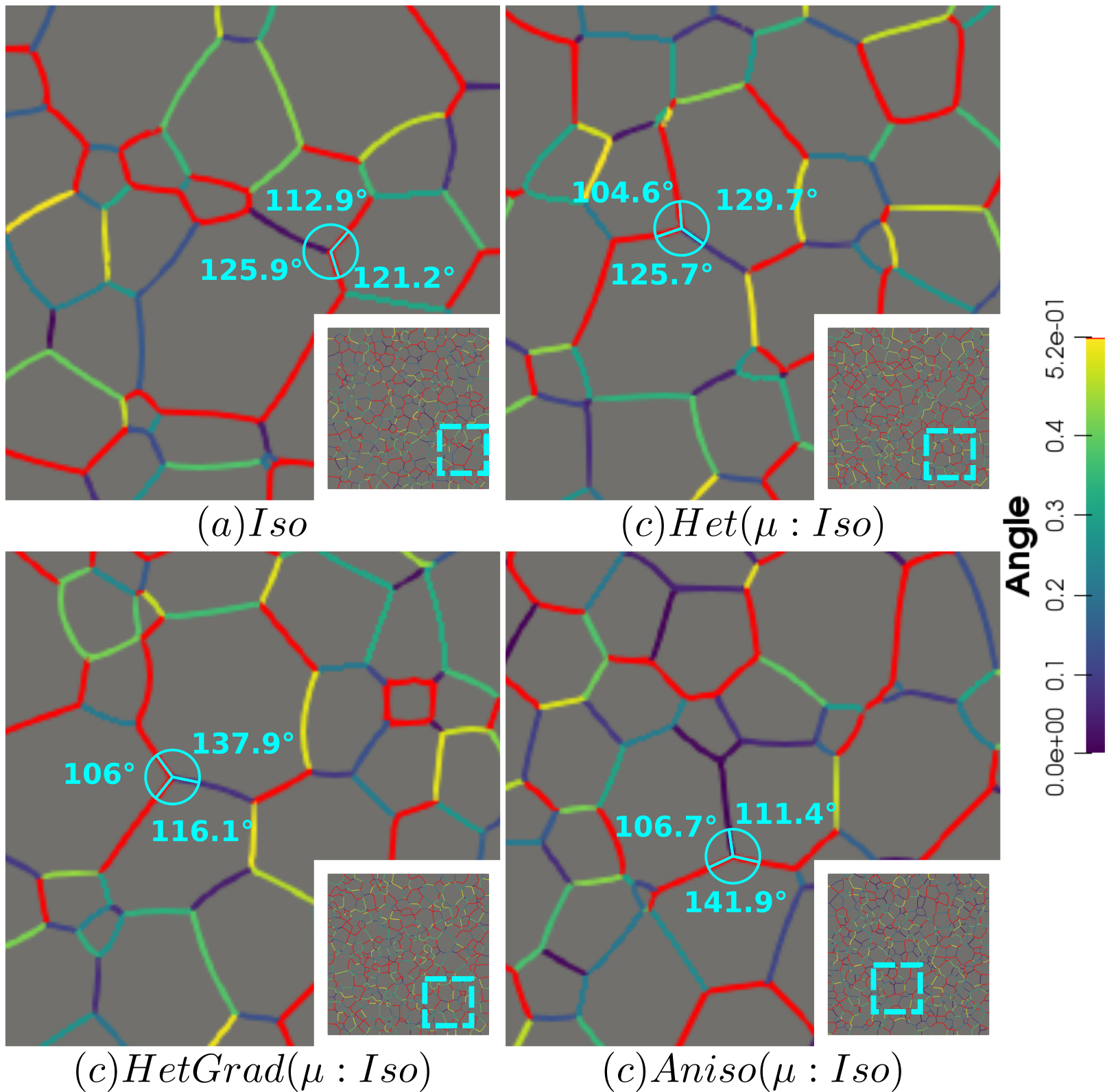}
  \caption{TJ dihedral angles among boundaries with high (red) and low (blue) GB energy. The disorientation of the boundaries is also depicted for the four formulations and homogeneous grain boundary mobility at $t=2h$. Boundaries with a disorientation higher than $30\degree$ are colored in red.}\label{fig:PX5000MuIsoFTBoundaryDis_TJ}
\end{figure}

Regarding both heterogeneous formulations (Het and HetGrad), the evolution of mean values and distributions are similar as illustrated in figures~\ref{fig:PX5000MuIsoFTMeanV} and \ref{fig:PX1000MuIsoFTBoundary}. First, both predicted distributions have similar groups with $n=4,5,6$ and second, the predicted microstructures show mostly equiaxed grain with a similar distribution of GB disorientation. In Fig.\ref{fig:PX1000MuIsoFTBoundaryDis}, one can see similar clusters of GBs with high values of disorientation depicted in red. From the morphology of grain boundaries (Fig.\ref{fig:PX1000MuIsoFTBoundaryDis}) the formulation that respects the most, on average, the triple junction angles is the Anisotropic one. This is illustrated in figure~\ref{fig:PX5000MuIsoFTBoundaryDis_TJ} where the dihedral angles of a triple junction formed by GBs with low and high disorientation angles are shown. For this particular example, from figure~\ref{fig:PX5000MuIsoFTBoundaryDis_TJ}, blue and red boundaries have values of $\gamma$ of about $0.25\cdot 10^{-7} J mm^{-2}$ and $6\cdot 10^{-7} J mm^{-2}$, respectively. One can estimate an approximated value of the dihedral angle opposite to the blue interface using equation~\ref{eqn:xi3} which is about $177^{\circ}$ with $r=6/0.25=24$. The results described in Fig.\ref{fig:PX5000MuIsoFTMeanV} show that while promoting a slower evolution of the microstructure, the (Aniso($\mu$:Iso)) formulation exhibits a better behaviour concerning the GB energy decreasing. 

\begin{figure}[h]
  \centering
  \begin{subfigure}[t]{0.48\textwidth}
    \centering
    \includegraphics[scale=0.25]{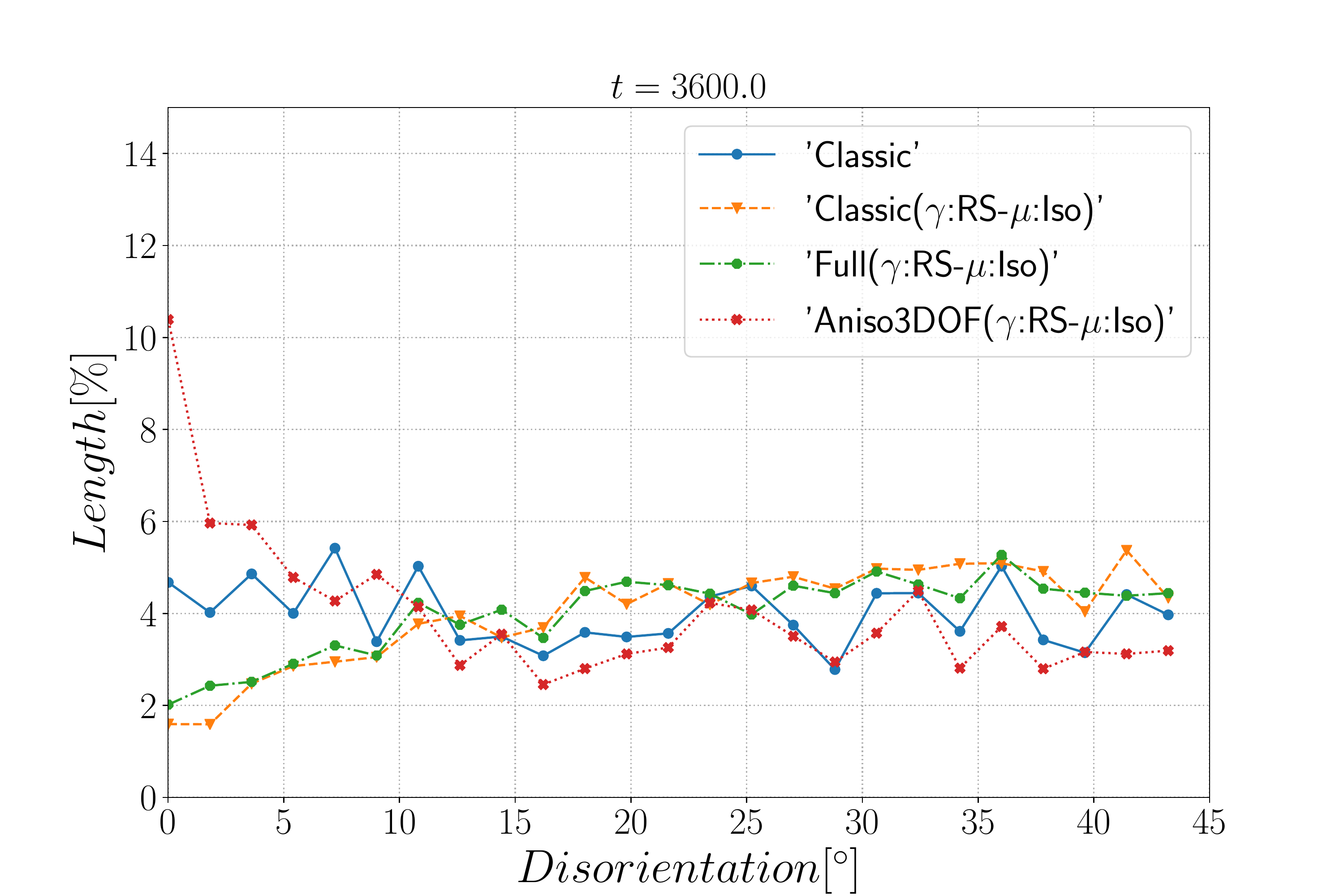}
    \caption{Distribution of $\theta$}
  \end{subfigure}
  \begin{subfigure}[t]{0.48\textwidth}
    \centering
    \includegraphics[scale=0.25]{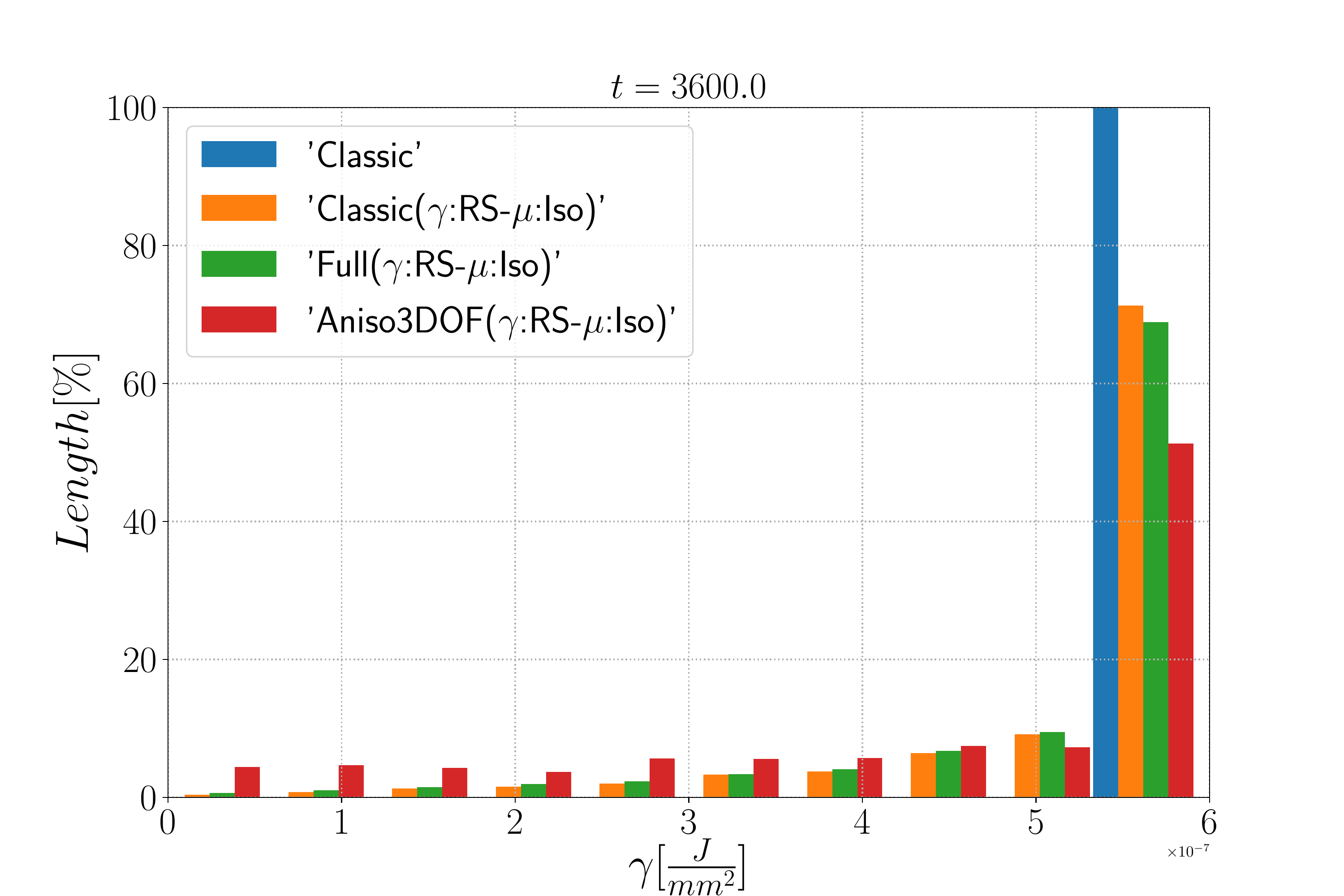}
    \caption{Distribution of $\gamma$}
  \end{subfigure}
  \caption{Grain boundary characteristics distributions at $t=1 h $.}
  \label{fig:PX1000CrysCharMuIsoFT}
\end{figure}

\pagebreak

\subsubsection{Heterogeneous grain boundary energy and mobility}
\label{sssec:PX1000GRSMuSFT}

\begin{figure}[h]
  \centering
  \begin{subfigure}{0.45\textwidth}
    \centering
    \includegraphics[scale=0.25]{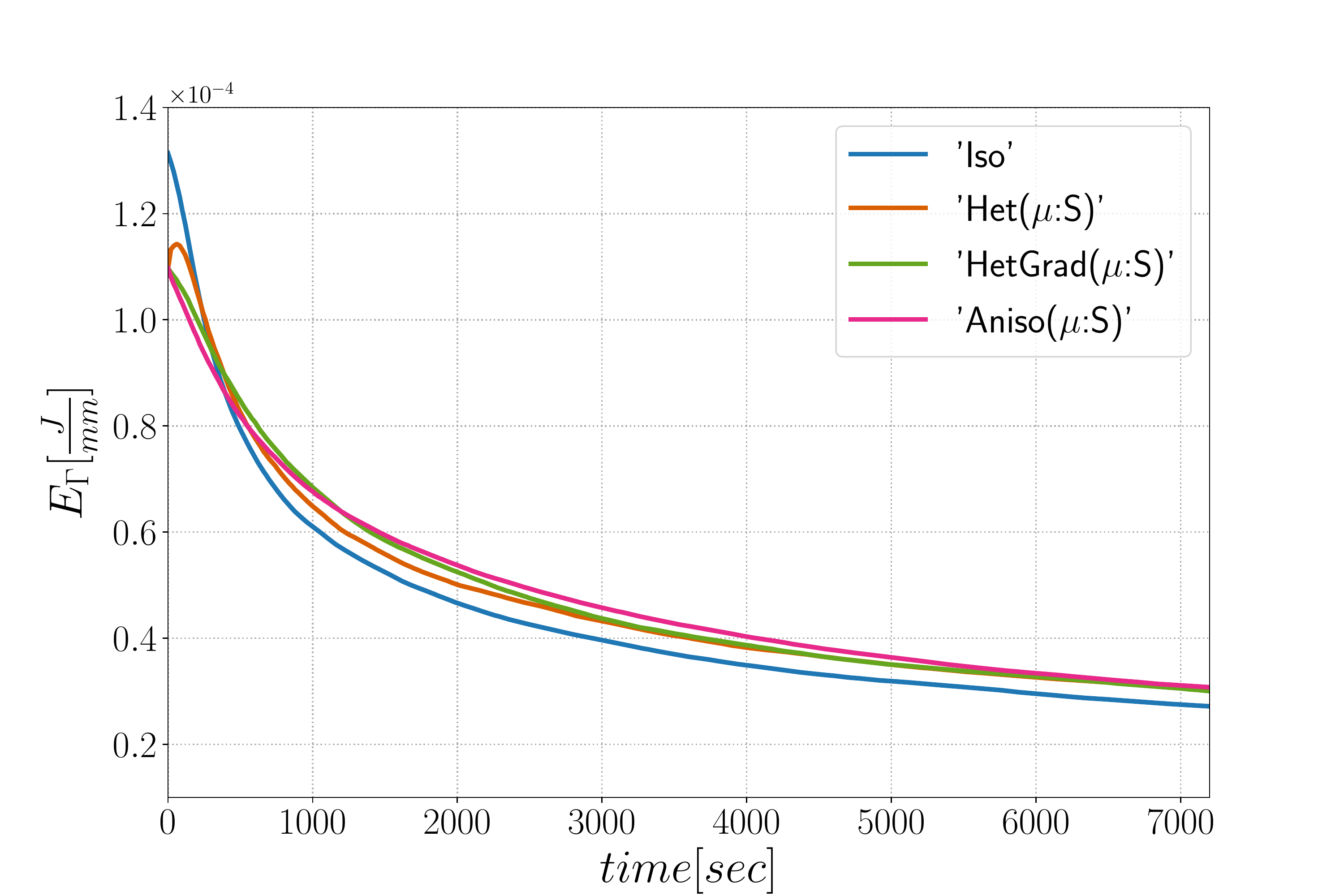}
    \caption{$E_{\Gamma}=f(t)$}
  \end{subfigure} 
  \begin{subfigure}{0.45\textwidth}
    \centering
    \includegraphics[scale=0.25]{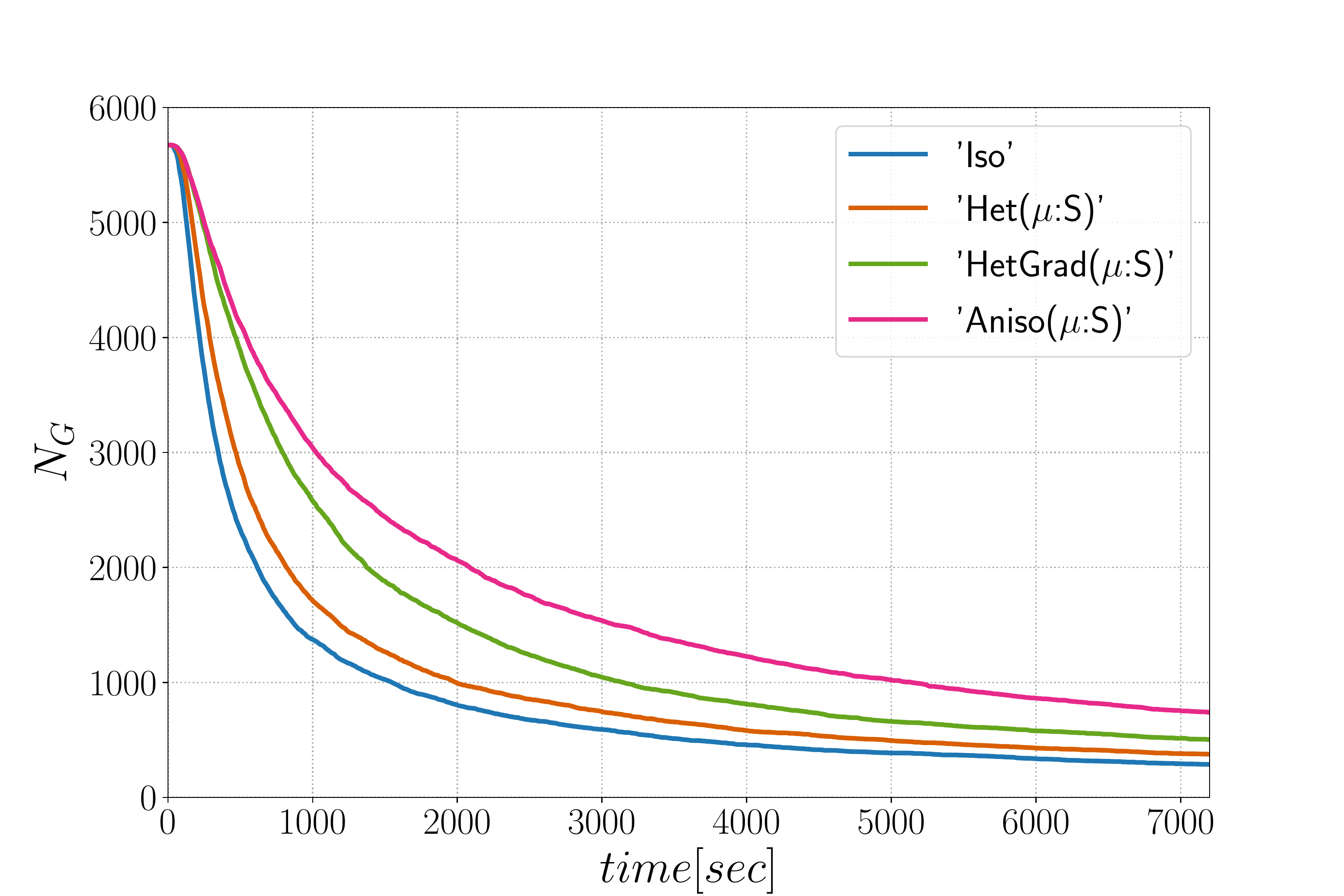}
    \caption{$N_G=f(t)$}
  \end{subfigure} \\
  \begin{subfigure}{0.45\textwidth}
    \centering
    \includegraphics[scale=0.25]{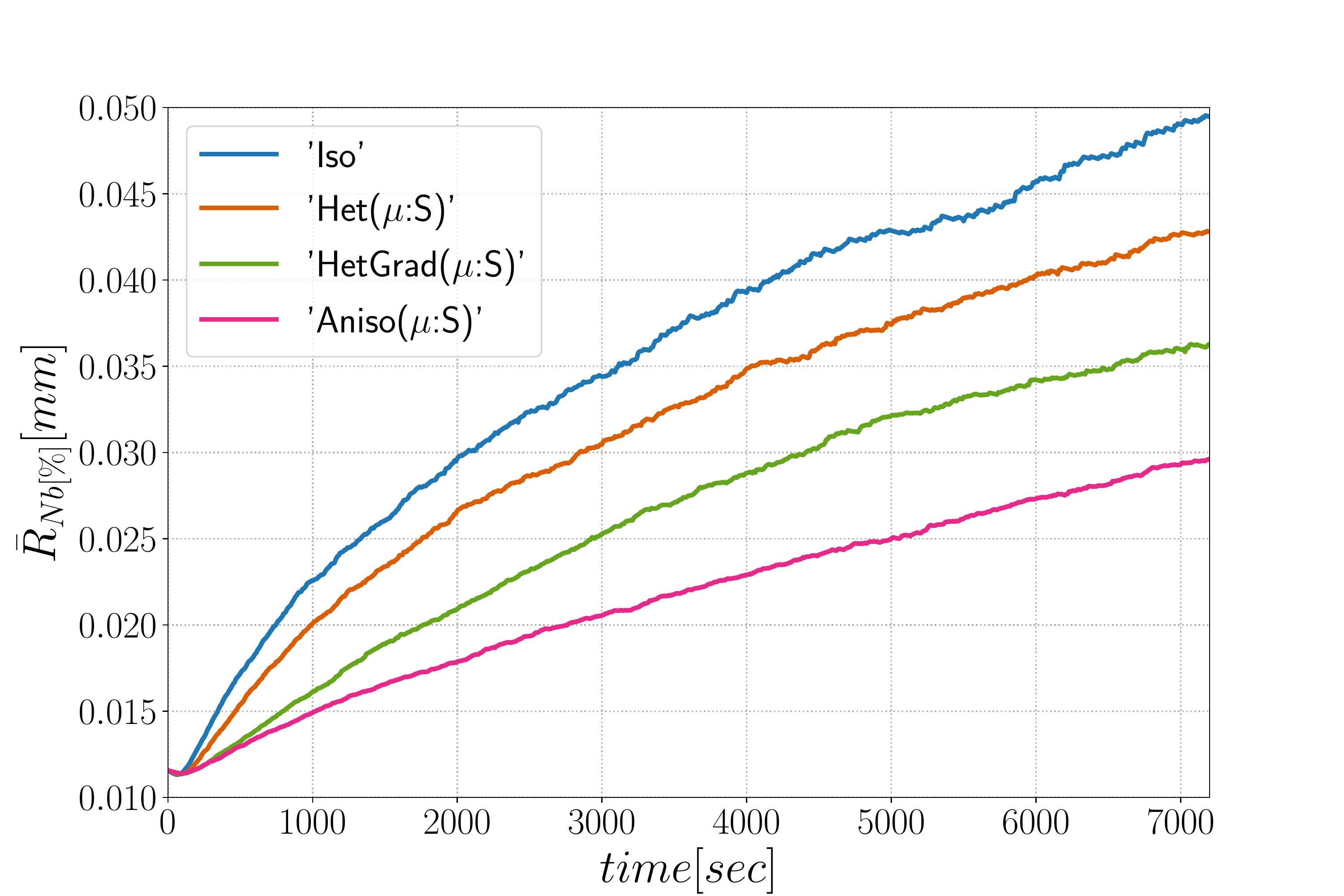}
    \caption{$\bar{R}_{Nb[\%]}=f(t)$}
  \end{subfigure}
  \begin{subfigure}{0.45\textwidth}
    \centering
    \includegraphics[scale=0.25]{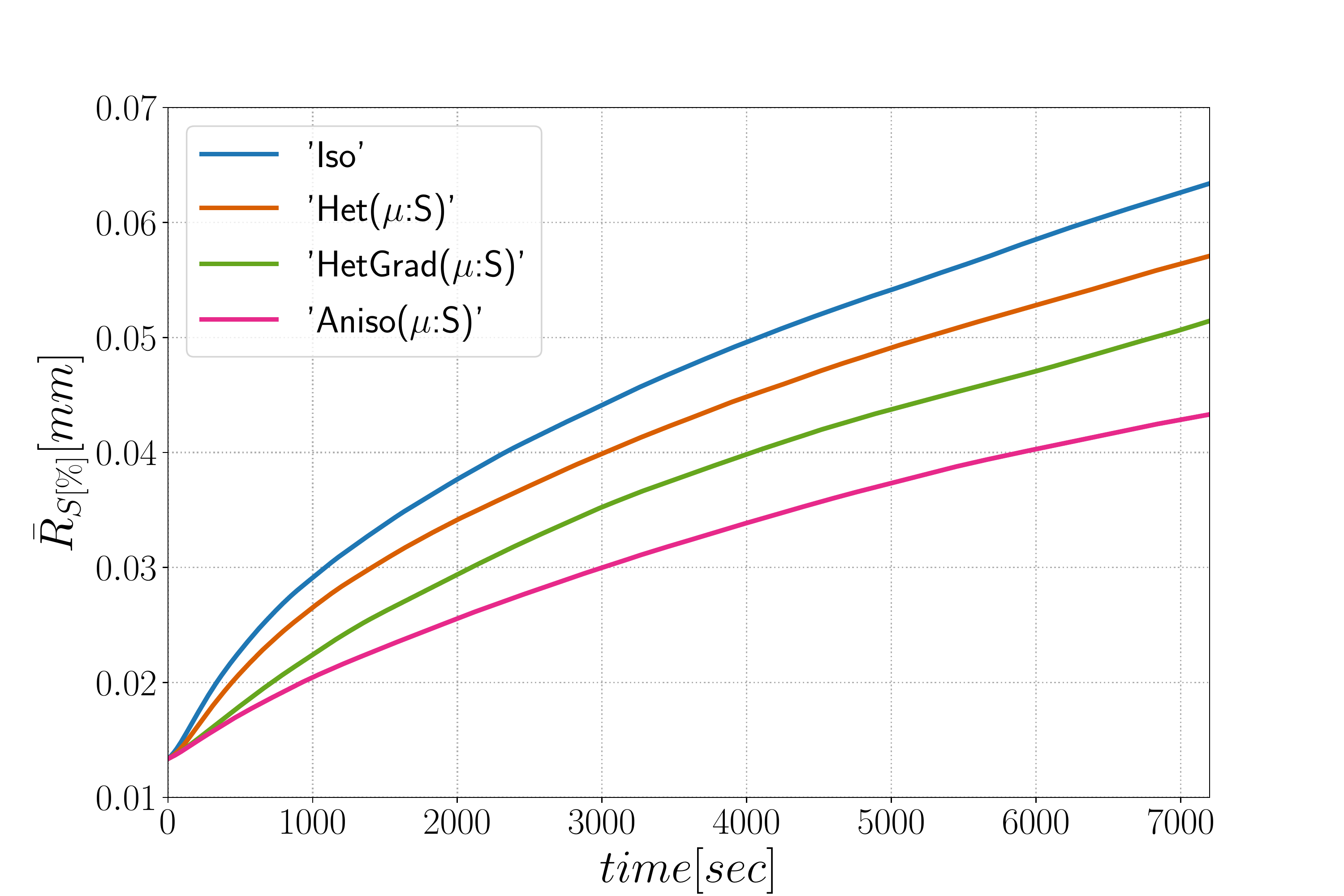}
    \caption{$\bar{R}_{S[\%]}=f(t)$}
  \end{subfigure}
  \caption{Time evolution for the different formulations: (a) the total GB energy, (b) the number of grains, (c) the mean radius weighted in number and (d) the mean radius weighted in surface.}\label{fig:PX5000MuSFTMeanV}
\end{figure}

If heterogeneous GB mobility is added, the evolution of the microstructures can vary significantly. The results presented in figure~\ref{fig:PX5000MuSFTMeanV} show two regimes for the Het($\mu$:S) formulation. First, one can infer that the Het formulation presents issues to reduce the interfacial energy and presents a peak which is the result of an evolution dominated by curvature flow without any effect of the heterogeneity. If we compare the results shown in figures~\ref{fig:PX5000MuIsoFTMeanV} and \ref{fig:PX5000MuSFTMeanV}, one can see the retarding effect of using a heterogeneous GB mobility. This effect is more stronger on the HetGrad and Anisotropic formulations due to the gradients introduced by heterogeneous fields and is completely natural because technically the effect of the crystallography is taken into account twice in the $\mu\gamma$ product. Thus the Isotropic case evolves faster than the other formulations. 

\begin{figure}[h]
  \centering
  \begin{subfigure}{0.48\textwidth}
    \centering
    \includegraphics[scale=0.25]{PX5kFT/RadCooHistIso_00002.pdf}
    \caption{Iso}
  \end{subfigure} 
  \begin{subfigure}{0.48\textwidth}
    \centering
    \includegraphics[scale=0.25]{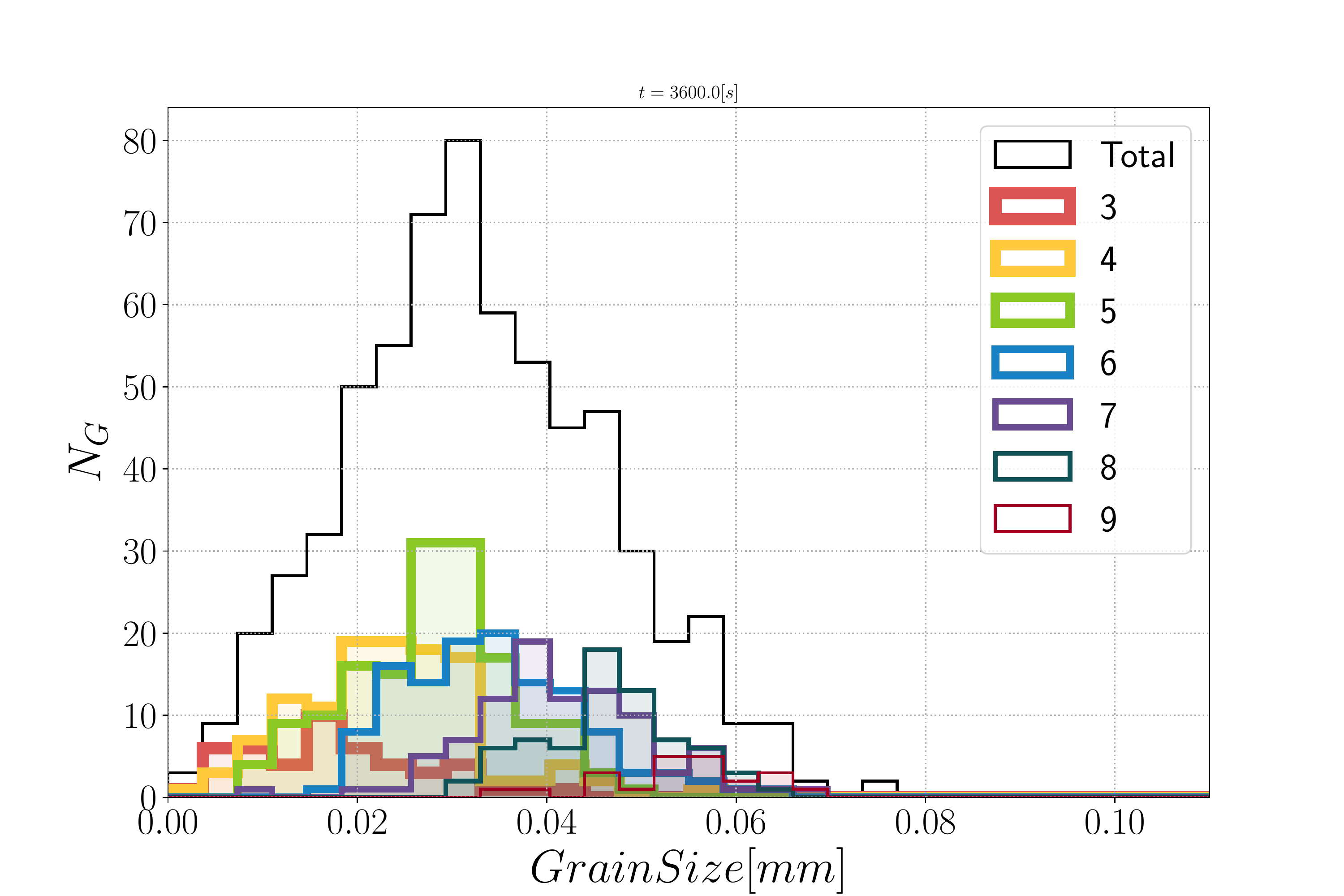}
    \caption{Het($\mu$:S) }
  \end{subfigure} \\
  \begin{subfigure}{0.48\textwidth}
    \centering
    \includegraphics[scale=0.25]{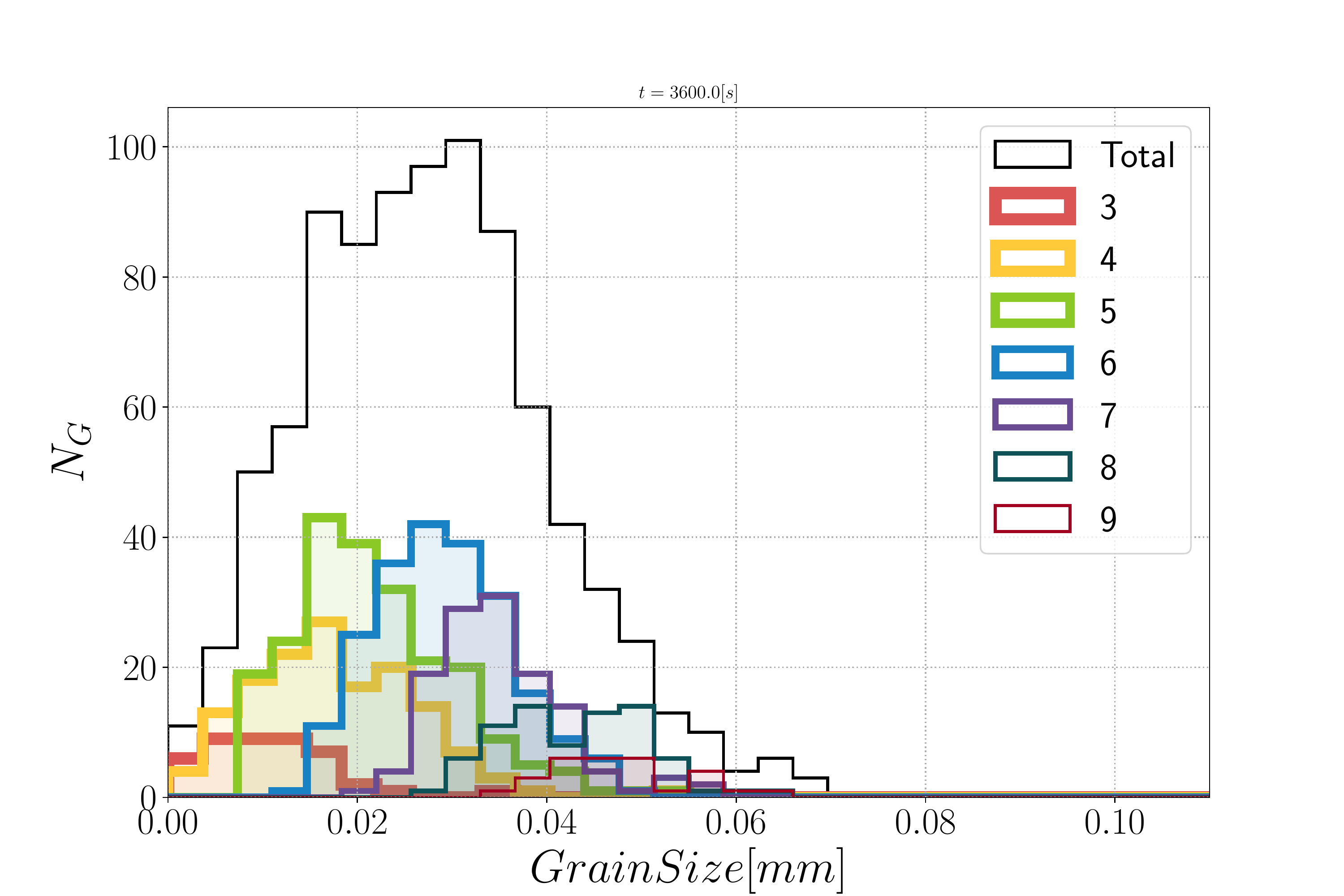}
    \caption{HetGrad($\mu$:S) }
  \end{subfigure}
  \begin{subfigure}{0.48\textwidth}
    \centering
    \includegraphics[scale=0.25]{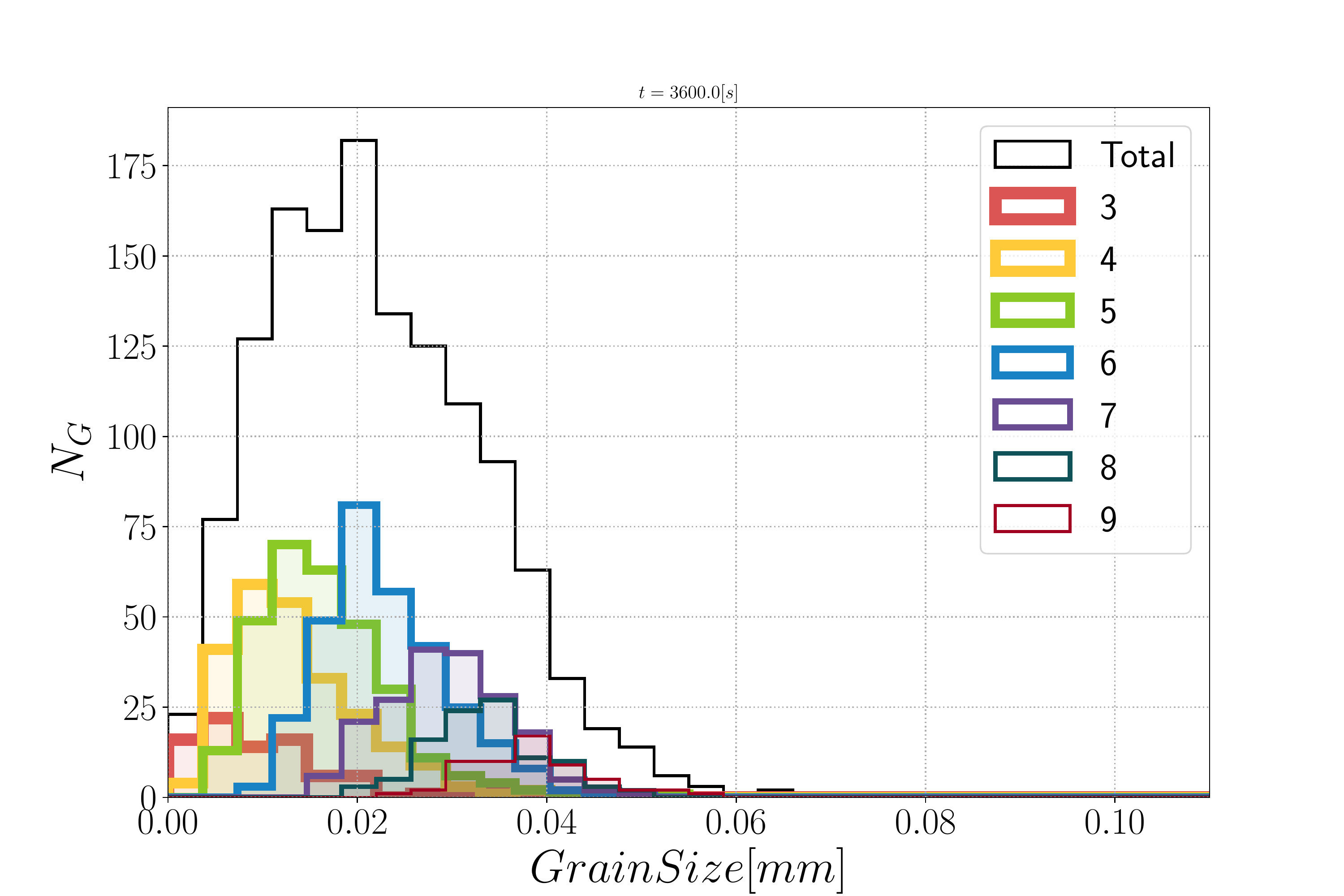}
    \caption{Aniso($\mu$:S) }
  \end{subfigure}
  \caption{Equivalent radius distribution and contribution from every group of grains with coordination number from 3 to 9 at $t=1h$.}\label{fig:PX1000MuSFTBoundary}
\end{figure}

\begin{figure}[h]
  \centering
  \includegraphics[scale=0.25]{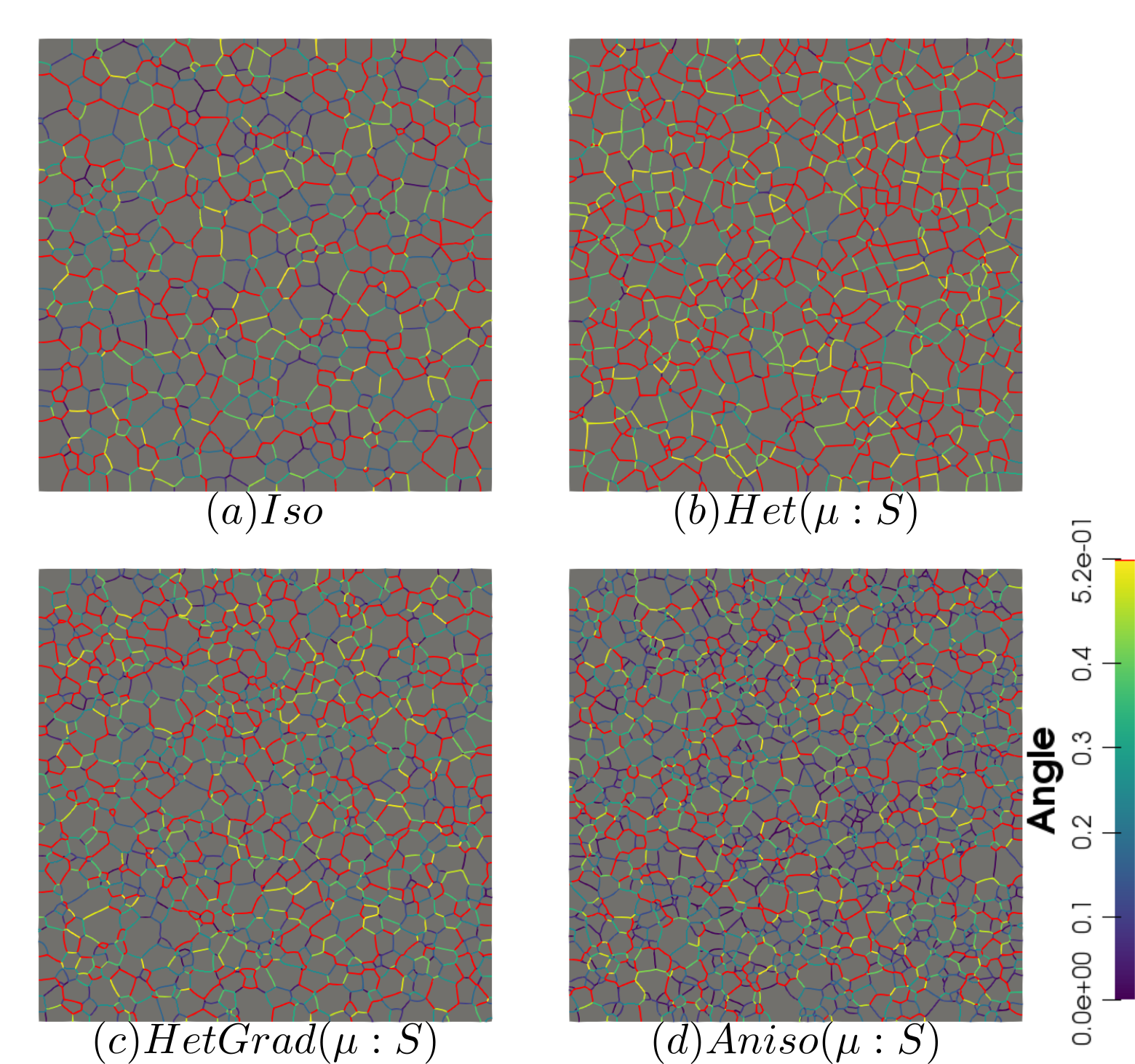}
  \caption{Disorientation of the boundaries using the four formulations with heterogeneous grain boundary mobility at $t=1h$, boundaries with a disorientation higher than $30\degree$ are colored in red.}\label{fig:PX1000MSFTBoundary}
\end{figure}

The results presented in figure~\ref{fig:PX1000MuSFTBoundary} and figure~\ref{fig:PX1000MSFTBoundary} show that, at $t=1h$, the Het and HetGrad formulations present more grains within the classes $n=4$ and $n=7$. On the other hand, the Anisotropic case didn't evolve enough to compare it to the other cases.

\begin{figure}[h]
  \centering
  \begin{subfigure}[t]{0.48\textwidth}
    \centering
    \includegraphics[scale=0.25]{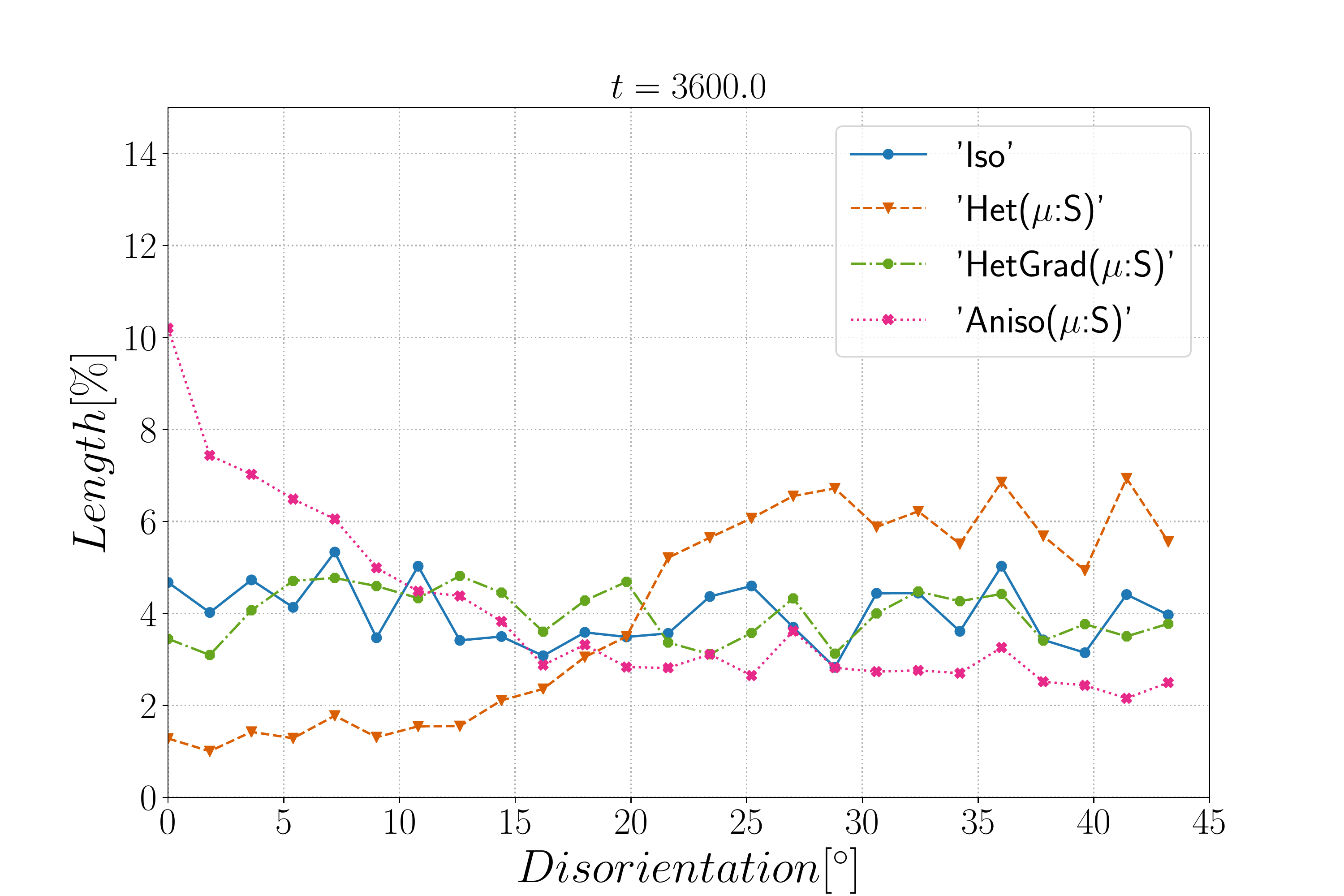}
    \caption{Distribution of $\theta$}
  \end{subfigure}
  \begin{subfigure}[t]{0.48\textwidth}
    \centering
    \includegraphics[scale=0.25]{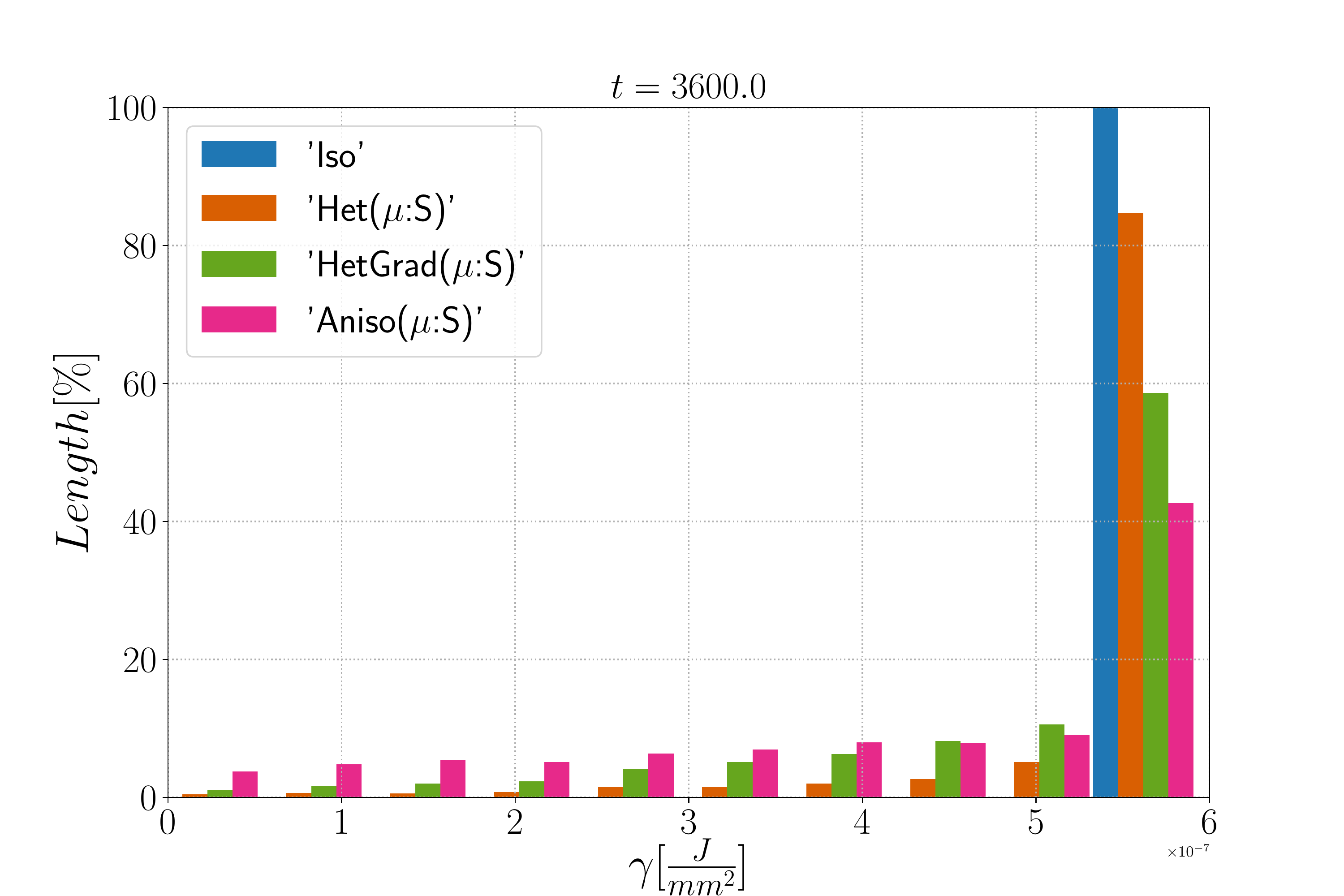}
    \caption{Distribution of $\gamma$}
  \end{subfigure}
  \caption{Grain boundary characteristics distributions at $t=1 h $.}
  \label{fig:PX1000CrysCharMuSFT}
\end{figure}

Interestingly, the DDF of the Het formulation disagrees with the results presented in \citep{elsey2013simulations}. Here, the evolution of the DDF evolves in the opposite way to the expected results (see Fig.\ref{fig:PX1000CrysCharMuSFT}). Indeed, the DDF tends to increase the percentage of interfaces with $\theta > \theta_0$ and decrease those with $\theta < \theta_0$ which clearly seems unphysical. Moreover, the Iso and HetGrad formulations do not exacerbate a particular disorientation. Finally, the Anisotropic formulation seems to exhibit a more physical behavior by promoting a higher percentage of boundaries with lower values of disorientation.  

\subsection{CPU Time}

All the simulations presented here were performed on 20 cores with the same mesh size, $h_n=1 \mu m$ at the normal direction of the interface and $h_t=5 \mu m$ in the tangential direction of the interface and far from the interface. As expressed before both heterogeneous formulations and the anisotropic formulation have additional terms which can be synonymous of more complex resolutions. This aspect if not significant when moderated anisotropy is considered as illustrated by the first line of table~\ref{tab:CPU-Time}. 

However, the CPU-time changes significantly for the textured case presented above. The HetGrad and Aniso formulations present, respectively, an increase of 35\% and of 74\% of the calculation time in comparison to the Isotropic formulation. 

\begin{table}[h]
\centering
\begin{tabular}{l | *{4}{c}}
Case              & Iso & Het & HetGrad & Aniso \\
\hline
Random            & 5.4 & 5.5 & 5.5 & 5.6  \\
Textured          & 5.4 & 5.5 & 7.3 & 9.4 \\
\end{tabular}
\caption{ CPU-time in hours for the four formulations with heterogeneous GB energy and mobility. }
\label{tab:CPU-Time} 
\end{table}

\section{Accounting for misorientation and inclination}
\label{sec:Torque}

The formulations presented by now have dealt with heterogeneous GB properties. However, we know that the nature of the GB is described in a 5D space generated by the inclination and the misorientation. The effect of the normal direction has been described by C. Herring in \cite{herring1951physics} as a torque term. Hence, a triple junction should respect a condition frequently known as Herring's equation (Eq.\ref{eqn:Herring}).\\

Due to the high dimensional space of GBs, many researchers have attempted to propose metrics that represent symmetries properly \cite{sutton2015five, morawiec1995misorientation, cahn2006metrics, morawiec2009models, patala2013symmetries, homer2015grain, OLMSTED20092793}. With these metrics, one can compare and compute the shortest paths (geodesics) between GBs. As the misorientation and the normal can change during the microstructure evolution due to grain rotation or grain disappearance/appearance, the evolution of the metric could reveal important informations about the structure-property relationship. Recent works by Chesser et al. and Francis et al. have proposed new metrics using octonions \cite{francis2019geodesic, chesser2020learning}, revealing good predictions of GB energy of the data published by Olmsted in \cite{olmsted2009survey}. To the authors' knowledge, the effect of the GB normal is not clear and more experimental, numerical and theoretical works are needed. Here we define the effect of the normal using a model of GB energy proposed for fcc metals by Bulatov et al. \cite{bulatov2014grain} and available in the GB5DOF code. When $\gamma$ is defined using the GB5DOF code, both the effect of the misorientation and inclination are taken into account using the crystallographic orientations of the two adjacent grains and the local coordinate system of the corresponding GB \cite{bulatov2014grain}.

\subsection{Triple junction}
\begin{figure}[h]
  \centering
  \includegraphics[scale=0.65]{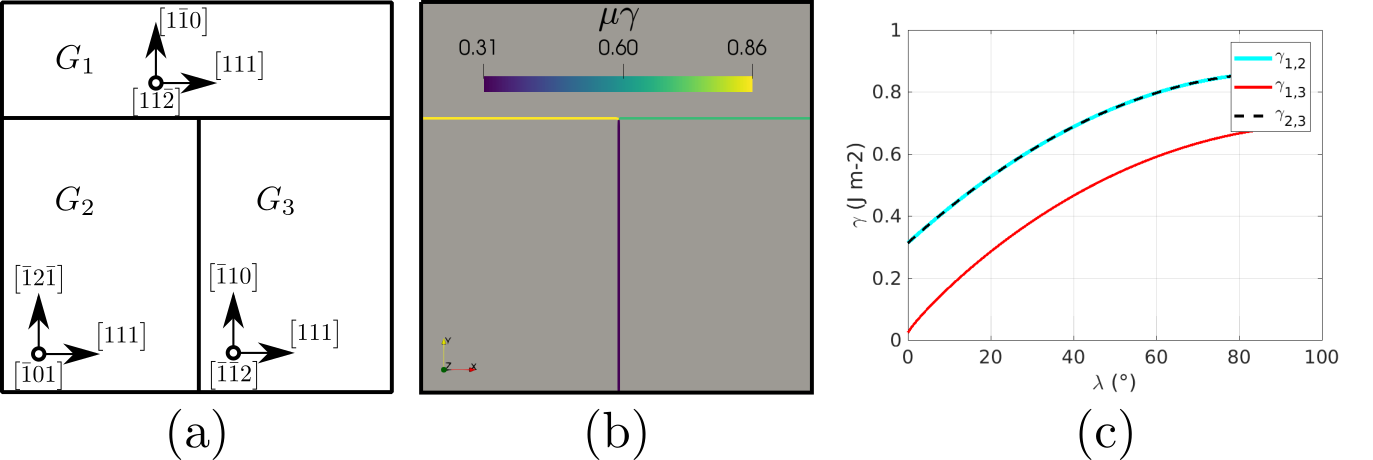}
  \caption{(a) Grain orientations. (b) Initial reduced mobility $[mm^2/s]$. (c) Change of GB energy as a function of the GB inclination $\lambda$ with respect to the x-axis evaluated using the code GB5DOF \cite{bulatov2014grain}.}\label{fig:AnisoNormalTJ}
\end{figure}

\begin{figure}[h]
  \centering
  \includegraphics[scale=0.75]{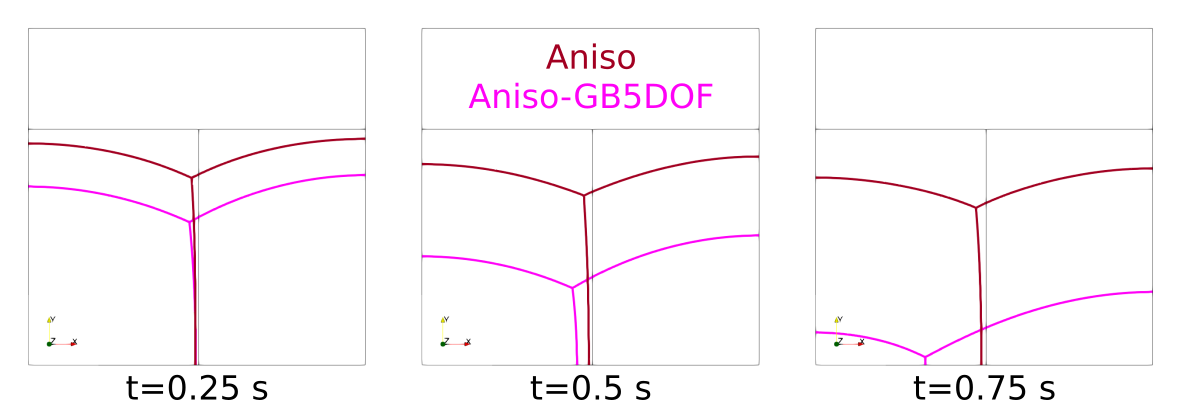}
  \caption{Interface evolution for three different times $t=0.25, \ 0.5, \ 0.75 \ s$.}\label{fig:AnisoNormalTJ_evo}
\end{figure}

This case consists again in a triple junction describes by the figure~\ref{fig:AnisoNormalTJ}. We performed simulations with a constant GB mobility set to $\mu=1e6 \ mm^4 J^{-1} s^{-1}$ taken from \citep{CHESSER201819}, a domain of $1 \times 1 \ mm^2$ and a time step $\Delta t= 5e-5 \ s$. The Aniso formulation is used by considering $\gamma$ as only initially defined by the misorientation and then also dependent of the inclination (obtained through the GB5DOF code and denoted as Aniso-GB5DOF).  The Iso, Het and HetGrad are not presented here because they evolve in the wrong direction (the expected movement should reduce the length of the interface between grain $G_1$ and $G_2$ depicted in yellow). The evolution of the interfaces shown in Fig.~\ref{fig:AnisoNormalTJ_evo} presents similar tendencies to the cases presented by Garcke in \citep{garcke1999anisotropy} and Hallberg in \citep{hallberg2019modeling}. If both evolutions (without or with the inclination dependence) seem promote similar triple junction evolution, the Aniso-GB5DOF case exhibits a much faster evolution which illustrates the importance to take into account the inclination in the reduced mobility description.

\subsection{Coherent and incoherent twin boundary}

\begin{figure}[h]
  \centering
  \includegraphics[scale=0.65]{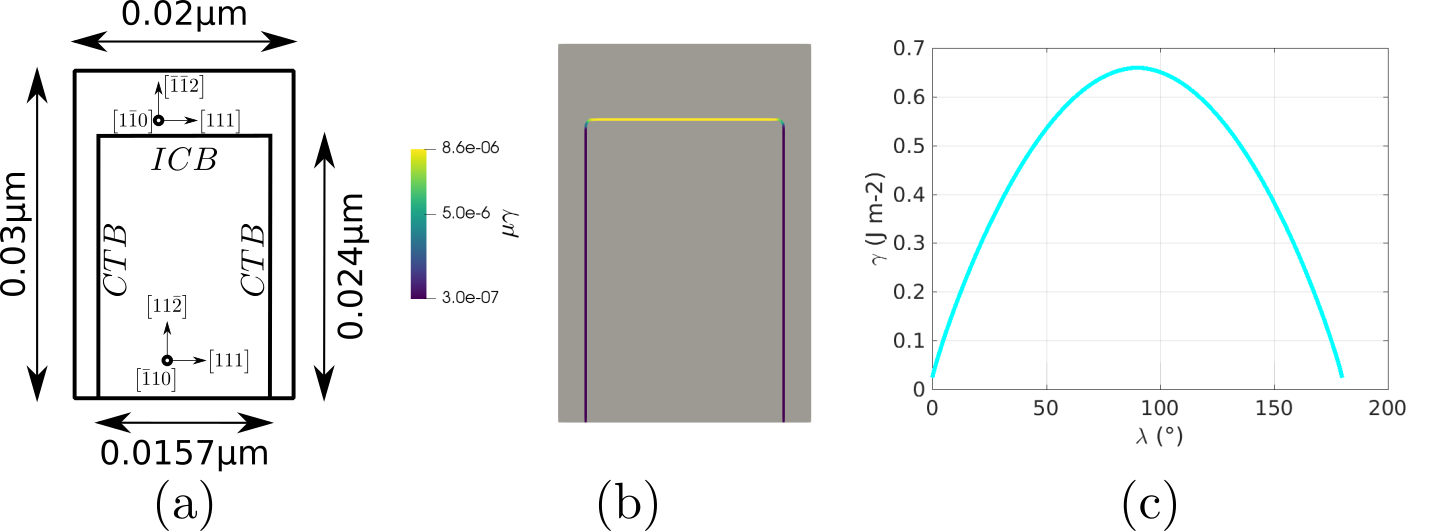}
  \caption{(a) Grain orientations. (b) Initial reduced mobility $[\mu m^2/s]$. (c) Change of GB energy as a function of the GB inclination $\lambda$ with respect to the x-axis evaluated using the GB5DOF code.}\label{fig:AnisoNormalTwin}
\end{figure}

\begin{figure}[h]
  \centering
  \includegraphics[scale=0.65]{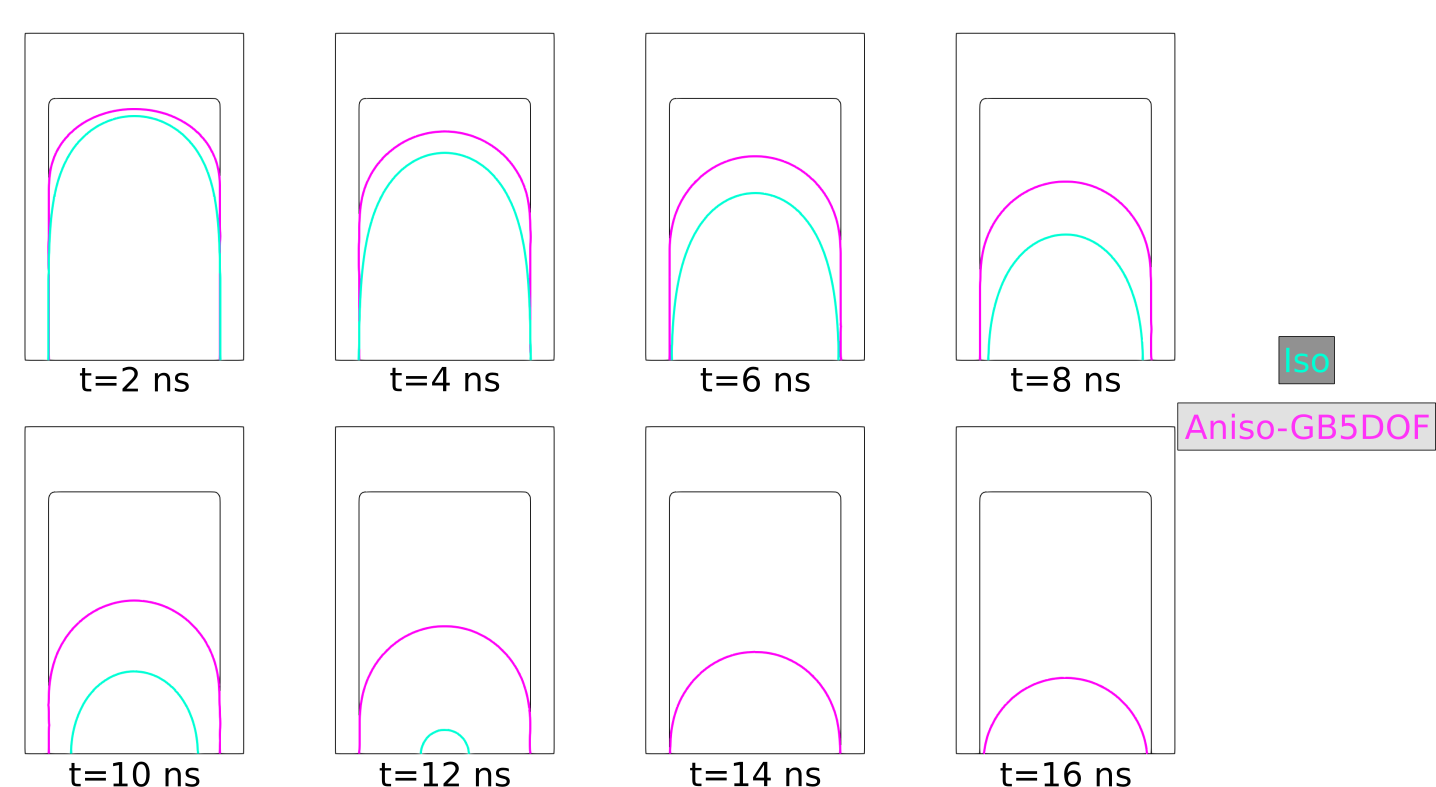}
  \caption{Interface evolution at diferent times $t=2, \ 4, \ 6, \ 8, \ 10, \ 12, \ 14, \ 16 \ ns$.}\label{fig:AnisoNormalTwin_evo}
\end{figure}

The main advantage of the GB5DOF code is that it is possible to characterize coherent and incoherent twin boundaries. These special GBs play an important role on polycrystalline microstructures and their modeling is not frequently discussed at the mesoscopic scale. The next example was firstly proposed by Brown and Ghoniem in \cite{BROWN20094454} and also reproduced at the mesoscopic scale in \cite{hallberg2019modeling}. It consists of two grains composed of two coherent twin boundaries (CTB) and one incoherent boundary (ICB). Fig.~\ref{fig:AnisoNormalTwin} shows the crystallographic orientation, the initial GB energy and the variation of the GB energy as a function of the GB inclination. The Iso and Aniso formulations were used to model the GB movement. For the Aniso formulation the GB5DOF code was used to compute the GB energy all along the simulation. On the other hand, the GB Energy of the Iso case is constant and set to $\gamma = 0.65969 \ J \ m^{-2}$. The evolution of the GB is shown in fig.~\ref{fig:AnisoNormalTwin_evo}. The time step was set to $\Delta t=0.1 \ ns$ and GB mobility was set to $\mu= 1.3e7 \ \mu m^4 \ J^{-1} \ ns^{-1}$ in order to reproduced the velocity of the ICB found by Brown and Ghoniem in \cite{BROWN20094454} $v_{ICB}=1.2 \ m \ s^{-1}$. The movement of the ICB should be uniform and it should respect the flatness of the CTB. The Aniso-GB5DOF simulation enables to respect the expected behavior.

\section{Conclusion}

Different level-set finite element formulations to simulate grain growth were presented and compared in this text, the Isotropic formulation in which the grain boundary mobility and energy are assumed constants being the most used framework in the literature. The Isotropic formulation is able to reproduce mean grain size and grain size distribution evolutions when a moderated anisotropy is involved. 

From the results presented using the triple junction cases, the Anisotropic formulation was the more accurate. The triple junction velocity predictions were the closest to the theoretical values while predicting accurate dihedral angles. In addition, the interfacial energy was always minimized and faster that the other approaches. 

Additionally to these academic configurations, simulations using two different polycrystalline microstructures were performed. First, the initial orientation were generated using an uniform distribution producing an initial Mackenzie-like disorientation distribution. And finally, another example with a textured orientation was considered. It was then illustrated that for a simple microstructure with initial random orientation, an Isotropic formulation can be used and that for a textured configuration, the Anisotropic formulation presents the best behavior in terms of grain morphology, DDF and interfacial energy evolution predictions while keeping a reasonable  efficiency comparatively to the isotropic formulation.

It was also illustrated that the Anisotropic approach is the most versatile approach enabling to take into account the inclination dependence. Future works will be focused on the use of 2D and 3D experimental results concerning 304L and 316L which are currently capitalized. These experimental results will be used to validate the Anisotropic formulation with more complex datasets.

\section*{Acknowledgements}
The authors thank the ArcelorMittal, ASCOMETAL, AUBERT \& DUVAL, CEA, SAFRAN, FRAMATOME, TIMET, Constellium and TRANSVALOR companies and the ANR for their financial support through the DIGIMU consortium and ANR industrial Chair (Grant No. ANR-16-CHIN-0001).

\section*{Data availability}
The raw data required to reproduce these findings cannot be shared at this time as the data also forms part of an ongoing study. The processed data required to reproduce these findings cannot be shared at this time as the data also forms part of an ongoing study.


\bibliographystyle{unsrtnat}
\bibliography{Manuscript}

\end{document}